\documentclass[11pt]{article}
\usepackage[dvipdfmx]{color}
\usepackage{graphicx}
\usepackage{xcolor}
  \usepackage{amsthm,amsfonts}
  \usepackage{amsmath}
\usepackage{chngcntr}
\usepackage{mathabx}
\usepackage{slashed}
\usepackage{ulem}
\usepackage{aliascnt}
\usepackage{hyperref}
\usepackage{cleveref}
\usepackage{cite}
\usepackage{titlesec}

\counterwithin{equation}{section}

\newcommand{\bea}   {\begin{eqnarray}}
\newcommand{\eea}   {\end{eqnarray}}
\def\zzg{${\mathbb Z}_2\times{\mathbb Z}_2$-graded }

\begin{document}
\renewcommand{\thefootnote}{\fnsymbol{footnote}}

\thispagestyle{empty}

\title{Inequivalent ${\mathbb Z}_2^n$-graded brackets, $n$-bit parastatistics and statistical transmutations of supersymmetric quantum mechanics}

\author{M. M. Balbino\thanks{{E-mail: {\it matheusdemb@cbpf.br}}}, \quad I. P. de Freitas\thanks{{E-mail: {\it isaquepfreitas@cbpf.br}}}, \quad R. G. Rana\thanks{{E-mail: {\it ranarg@cbpf.br}}}\quad and\quad
F. Toppan\thanks{{E-mail: {\it toppan@cbpf.br}}}
\\
\\
}
\maketitle

{\centerline{
{\it CBPF, Rua Dr. Xavier Sigaud 150, Urca,}}\centerline{\it{
cep 22290-180, Rio de Janeiro (RJ), Brazil.}}
~\\
\maketitle

\begin{abstract}

Given an associative ring of ${\mathbb Z}_2^n$-graded operators, the number of inequivalent brackets of Lie-type which are compatible with the grading and satisfy graded Jacobi identities is $b_n= n+\lfloor n/2\rfloor+1 $. This follows from the Rittenberg-Wyler and Scheunert analysis of ``color" Lie (super)algebras which is revisited here in terms
of Boolean logic gates. \\The inequivalent brackets, recovered from  ${\mathbb Z}_2^n\times {\mathbb Z}_2^n\rightarrow {\mathbb Z}_2$ mappings, are defined by consistent sets of commutators\slash anticommutators describing particles accommodated into an $n$-bit parastatistics (ordinary bosons/fermions correspond to $1$ bit).
Depending on the given graded Lie (super)algebra, its graded sectors can fall into different classes of equivalence
expressing different types of particles (bosons, parabosons, fermions, parafermions). As a consequence,  the assignment of certain ``marked" operators to a given graded sector is a further mechanism to induce inequivalent graded Lie (super)algebras (the basic examples of quaternions, split-quaternions and biquaternions illustrate these features). \\
As a first application we construct ${\mathbb Z}_2^2$ and ${\mathbb Z}_2^3$-graded quantum Hamiltonians which respectively admit $b_2=4$ and $b_3=5$ inequivalent multiparticle quantizations  
(the inequivalent parastatistics are discriminated by measuring the eigenvalues of certain observables in some  given states).  The extension to ${\mathbb Z}_2^n$-graded quantum Hamiltonians for $n>3$ is immediate.\\
As a main physical application we prove that the ${\cal N}$-extended, one-dimensional supersymmetric and superconformal quantum mechanics, for ${\cal N}=1,2,4,8$,  are respectively described by $s_{\cal N}=2,6,10,14 $  alternative formulations based on the inequivalent graded Lie (super)algebras.  The $s_{\cal N}$ numbers correspond to all possible ``statistical transmutations" of a given set of supercharges
which, for  ${\cal N}=1,2,4,8$, are accommodated into a ${\mathbb Z}_2^n$-grading with $n=1,2,3,4$ (the identification is  ${\cal N}= 2^{n-1}$).\\
In the simplest ${\cal N}=2$ setting (the $2$-particle sector of the de Alfaro-Fubini-Furlan deformed oscillator with $sl(2|1)$ spectrum-generating superalgebra), the ${\mathbb Z}_2^2$-graded parastatistics imply a degeneration of the energy levels which cannot be reproduced by ordinary bosons/fermions statistics. 
\end{abstract}
\vfill
\rightline{CBPF-NF-002/23}

\section{Introduction}

The main physical application of this paper is the computation of the ``statistical transmutations" of the ${\cal N}$-extended one-dimensional Supersymmetric Quantum Mechanics.  For ${\cal N}=1,2,4,8$ the ${\cal N}$ fermionic supercharges can be expressed as (para)bosons/(para)fermions entering an $n$-bit parastatistics (the identification being ${\cal N}=2^{n-1}$ for $n=1,2,3,4$). It is further shown that in the simplest ${\cal N}=2$ setting (the multiparticle sector of a statistically transmuted Superconformal Quantum Mechanics with de Alfaro-Fubini-Furlan oscillator term),
the degeneracy of the energy spectrum of the paraparticles cannot be reproduced by ordinary bosons/fermions.\par
~\par
The starting mathematical framework is an associative ring of ${\mathbb Z}_2^n$-graded operators which possesses addition and multiplication. Several questions are addressed. The first one concerns the $b_n$ number of inequivalent graded Lie brackets which are compatible with the ${\mathbb Z}_2^n$ associative grading (the brackets are defined in terms of commutators/anticommutators which describe particles accommodated into an $n$-bit parastatistics, with ordinary bosons/fermions corresponding to $1$ bit). Consistency conditions require the brackets
to satisfy graded Jacobi identities, while the operators of the associative ring obey a graded Leibniz rule.
The formula for $b_n$ can be recovered from the analysis of Rittenberg-Wyler in \cite{{riwy1},{riwy2}} and Scheunert in  \cite{sch}. In those works the notions of
${\mathbb Z}_2^n$-graded ``color" Lie algebras and superalgebras were introduced as extensions and generalizations of the ordinary ${\mathbb Z}_2$-graded Lie superalgebras defined in \cite{kac} (the Rittenberg-Wyler construction is here revisited for our scopes and in particular, for motivations discussed in the following,  the inequivalent graded Lie (super)algebras are reformulated in terms of Boolean logic gates).\par
~\par
The $b_n$  number of inequivalent, ${\mathbb Z}_2^n$-graded compatible, brackets is expressed in terms of the floor function. One gets
\bea\label{introfloor}
b_n&=& n+\lfloor n/2\rfloor+1,\qquad {\textrm{so that}} \quad b_0=1,~b_1= 2, ~b_2=4,~b_3=5, ~b_4=7, \ldots .
\eea
The $b_n$ values represent a model-independent {\it lower bound} on the number of inequivalent graded Lie (super)algebras induced by a given set of ${\mathbb Z}_2^n$-graded operators. For a specific model, the number of
induced graded Lie (super)algebras is $c_n$, where
\bea\label{introbound}
c_n&\geq &b_n.
\eea 
This is a consequence of the fact that the graded sectors of Lie-type brackets define different types of particles  (bosons, parabosons, fermions, parafermions). If the original set of graded operators falls into different classes of equivalence, the assignment of certain ``marked" operators to a given type of particle is a further mechanism to induce  inequivalent graded Lie (super)algebras. This feature is illustrated with the computations of the inequivalent graded Lie (super)algebras induced by quaternions and split-quaternions (for $n=2$) and biquaternions (for $n=3$), the result being 
\bea\label{introquat}
\begin{array}{ll}
{\textrm{for quaternions:}}&~~~~~ c_2=~4=b_2;\\
{\textrm{for split-quaternions:}}&~~~~~ c_2=~6>b_2=4;\\
{\textrm{for biquaternions:}} &~~~~ ~c_3=16>b_3=5
\end{array}
&&
\eea
(in the case of quaternions the three imaginary quaternions are on equal footing, while for split-quaternions one generator is single out, i.e. ``marked", since its square is negative).\par
The computations, presented in Section {\bf 6}, of the statistical transmutations of the ${\cal N}=1,2,4,8$ Supersymmetric Quantum Mechanics  make use of the fact that the operators entering the ${\mathbb Z}_2^n$-graded associative ring are split
into two classes of equivalence: the supercharges $Q_i$'s and the ``empty slots $\emptyset$". It follows that the $s_{\cal N}$ numbers of induced inequivalent graded Lie (super)algebras are
\bea\label{introsusy}
s_{{\cal N}=1} &:=& c_1 =~2 =b_1,\nonumber\\
s_{{\cal N}=2} &:=& c_2 =~6 >b_2,\nonumber\\
s_{{\cal N}=4} &:=& c_3 =10 >b_3,\nonumber\\
s_{{\cal N}=8} &:=& c_4 =14 >b_4.
\eea
~\par
The next topic consists in investigating whether the induced inequivalent graded Lie (super)algebras imply
{\it physically inequivalent} - that is, theoretically distinguishable by performing measurements
 leading to unambiguous results - 
parastatistics. This question became relevant when the first example of a quantum Hamiltonian invariant under the ${\mathbb Z}_2^2$-graded worldline super-Poincar\'e algebra was produced \cite{brdu} (the Hamiltonian under consideration being also invariant under an ``ordinary" ${\cal N}=2$ supersymmetry). The positive answer was given in \cite{top1}. It was shown that, within the Majid's \cite{maj} framework of graded Hopf algebras endowed with a braided tensor product, in the multiparticle sector of a ${\mathbb Z}_2^2$-graded quantum model an observable
(different from the Hamiltonian) could be selected. Measuring  the eigenvalues of this observable on certain eigenstates allows to discriminate the ${\mathbb Z}_2^2$-graded parafermionic
statistics from the ordinary bosons/fermions statistics. This scheme was further applied in \cite{top2} to prove the detectability of the ${\mathbb Z}_2^2$-graded parabosons.\par
~\par
The first physical application of this paper is presented in Section {\bf 5}. It consists in the construction, for any integer $n\in{\mathbb N} $, of a class of ${\mathbb Z}_2^n$-graded quantum Hamiltonians which possess $b_n$ inequivalent parastatistics - that is, they satisfy the (\ref{introbound}) lower bound. For $n=2,3$ the detectability
of all $b_2=4$ and $b_3=5$ parastatistics is explicitly computed. In the framework of graded Hopf algebras endowed with a braided tensor product, observables (different from Hamiltonians)  are introduced; measuring their eigenvalues determines which type of (para)particles are present in a multiparticle sector of the quantum model (the extension of this result to $n>3$ is straightforward). To our knowledge, this is the first time that all $b_3=5$ parastatistics of a ${\mathbb Z}_2^3$-graded quantum Hamiltonian are considered.\par
~\par
After the computation in Section {\bf 6} of the statistical transmutations of the one-dimensional, 
${\cal N}=1,2,4,8$-extended, Supersymmetric Quantum Mechanics, we present in Section {\bf 7} the preliminary investigation concerning the detectability of the induced parastatistics. The framework of graded Hopf algebras endowed with a braided tensor product is applied to, as already mentioned, the ${\cal N}=2$  Superconformal Quantum Mechanics with de Alfaro-Fubini-Furlan oscillator term. This system is chosen for its simplicity since it possesses a unique vacuum and a discrete set of energy eigenvalues, so that the involved combinatorics produce neat results. In the original superconformal setting (before applying the statistical transmutations) the spectrum-generating superalgebra of the model is $sl(2|1)$.  The $s_{{\cal N}=2}= 6$ parastatistics from (\ref{introsusy})
are split into $3$ ordinary bosons/fermions statistics and $3$ parastatistics involving ${\mathbb Z}_2^2$-graded paraparticles. We prove that, at the second excited level of the $2$-particle states,  the $3$ parastatistics involving paraparticles imply a degeneracy of the energy eigenvalue which is not reproduced by the $3$ ordinary statistics.
The consequence is that ${\mathbb Z}_2^2$-graded paraparticles have to be introduced in order to reproduce this behaviour. We point out that this is the first time that  ${\mathbb Z}_2^2$-graded paraparticles are shown to directly affect the energy spectrum of a quantum model.  In the quantum theories presented in \cite{{top1},{top2}} and those in Section {\bf 5}, as mentioned before,  extra observables besides the Hamiltonian have to be measured. The key difference is that the raising operators applied to the \cite{{top1},{top2}} and Section {\bf 5} models are nilpotent, while 
the raising operators of the superconformal quantum mechanics are not.
We postpone to the Conclusions a more thorough assessment of our results and of the works in progress that
the analysis here presented generate. \par
~\par
Before briefly sketching the state of the art concerning mathematical and physical aspects of ${\mathbb Z}_2^n$-graded structures and the context of the present investigation we point out that,
on a separate development, paraparticles  (different from the ones here described) can now be quantum simulated (see, e.g. \cite{parasim}) and
even engineered in the laboratory as described in \cite{paraexp}. It is in the light of these experimental developments that in Section {\bf B} the inequivalent graded brackets are presented in terms of Boolean logic gates. They can be possibly used as a blueprint to engineer ${\mathbb Z}_2^n$-graded paraparticles in the laboratory.\par
~\par
Concerning the mathematical aspects of ${\mathbb Z}_2^n$-graded structures and their physical applications we mention that these topics received, in recent years, increased attention. This recent boost of activity finds its motivation in advances obtained from  different directions (for earlier works based on the \cite{{riwy1},{riwy2},{sch}}
color Lie (super)algebras see \cite{{luri},{vas},{jyw}}). It was recognized in \cite{{aktt1},{aktt2}}  that ${\mathbb Z}_2^2$-graded Lie superalgebras appear as dynamical symmetries of relevant physical  systems such as the non-relativistic L\'evy-Leblond spinors. Classical and quantum models invariant under ${\mathbb Z}_2^2$-graded 
superalgebras have been constructed \cite{{akt1},{brusigma},{akt2},{brdu},{aad},{kuto},{que}} and ${\mathbb Z}_2^2$-graded superspace formulations investigated \cite{{pon},{brusg},{aido1},{aido2},{aikt}}. The connection with parastatistics has been discussed in \cite{{yaji},{tol1},{stvj1},{top1},{top2},{stvj2}}.
In \cite{z2z2sdiv} ${\mathbb Z}_2^2$-graded superdivision algebras were classified and a parafermionic Hamiltonian was described in this context. In the meanwhile, mathematical aspects of the graded (super)algebras continue to be investigated, see e.g. \cite{{isv},{stvdjclass}}. The majority of the papers mentioned above discuss theories invariant under ${\mathbb Z}_2^2$-graded superalgebras; the reason for that is that such theories involve parafermions and, in most cases, correspond to generalizations of supersymmetry. On the other hand, it was pointed out in 
\cite{kuto} that invariance under ${\mathbb Z}_2^2$-graded Lie algebras produces models with only bosonic and parabosonic particles. It was shown in \cite{top2} that a quantum Hamiltonian invariant under different choices of ${\mathbb Z}_2^n$-graded Lie (super)algebras (for $n=0,1,2$ in the given example)  admits several particles' assignments which are compatible with the gradings (bosons for $n=0$, different mixings of bosons/fermions for $n=1$, parabosons or parafermions for $n=2$). Each assignment corresponds to a different set of (para)particles with measurable consequences. This is so because, in the multiparticle sectors, the measurements of chosen observables can discriminate the different cases.
Concerning $n\geq 3$, ${\mathbb Z}_2^n$-graded mathematical structures have been investigated in \cite{{ckp},{brgr1},{brgr2},{luta}} and the first invariant quantum models have started being constructed in \cite{{doai},{aad2}}.\par 
These mathematical, theoretical and experimental advances provided us the motivations to systematically revisit the theory of (inequivalent) ${\mathbb Z}_2^n$-graded Lie brackets and in pointing out their physical consequences. \par
~\par

The scheme of the paper is the following:\par
in Section {\bf 2} we revisit the concept of the graded Lie (super)algebras which are compatible with a ${\mathbb Z}_2^n$-grading assignment of the operators belonging to an associative ring. In Section {\bf 3} we discuss the classes of equivalence of the operators belonging to graded sectors and the notion of marked operators is introduced. The detectability of $n$-bit parastatistics within the framework of graded Hopf algebras endowed with a braided tensor product is discussed in Section {\bf 4}. The first physical application, the construction of a class of ${\mathbb Z}_2^n$-graded quantum Hamiltonians with $b_n$ inequivalent multiparticle quantizations, is presented in Section {\bf 5}. Section {\bf 6} introduces the algebraic statistical transmutations of the ${\cal N}=1,2,4,8$ Supersymmetric Quantum Mechanics. In Section {\bf 7} the detectability of the induced parastatistics in the setting of the Superconformal Quantum Mechanics with $DFF$ oscillator term is shown. The obtained results and an outline of future works are discussed here and in the Conclusions.  Five mathematical Appendices complement the paper. In Appendix {\bf A} we present, for $n=1,2,3,4$, the tables of the $b_n$ inequivalent graded Lie (super)algebras which are compatible with an associative  ${\mathbb Z}_2^n$-graded ring of operators. In Appendix {\bf B}  the graded Lie brackets are presented in terms of Boolean logic gates.  Appendix {\bf C} presents the inequivalent graded Lie (super)algebras induced by quaternions and split-quaternions. Appendix {\bf D} illustrates the connection of the  ${\mathbb Z}_2^3$ grading with the Fano's plane. Appendix {\bf E} presents the $16$ inequivalent graded Lie (super)algebras induced by biquaternions.
\par

\section{Revisiting the construction of ${\mathbb Z}_2^n$-graded Lie brackets}

In this Section we revisit, adapting to our goals and introducing definitions suitable for our scope, the Rittenberg-Wyler \cite{{riwy1},{riwy2}} and Scheunert \cite{sch} constructions of the inequivalent graded Lie brackets based on a ${\mathbb Z}_2^n$ grading.\par
The starting point is an associative ring of ${\mathbb Z}_2^n$-graded operators which possesses addition and multiplication (the standard $+$ and $\cdot$ symbols are used). The $A,B,C,\ldots$ operators are associated with
the $n$-bit gradings $[A]={\alpha}, [B]={\beta},[C]= {\gamma},\ldots$. \par
The bits are $\{0,1\}$ for $n=1$, 
$\{00, 10,01,11\}$ for $n=2$, $\{000, 100, 010, 001, 110, 101, 011, 111\}$ for $n=3$ and so on. The multiplication satisfies
\bea
[A\cdot B] = \alpha+\beta &&{\textrm{with $mod~2$ addition}}.
\eea
The grading  of the identity operator ${\mathbb I}$ is  
\bea\label{zero}
[{\mathbb I}]={\underline{0}}, &&{\textrm{so that ~~${\underline{0}}+\alpha=\alpha+{\underline{0}}=\alpha$ ~~ for any~~ $\alpha$.}}
\eea
For $n=1,2,3,\ldots $,  the ${\underline{0}} $ grading is respectively given by $0, ~00,~000, ~\ldots$.\par
A graded Lie bracket, denoted for $A,B$ as $(A,B)$, is either a commutator or an anticommutator defined as
\bea\label{gradedbrackets}
(A,B)&:=&A\cdot B -(-1)^{\langle\alpha,\beta\rangle} B\cdot A,
\eea
where the bilinear mapping $\langle\alpha,\beta\rangle$ takes values $0,1$ $mod~2$.\par
The bracket is (anti)symmetric in accordance with
\bea
(B,A) = (-1)^{\langle\alpha,\beta\rangle+1}(A,B).
\eea
In \cite{{riwy1},{riwy2}} two conditions on the bilinear mapping are imposed. The first condition simply states that
(\ref{gradedbrackets}) is a commutator or an anticommutator; it implies, for the $n$-bit grading,
\bea 
\langle\beta,\alpha\rangle &=&\pm\langle\alpha,\beta\rangle\quad mod~2.
\eea 
Without loss of generality one can assume a symmetric choice for $\langle\alpha,\beta\rangle$, so that 
\bea\label{symmpres}
\langle ,\rangle &:& {\mathbb Z}_2^n\times{\mathbb Z}_2^n\rightarrow {\mathbb Z}_2
\eea
is represented by $2^n\times 2^n$ symmetric matrices with $0,1$ entries. \par
It is worth pointing out that in the original \cite{{riwy1},{riwy2}} papers
the brackets of color Lie algebras were presented, using a different convention, as antisymmetric matrices with $0,\pm 1$ entries. As shown later on, the convenience of the (\ref{symmpres}) symmetric choice is that it allows to present the inequivalent graded Lie brackets in terms of Boolean logic operators.\par
The second \cite{{riwy1},{riwy2}} consistency condition for the $\langle,\rangle$ scalar product, under the (\ref{symmpres}) assumption with the symmetric choice, simply reads
\bea\label{condition2}
\langle \alpha , \beta+\gamma\rangle &=& \langle \alpha,\beta\rangle +\langle \alpha,\gamma\rangle \quad ~ mod ~~ 2.
\eea
The justification for imposing this condition is the requirement that a graded Leibniz rule should be satisfied.  Indeed, by
assuming
\bea\label{leibnizrule}
(A,BC) &=& (A,B)\cdot C + (-1)^{\varepsilon_{AB} } B\cdot(A,C)\qquad {\textrm{with}}\quad \varepsilon_{AB}=0,1,
\eea
straightforward manipulations prove that (\ref{leibnizrule}) implies  both the identification
\bea
\varepsilon_{AB} &=& \langle \alpha,\beta\rangle
\eea
and the condition (\ref{condition2}) to be fulfilled.\par
The associativity of the multiplication implies that the graded brackets (\ref{gradedbrackets}), expressed in terms of the symmetric scalar product $\langle,\rangle$ which satisfies (\ref{symmpres}) and (\ref{condition2}), obey the graded Jacobi identities
\bea\label{jacobi}
(-1)^{\langle \gamma,\alpha\rangle} (A,(B,C))+(-1)^{\langle \alpha,\beta\rangle} (B,(C,A))+(-1)^{\langle \beta,\gamma\rangle} (C,(A,B))&=&0.
\eea
Therefore, the $(,)$ brackets define a graded Lie (super)algebra which is compatible with the ${\mathbb Z}_2^n$ grading of
the ring of operators.\par
For the cases under consideration here $\alpha,\beta$ are the $n$-component vectors $\alpha^T =(\alpha_1,\alpha_2,\ldots,\alpha_n)$ and $\beta^T =(\beta_1,\beta_2,\ldots,\beta_n)$. Their respective $i$-th components $\alpha_i$ and $\beta_i$ correspond to $1$-bit (either $0$ or $1$). In this setting the following definitions can be conveniently introduced:\\
~\\
- {\it definition I}: a ``${\mathbb Z}_2^n$-graded {\it compatible}" Lie algebra is defined by the (\ref{gradedbrackets}) bracket;  the symmetric scalar product $\langle,\rangle$ satisfies (\ref{symmpres}) and (\ref{condition2}) so that, as a consequence, the graded Jacobi
identities (\ref{jacobi})  hold. The further requirement is that the $2^n\times 2^n$ symmetric matrix (\ref{symmpres}) has all vanishing diagonal elements: $\langle\alpha,\alpha\rangle =0$ for any $\alpha$;\\
~\\
- {\it definition II}: a ``${\mathbb Z}_2^n$-graded {\it compatible}" Lie superalgebra satisfies all the above properties with the extra condition that at least one diagonal element of the scalar product is nonzero: $\langle\alpha,\alpha\rangle =1$ for at least one vector $\alpha$.\par
~\\
{\it Comments}: the difference between graded Lie algebras and superalgebras is relevant in physical applications. Graded Lie algebras only describe (para)bosonic particles, while graded Lie superalgebras include (para)fermions which obey the
Pauli exclusion principle.
\\The motivation for introducing the term ``{compatible}" in definitions $I$ and $II$ will be presented in the next subsection.

\subsection{On the inequivalent, ${\mathbb Z}_2^n$-graded compatible, Lie (super)algebras}

The \cite{{riwy1},{riwy2},{sch}} analysis (which can be easily reproduced) allows to define the inequivalent,
${\mathbb Z}_2^n$-graded compatible, Lie (super)algebras which satisfy the above definitions $I$ and $II$. The scalar products
$\langle,\rangle$ can be brought to a convenient set of canonical forms. Other equivalent presentations are obtained by exchanging the $n$-bit graded sectors (this is tantamount to permute rows and columns in the scalar product symmetric matrix (\ref{symmpres}) in such a way to bring the matrix into one of the canonical forms). \par
The complete list of inequivalent canonical forms for $\langle\alpha,\beta\rangle$ can be expressed as follows.\par
~\\
$I$ - For ${\mathbb Z}_2^n$-graded compatible Lie algebras:\\
~\\
$\langle\alpha,\beta\rangle _0=0$ (it is an ordinary Lie algebra induced by a vanishing scalar product matrix) and\\
~\\
$\langle\alpha,\beta\rangle _{n+1+k}=\sum_{j=0}^{j=k} (\alpha_{2j+1}\beta_{2j+2} +\alpha_{2j+2}\beta_{2j+1}) \quad mod ~ 2$, $~~$ for $k=0,1,2, \ldots, \lfloor n/2\rfloor-1$,
\bea\label{algebras}{\textrm{\qquad ~~~~~ where the maximal value for $k$ is expressed in terms of the floor function.}}&&\eea
$II$ - For ${\mathbb Z}_2^n$-graded compatible Lie superalgebras:
\bea\label{superalgebras} {\textrm{
$\langle\alpha,\beta\rangle _k=\sum_{j=1}^{j=k} (\alpha_{j}\beta_{j})\quad mod ~ 2$, $~~$ for $k$ taking the values $k= 1,2,\ldots n$.}}&&
\eea
Therefore, the total number $b_n$ of inequivalent, ${\mathbb Z}_2^n$-graded compatible, Lie (super)algebras which obey definitions $I$ and $II$ is
\bea\label{floor}
b_n&=& n+\lfloor n/2\rfloor+1 .
\eea
From (\ref{algebras},\ref{superalgebras}) the corresponding scalar products are labeled as $\langle,\rangle_r$ with
$r=0,1,2,\ldots, n+\lfloor n/2\rfloor$.\par
For $n=1,2,3,\ldots$, the series produced by $b_n$ is
\bea
&2,4,5,7,8,10,\ldots&.
\eea
Tables of the representatives of the inequivalent $2^n\times 2^n$ scalar products symmetric matrices are given in Appendix {\bf A} for $n=1,2,3,4$.\par
~\par
The set of the inequivalent ${\mathbb Z}_2^n$-graded compatible Lie (super)algebras is given by the union of ${\mathbb Z}_2^m$-graded Lie (super)algebras for $m=0,1,2,\ldots, n$:
\bea
\{ {\textrm {${\mathbb Z}_2^n$-graded compatible L(S)A's}}\} &=& \bigcup_{m=0,\ldots, n}\{ {\textrm{${\mathbb Z}_2^m$-graded L(S)A's}}\}.
\eea
The notion of a ${\mathbb Z}_2^m$-graded Lie (super)algebra ``embedded" into an $n$-bit decomposition  is based on the following property: for the minimal integer $m$, its graded sectors defining  the $2^n\times 2^n$ symmetric  $\langle,\rangle$ scalar product  matrix (\ref{symmpres}) are consistently grouped into $n-m$ bits which produce a reduced $2^m\times 2^m$ symmetric matrix.\par
It follows that
a ${\mathbb Z}_2^0$-graded Lie algebra ($m=0$) is an ordinary Lie algebra embedded into $n$-bit via the
$\langle,\rangle_0$ scalar product which corresponds to a symmetric matrix with only $0$  entries; similarly, a
${\mathbb Z}_2^1$-graded Lie superalgebra is an ordinary Lie superalgebra recovered from the $\langle,\rangle_1$ scalar product.\par
An example of a ${\mathbb Z}_2^2$-graded Lie superalgebra for a $3$-bit assignments is given by the $3_4$ case in formula (\ref{inequivalent3}).  Its graded sectors can be consistently grouped as
\bea
&\{000,001\}\rightarrow\{ 00\},\quad \{100,101\}\rightarrow \{10\},\quad\{010,011\}\rightarrow \{01\},\quad\{110,111\}\rightarrow \{11\};&
\eea
the produced output is the $2$-bit graded superalgebra $2_4$ of formula (\ref{2cases}).\par
For $m=n$ the $2^n\times 2^n$ symmetric matrix (\ref{symmpres}) cannot be reduced. This happens in the following
cases:\\
{\it i}) for any integer $n$, a unique ${\mathbb Z}_2^n$-graded Lie superalgebra is obtained from the scalar product $\langle,\rangle_n$ of formula (\ref{superalgebras});\\
{\it ii}) for odd integers $n$ there is no ${\mathbb Z}_2^n$-graded Lie algebra; for even integers $n=2s$, a unique ${\mathbb Z}_2^n$-graded Lie algebra is obtained from the scalar product $\langle,\rangle_{\frac{3}{2}n}=\langle,\rangle_{3s}$ of formula (\ref{algebras}).\par
~\par
{\it Remark}: the usefulness of the introduced notion of {\it $Z_2^n$-graded compatible Lie (super)algebras } lies on the fact that,
in the physical applications discussed in the following (inequivalent multiparticle quantizations, statistical transmutations of supersymmetric quantum mechanics), the contributions coming from each one of the $b_n$ compatible graded Lie (super)algebras must be added up.

\subsection{The $n$-bit Boolean logic gates presentation of the graded Lie brackets}

The (\ref{symmpres}) scalar product which defines the inequivalent graded Lie brackets admits a presentation in
terms of Boolean logic gates. This alternative formulation can  in principle offer a practical set of instructions  for simulating or engineering in the laboratory the desired set of graded brackets (we already recalled in the Introduction the recent experimentalists' advances, see \cite{{parasim},{paraexp}}, in manipulating parastatistics).  This presentation can then be potentially applied to the two main physical results obtained in the following: quantum toy models with theoretically detectable paraparticles and the statistical transmutations of supersymmetric quantum mechanics. \par
The formulation of the inequivalent graded Lie brackets in terms of Boolean logic gates requires a few steps. At first the graded sectors entering the scalar product tables are rearranged in a Gray code presentation (only one bit  changes from one graded sector to the next one). This rearrangement permits to make use of the Karnaugh maps \cite{kar} which, further simplified, allow to express the graded-bracket tables in terms of the logical gates ``AND", ``OR" and ``XOR". The details of the construction are given in Appendix {\bf B}. 

\section{Classes of equivalence of the graded sectors}

The $2^{n}-1$ nonzero graded sectors of an associative ${\mathbb Z}_2^n$-graded ring of operators are all on equal footing and belong to the same class of equivalence. This feature is well illustrated, e.g., in the $n=3$ case by the assignment of
the nonvanishing entries of the $8\times 8$ matrices as shown in (\ref{3bit8x8}). Its $7$ nonzero graded sectors can also be
represented, see figure (\ref{fano}), as labeled vertices of a Fano's plane.\par
The introduction of a ${\mathbb Z}_2^n$-graded bracket as defined in (\ref{gradedbrackets}) implies,
on the other hand, that the nonzero graded sectors not necessarily belong, with respect to the brackets,  to the same class of equivalence.   By inspecting,
for instance, the brackets induced by the $3_4$ scalar product  given in (\ref{inequivalent3}) one can notice that
the column/row of the $001$ graded sector is given by all $0$'s. It follows from (\ref{gradedbrackets}) that the
$001$-graded particles are bosons (they commute with every other particle).  Concerning the remaining nonzero graded sectors of the $3_4$ scalar product one can further notice that $110, 111$ define parabosons, while $101, 011, 100, 010$ define parafermions. Therefore, the nonzero graded sectors of the $3_4$ superalgebra fall into $3$ distinct classes of equivalence describing bosons, parabosons and parafermions.\par
This analysis can be repeated for all $n=1,2,3,4$-bit inequivalent graded Lie (super)algebras presented in Appendix {\bf A}. Their respective numbers of distinct classes of equivalence are:
\bea
&\begin{array}{ccccccc}
1_1:~ {\bf 1},~&1_2: ~{\bf 1};~&~&~&~&~&\\
2_1:~{\bf 1},~&2_2:~{\bf 2},~&2_3:~{\bf  1},~&2_4:~{\bf 2};~&&&\\
3_1:~ {\bf 1},~&3_2: ~{\bf 2},~&3_3:~{\bf 2},~&3_4:~{\bf 3},~&3_5:~{\bf  2};~&~&\\
4_1:~ {\bf 1},~&4_2: ~{\bf 2},~&4_3:~{\bf 1},~&4_4:~{\bf 2},~&4_5:~{\bf  3},~&4_6:~{\bf 3 },~&4_7:~{\bf 2}.
\end{array}&
\eea 
These classes of equivalence, defined for the (\ref{symmpres}) scalar product, characterize the types of
particles as
(para)bosons or (para)fermions.  \par
~\par
Concerning a given ${\mathbb Z}_2^n$-graded ring of operators, classes of equivalence can be introduced for the operators belonging to the nonzero graded sectors. The operators are on equal footing if their table of multiplications is preserved when they are interchanged (possibly, up to a sign normalization). In this case the operators all belong to the same equivalence class. On the other hand,  two or more classes of equivalence accommodate the operators if some of them are distinguished (they can be referred to as the ``marked" operators).\par
If the ${\mathbb Z}_2^n$-graded operators are on equal footing and belong to the same class of equivalence it makes no difference which operator is assigned to which nonzero graded sector. Therefore, the number of the induced, inequivalent, compatible graded Lie (super)algebras coincides with $b_n$ given in (\ref{floor}). On the other hand if the operators are accommodated into two or more classes of equivalence, the number of induced, inequivalent, ${\mathbb Z}_2^n$-graded compatible Lie  (super)algebras can be larger. It will be denoted as $c_n$, with $c_n\geq b_n$.\par
~\par
These features are exemplified by the computations of the induced inequivalent graded Lie (super)algebras  presented in Appendix {\bf C} (for the ${\mathbb Z}_2^2$-graded quaternions and split-quaternions) and  Appendix
{\bf E} (for the ${\mathbb Z}_2^3$-graded biquaternions).  In all these cases the identity operator is assigned to the zero grading ${\underline 0}$. Concerning the remaining operators associated with the nonzero gradings we have:
\\
~\\
- $1$ class of equivalence for the quaternions since the three imaginary quaternions are on equal footing;\\
- $2$ classes of equivalence for the split-quaternions since the three generators, besides the identity, are split as $1+2$ (the square of the marked operator is $-{\mathbb I}$, while the squares of the $2$ remaining operators are $+{\mathbb I}$);\\
- $3$ classes of equivalence for the biquaternions (the seven generators besides the identity being split as $1+3+3$). 
\bea&&\eea

The results of the computations  reported in Appendices {\bf C} and {\bf E} are
\bea
\begin{array}{ll}
{\textrm{for quaternions:}}&~~~~~ c_2=~4=b_2;\\
{\textrm{for split-quaternions:}}&~~~~~ c_2=~6>b_2=4;\\
{\textrm{for biquaternions:}} &~~~~ ~c_3=16>b_3=5.
\end{array}
&&
\eea
~\par
This analysis is our starting point for proceeding at physical applications. \par
In Section {\bf 5} we construct, for any integer $n$, ${\mathbb Z}_2^n$-graded quantum Hamiltonians obtained in terms of $2^n$ pairs of creation/annihilation operators.  The pairs are chosen to be on equal footing so that the number of induced inequivalent graded Lie (super)algebras is $b_n$. This is also the number of parastatistics supported by these Hamiltonians. Explicit $n=2,3$ computations prove that $b_2$ and respectively $b_3$ are the inequivalent mutiparticle quantizations (with detectable paraparticles) for these Hamiltonians.\par
A second application presented in Section {\bf 6} (the counting of the statistical transmutations of the ${\cal N}$-extended, one-dimensional, supersymmetric and superconformal quantum mechanics) is also based on this analysis. The ${\cal N}$ supercharges $Q_i$, for ${\cal N}=1,2,4,8$, are respectively accommodated, for $n=1,2,3,4$, into a ${\mathbb Z}_2^n$-graded associative ring of operators. The Hamiltonian $H$, being an observable, is assigned to the zero sector ${\underline 0}$. Concerning the $2^n-1$ nonzero sectors, $2^{n-1}$ are occupied by the supercharges while the
remaining $2^{n-1}-1$ sectors are originally left unoccupied by the supersymmetry generators (they are later occupied by
the descendant operators $Q_iQ_j$ for $i\neq j$). The grading assignment of the supercharges produces inequivalent classes of equivalence (one can loosely says that the unoccupied graded sectors are marked).  The computations presented in Section {\bf 6} give that the induced inequivalent graded Lie (super)algebras, denoted as  $s_{\cal N}$, are
\bea
&s_{{\cal N}=1} := c_1 = 2,\qquad s_{{\cal N}=2} := c_2= 6, \qquad s_{{\cal N}=4} := c_3 = 10,\qquad 
s_{{\cal N}=8} := c_4 = 14. &
\eea

\section{Detectable $n$-bit parastatistics}

A single-particle quantum Hamiltonian belonging to a ${\mathbb Z}_2^n$-graded associative ring of operators
admits, following the construction presented in Section {\bf 3}, a total number of
\bea\label{lower}
c_n&\geq & b_n
\eea
inequivalent, graded Lie (super)algebras formulations. The number $c_n$ is model-dependent, with the lower bound $b_n$ given in (\ref{floor}).\par
For the single-particle theory no measurement can discriminate the different alternatives; we end up with $c_n$ {\it physically equivalent} descriptions of the same quantum model; 
choosing one description instead of another one is just a matter of taste and/or convenience.\par
On the other hand, in the First Quantized formulation, the multiparticle sector of the graded quantum Hamiltonian allows discriminating the different alternatives. The different choices of (anti)commutation relations among particles
imply several consistent $n$-bit parastatics. The alternative parastatatistics produce physically measurable consequences.\par
 In this Section we present the general framework which is later applied to analyze ${\mathbb Z}_2^n$-graded quantum models. This approach  was used in \cite{top1} and \cite{top2} to prove the detectability of, respectively, the ${\mathbb Z}_2^2$-graded parafermions and the ${\mathbb Z}_2^2$-graded parabosons.  Following Majid, see\cite{maj}, the parastatistics is encoded in a graded Hopf algebra endowed with a braided tensor product (the connection between this formulation and the more traditional  approach to parastatistics based on the  \cite{gre}
trilinear relations has been discussed in \cite{{anpo},{kada}}).\par
Let  $A,B,C,D$ be ${\mathbb Z}_2^n$-graded operators whose respective $n$-bit gradings are $\alpha,\beta,\gamma,\delta$. The braided tensor product, conveniently denoted as ``${\otimes}_{br}$", is defined to satisfy
the relation
\bea\label{braidedtensor}
(A\otimes_{br} B)\cdot (C\otimes_{br} D) &=& (-1)^{\langle\beta,\gamma\rangle}(AC)\otimes_{br} (BD),
\eea
where the $(-1)^{\langle\beta,\gamma\rangle}$ sign on the right hand side depends on the symmetric scalar product
introduced in (\ref{symmpres}). Due to the presence of the scalar product, the braided tensor product can be consistently applied to a ${\mathbb Z}_2^n$-graded compatible Lie (super)algebra ${\mathfrak g}$ and to its Universal Enveloping Algebra $U:={\cal U} ({\mathfrak g})$. The Universal Enveloping Algebra $U$ is a graded Hopf algebra.  \par
Among the graded Hopf algebra structures and costructures, the coproduct $\Delta$ is the relevant operation which allows, in physical applications, to construct multiparticle states.
The coproduct map
\bea\label{coproduct}
\Delta &:& U\rightarrow U\otimes_{br} U
\eea 
satisfies the coassociativity property
\bea\label{coassoc}
 &\Delta^{m+1}:=  (\Delta\otimes_{br} {\mathbf 1})\Delta^m=({\mathbf 1}\otimes_{br} \Delta)\Delta^m \qquad ~ {\textrm{(where~ $\Delta^1\equiv \Delta$)}}
\eea
and the comultiplication
\bea\label{comult}
\Delta(u_1u_2) &=& \Delta(u_1)\cdot \Delta(u_2) \qquad {\textrm{for any ~$u_1,u_2\in U$.}}
\eea
The action of the coproduct on the identity ${\bf 1}\in {\cal U}({\mathfrak{g}})$ and on the primitive elements $g\in{\mathfrak{g}}$ is given by
\bea\label{coproductaction}
\Delta({\bf 1})={\bf 1}\otimes_{br}{\bf 1}, \quad&&\quad
\Delta(g) = {\bf 1}\otimes_{br} g+g\otimes_{br} {\bf 1}.
\eea 
The result of the map $\Delta(u)$ on any given element $u\in U$ is recovered from  the operations (\ref{coproductaction}) and from the comultiplication (\ref{comult}).\par
In physical applications, typical primitive elements satisfying the second equation of (\ref{coproductaction}) are the Hamiltonians and the creation/annihilation operators. The coproduct $\Delta=\Delta^1$ is used to construct $2$-particle states, while $\Delta^m$ from (\ref{coassoc}) is employed to construct $(m+1)$-particle states. In the quantum models discussed in the following Sections of the paper, an $m$-particle Hilbert space ${\cal H}_m$ possesses a Fock vacuum $|vac\rangle_m$ (in the single-particle sector $|vac\rangle\equiv|vac\rangle_1$). We have
\bea
{\cal H}_m &\subset& {\cal H}_1^{\otimes_{br}^m}.
\eea
Let $R$ be a representation
\bea
R &:& {\cal U}({\mathfrak g}) \rightarrow End({\cal H}_1).
\eea
We denote with a hat the represented operators:
\bea
{\widehat u} &:=& R(u)\in End({\cal H}_1).
\eea
This notation is extended to the $m$-particle Hilbert spaces so that, e.g., ${\widehat{\Delta (u)}}\in End({\cal H}_2)$.
\\
~\\
Let us now now assume $a^\dagger\equiv A\in {\mathfrak{g}}$ to be a nilpotent creation operator with $n$-bit grading $\alpha$:
\bea
A^2 &=& 0, \qquad {\textrm{where ~$[A] =\alpha$ ~and~ $\langle\alpha,\alpha\rangle=0,1$.}}
\eea
The corresponding $2$-particle creation operator is $\Delta(A)$ which, from (\ref{coproductaction}), satisfies
\bea
\Delta(A) &=& {\mathbf 1}\otimes_{br} A+ A\otimes_{br} {\mathbf 1}.
\eea
The comultiplication (\ref{comult}) implies
\bea
A^2=0& \Rightarrow & \Delta(A^2)=\Delta (A)\cdot \Delta(A) = \left(1+(-1)^{\langle\alpha,\alpha\rangle}\right) (A\otimes_{br}A).
\eea
It results, for (para)fermions with $\langle\alpha,\alpha\rangle=1$,
\bea\label{pauli}
{\textrm{ $A^2=0$ ~~and ~~ $\langle\alpha,\alpha\rangle=1$}} ~&\Rightarrow & ~\Delta(A^2)=0.
\eea 
Formula (\ref{pauli}) encodes, in the graded Hopf algebra formalism endowed with a braided tensor product, the Pauli exclusion principle of (para)fermions in the multiparticle sector. \par
~\par
Some further comments are in order. The (\ref{braidedtensor}) signs resulting from the braided tensor product
enter the construction of the vectors $v\in {\cal H}_m$ belonging to the $m$-particle Hilbert spaces with $m=2,3,\ldots $. For a ${\mathbb Z}_2^n$-graded compatible Lie (super)algebra an admissible observable operator $\Omega\in End({\cal H}_m)$ must satisfy the following two necessary conditions:
\bea
~{\textrm{{\it ~ i}) }}&& {\textrm{to be hermitian: \quad $\quad {\Omega^\dagger =\Omega}$ \quad and}}
\nonumber\\ 
{\textrm{{\it ii})}}&&{\textrm{to be zero-graded: \quad ${[\Omega] ={\underline 0}}$.}} 
\eea
These conditions guarantee that the $\Omega$ eigenvalues are real.\par
For the specific quantum models investigated in Section {\bf 5} (the quantum Hamiltonians supporting $b_n$ inequivalent parastatistics)  we construct admissible observables which, applied on certain given multiparticle states, produce the (\ref{braidedtensor}) signs as eigenvalues. Therefore, the inequivalent parastatistics can be physically discriminated. For our scopes it is sufficient  to present the construction and the results for the $2$-particle sector.
A different mechanics appears in the statistical transmutations of the superconformal quantum mechanics with de Alfaro-Fubini-Furlan \cite{dff} oscillator terms: different contributions of physically inequivalent parastatistics are directly inferred from the degeneracy
of the energy eigenvalues.

\section{Construction of ${\mathbb Z}_2^n$-graded quantum Hamiltonians with \\$b_n$ inequivalent multiparticle quantizations}

In this Section we present a construction of a class of quantum Hamiltonians, belonging to an associative ${\mathbb Z}_2^n$-graded ring of operators, whose $b_n$ induced inequivalent graded Lie (super)algebras satisfy the lower bound (\ref{lower}).  We prove, with explicit $n=2,3$ computations, that the graded Lie (super)algebras imply for these Hamiltonians detectable parastatistics resulting in, respectively, $b_2=4$ and
$b_3=5$ inequivalent multiparticle quantizations.\par
For any integer $n$ we introduce $2^n$ pairs of annihilation/creation operators $a_{i;n}, a_{i;n}^\dagger$ labeled as
$i=0,1,2,\ldots, 2^n-1$. These annihilation/creation operators, given by $2^{n+1}\times 2^{n+1}$ constant matrices with $0,1$ entries, are expressed as tensor products of $2\times 2$ matrices. They are constructed in terms of
the {\footnotesize{$I=\left(\begin{array}{cc}1&0\\0&1\end{array}\right),~
Y=\left(\begin{array}{cc}0&1\\1&0\end{array}\right)$
}} matrices
introduced in (\ref{letters}) and by the $\beta,~\gamma$ matrices defined as
{\small{\bea
\beta=\left(\begin{array}{cc}0&1\\0&0\end{array}\right),&&
\gamma=\left(\begin{array}{cc}0&0\\1&0\end{array}\right).
\eea}}

For $n=1$ (i.e., the $1$-bit case) we set the $4\times 4$ matrices
\bea 
a_{0;1}^\dagger =I\otimes \gamma,&& a_{0;1}=(a_{0;1}^\dagger)^\dagger= I\otimes \beta,\nonumber\\
~a_{1;1}^\dagger = Y\otimes \gamma, && a_{1;1}=(a_{1;1}^\dagger)^\dagger =Y\otimes\beta.
\eea

For the $n=2$ ($2$-bit case) we set the $8\times 8$ matrices
\bea \label{creation2}
a_{0;2}^\dagger =I\otimes I\otimes\gamma,&& a_{0;2}=(a_{0;2}^\dagger)^\dagger= I\otimes I\otimes \beta,\nonumber\\
~a_{1;2}^\dagger = I\otimes Y\otimes \gamma, && a_{1;2}=(a_{1;2}^\dagger)^\dagger =I\otimes Y\otimes\beta,\nonumber\\
a_{2;2}^\dagger =Y\otimes I\otimes\gamma,&& a_{2;2}=(a_{2;2}^\dagger)^\dagger= Y\otimes I\otimes \beta,\nonumber\\
~a_{3;2}^\dagger = Y\otimes Y\otimes \gamma, && a_{3;2}=(a_{3;2}^\dagger)^\dagger =Y\otimes Y\otimes\beta.
\eea
The general construction of the $n$-bit annihilation/creation operators  $a_{i;n}, ~ a_{i;n}^\dagger$ now becomes obvious: the annihilation operator $a_{i;n}$ is the hermitian conjugate of the creation operator ($a_{i;n}={ (a_{i;n}^\dagger})^\dagger$); the creation operators $a_{i;n}^\dagger$ are such that the matrix entering the $n+1$-th tensor product is always $\gamma$, while the first $n$ tensor products are given by all combinations of the $I,Y$ matrices. \par
Following these rules we have, for $n=3$, that the eight $16\times 16$ creation operators are
\bea \label{creation3}
&
a_{0;3}^\dagger =I\otimes I\otimes I\otimes\gamma,~
a_{1;3}^\dagger =I\otimes I\otimes Y\otimes\gamma,~
a_{2;3}^\dagger =I\otimes Y\otimes I\otimes\gamma,~
a_{3;3}^\dagger =I\otimes Y\otimes Y\otimes\gamma,\nonumber\\&
a_{4;3}^\dagger =Y\otimes I\otimes I\otimes\gamma,~
a_{5;3}^\dagger =Y\otimes I\otimes Y\otimes\gamma,~
a_{6;3}^\dagger =Y\otimes Y\otimes I\otimes\gamma,~
a_{7;3}^\dagger =Y\otimes Y\otimes Y\otimes\gamma.&\nonumber\\
&&
\eea
Due to the presence of the nilpotent operators $\beta,\gamma$ ($\beta^2=\gamma^2=0$), for any given $i$, the pair 
$a_{i;n}, a_{i;n}^\dagger$ defines a fermionic oscillator satisfying
\bea
&\{a_{i;n}, a_{i;n}\}=\{a_{i;n}^\dagger, a_{i;n}^\dagger\}=0, \qquad \{a_{i;n}, a_{i;n}^\dagger\} = {\mathbb I}_{2^{n+1}}.
&
\eea
Furthermore,  we have,
\bea\label{zerorelation}
{\textrm{for any $i,j=0,1,\ldots, 2^n-1$:}} \quad\qquad a_{i;n}^\dagger  a_{j;n}^\dagger&=&0.
\eea
For our purposes the ${\mathbb Z}_2^n$ grading is defined by assigning a $0$  ($1$) to any diagonal (antidiagonal) operator $I$ ($Y$)  entering the $a_{i;n}^\dagger, a_{i;n}^\dagger$ tensor products. We therefore have, for the $n=1,2$ creation operators:
\bea
[a_{0;1}^\dagger]=0, ~~ [a_{1;1}^\dagger]=1; &\quad& [a_{0;2}^\dagger]=00, ~~ [a_{1;2}^\dagger]=01, ~~[a_{2;2}^\dagger]=10, ~~ [a_{3;2}^\dagger]=11.
\eea
The extension to $n\geq 3$ is immediate, with the  $a_{0;n}^\dagger$ operators assigned to the respective zero-graded sectors:
\bea
[a_{0;n}^\dagger]&=&{\underline 0}.
\eea
The (\ref{zerorelation}) relation implies that, for any given $n$, the creation operators $ a_{i;n}^\dagger$ are all on equal footing and produce $b_n$ inequivalent, ${\mathbb Z}_2^n$-graded compatible, {\it abelian}, Lie (super)algebras, defined by the brackets 
\bea\label{apapzero}
(a_{i;n}^\dagger,a_{j;n}^\dagger)&=&0
\eea
which correspond to the respective graded (anti)commutators (for $n=1,2,3,4$ they can be read from the tables in Appendix {\bf A}).\par
The $n$-bit hermitian Hamiltonian operator $H_n$ is defined to be
\bea
&H_n := a_{0;n}^\dagger a_{0;n}, \qquad {\textrm{so that}}&\nonumber\\
&H_1 = diag(0,1), \quad H_2=diag(0,1,0,1),\quad H_3= diag(0,1,0,1,0,1,0,1),\quad \ldots .&
\eea
By construction $H_n=a_{i;n}^\dagger a_{i;n}$ for any $i$.\par
It follows, for any $i$, that
\bea
[H_n, a_{i;n}] = -a_{i;n}, && [H_n, a_{i;n}^\dagger]= +a_{i;n}^\dagger.
\eea
The single-particle $n$-bit Hilbert space ${\cal H}_{1;n}$ is spanned by the $a_{i;n}^\dagger$ creation operators acting on the
$n$-bit Fock vacuum $|vac\rangle_n$ which satisfies the condition
\bea
a_{i;n} |vac\rangle_n&=& 0 \qquad {\textrm{for any $i=0,1,\ldots, 2^n-1$.}}
\eea
The $ |vac\rangle_n$ vacuum is a $2^{n+1}$-component column vector given by
\bea
 |vac\rangle_n&=& r_1
\eea
(here and in the following $r_j$ denotes a column vector with entry $1$ in the $j$-th position and $0$ otherwise).\par
We introduce the vectors
\bea
v_{i;n} &=& a_{i;n}^\dagger |vac\rangle_n .
\eea
The order of the creation operators presented in (\ref{creation2}) for $n=2$ and  (\ref{creation3}) for $n=3$
is chosen  so that $v_{i;2}$ ($v_{i;3}$) is the $8$-component ($16$-component) column vector $v_{i;2}=r_{2i+2}$ for $i=0,1,2,3$ ($v_{i;3}=r_{2i+2}$ for $i=0,1,\ldots, 7$).\par
The ${\mathbb Z}_2^n$-graded $2^{n}+1$-dimensional Hilbert space ${\cal H}_{1;n} $ is spanned by
\bea
{\cal H}_{1;n}&=& \{  |vac\rangle_n, v_{i;n}\},
\eea
with $|vac\rangle_n$ and $v_{0;n}$ belonging to the ${\underline 0}$-graded sector. \par
The energy spectrum is given by $0,1$, with the excited state being $2^n$-degenerate:
\bea
H_n|vac\rangle_n=0, && H_nv_{i;n} = v_{i;n} \qquad {\textrm{for any $i$}}.
\eea
A more general ${\underline 0}$-graded diagonal hermitian operator  $H_{d;n}$ can be introduced through the position
\bea\label{diagop}
&H_{d;n} := diag(x_0,x_1,\ldots , x_{2^n-1})\otimes (\beta\gamma), \qquad \quad \textrm{so that}&\nonumber \\
&H_{d;1} = diag (0,x_0, 0,x_1), \quad H_{d;2} = (0,x_0,0,x_1,0,x_2,0,x_3), \quad \ldots .&
\eea
The real parameters $x_0,x_1, \ldots$ are eigenvalues  for the $v_{i;n}$ eigenvectors of $H_{d;n}$:
\bea
H_{d;n} |vac\rangle_n =0, && H_{d;n}v_{i;n} = x_iv_{i;n}.
\eea
In the following we make use of the graded exchange operators $X_{ij;n}$ introduced as
\bea
&{\overline X}_{ij;n} = e_{i,j} +e_{j,i}, \qquad X_{ij;n} = {\overline X}_{ij}\otimes I.
\eea
The $X_{ij;n}$'s are symmetric $2^{n+1}\times 2^{n+1}$ matrices. In the above formulas the $e_{i,j}$ symbols
indicate a matrix with entry $1$ at the crossing of the $i$-th row with the $j$-th column and $0$ otherwise. The
$i,j$ indices take values $0<i<j\leq 2^n-1$. \par
The $X_{ij;n}$ matrices are ${\mathbb Z}_2^n$-graded. Their gradings are
\bea
[X_{ij;n}] &=& [a_{i;n}^\dagger] + [a_{j;n}^\dagger]\quad mod~2.
\eea
For $n=2$ the three exchange operators are
{\footnotesize{\bea
&X_{12;2}= 
\left(\begin{array}{cccc}0&0&0&0\\0&0&1&0\\0&1&0&0\\0&0&0&0 
\end{array}\right)\otimes I, \quad X_{13;2}= 
\left(\begin{array}{cccc}0&0&0&0\\0&0&0&1\\0&0&0&0\\0&1&0&0 
\end{array}\right)\otimes I,\quad X_{23;2}= 
\left(\begin{array}{cccc}0&0&0&0\\0&0&0&0\\0&0&0&1\\0&0&1&0 
\end{array}\right)\otimes I.&
\eea
}}
Their respective gradings are $[X_{12;2}]=11,~[X_{13;2}]=01,~[X_{23;2}] =10$.  

\subsection{The inequivalent $2$-particle quantizations}
Since the construction of the multi-particle sectors from graded Hopf algebras has been outlined in Section {\bf 4}, we limit here to present the results. As mentioned at the end of that Section, for the system under consideration
here the proof of the distinguishability of the $b_n$ parastatistics only requires to analyze the $2$-particle sector. We explicitly discuss the $n=2,3$ cases. Extending both analysis and results to $n>3$ is straightforward.\par
~\par
The $n$-bit $2$-particle vacuum $|vac\rangle_n^{(2)}$ is given by the $2^{2n+2}$-component column vector
\bea
 |vac\rangle_n^{(2)}&=& |vac\rangle_n\otimes |vac\rangle_n.
\eea
The $2^n$ first-excited states (with energy level $1$) are denoted as $v_{i;n}^{(2)}$; they are given, for $i=0,1,\ldots, 2^n-1$, by
\bea
v_{i;n}^{(2)} &=& ( a_{i;n}^\dagger\otimes_{br} {\mathbb I}_{2^{n+1}}+{\mathbb I}_{2^{n+1}} \otimes_{br}a_{i;n}^\dagger) |vac\rangle_n^{(2)}\equiv  ( a_{i;n}^\dagger\otimes {\mathbb I}_{2^{n+1}}+ {\mathbb I}_{2^{n+1}} \otimes a_{i;n}^\dagger) |vac\rangle_n^{(2)}.
\eea
The maximal number of second-excited states (energy level $2$),  denoted as $v_{ij;n}^{(2)}$ for $0\leq i\leq j\leq 2^n-1$, is 
$2^{n-1}(2^n-1)$. These states are given by
\bea\label{secondexcited}
v_{ij;n}^{(2)} &=& ( a_{i;n}^\dagger\otimes_{br} {\mathbb I}_{2^{n+1}}+{\mathbb I}_{2^{n+1}} \otimes_{br}a_{i;n}^\dagger) ( a_{j;n}^\dagger\otimes_{br} {\mathbb I}_{2^{n+1}}+{\mathbb I}_{2^{n+1}} \otimes_{br}a_{j;n}^\dagger)|vac\rangle_n^{(2)}=\nonumber\\
&=&  ( a_{i;n}^\dagger\otimes a_{j,n}^\dagger+ (-1)^{\varepsilon_{ij}}a_{j,n}^\dagger \otimes a_{i;n}^\dagger) |vac\rangle_n^{(2)}.
\eea
The $(-1)^{\varepsilon_{ij}}= (-1)^{\langle a_{i;n}^\dagger,a_{j;n}^\dagger\rangle}$ sign depends on the mutual
(anti)commutation properties of the $i$-th and $j$-th particles.
The last equation in (\ref{secondexcited}) is implied by the (\ref{apapzero}) relation. This relation also guarantees that there are no third-excited states in the $2$-particle sector.\par
Since the parastatistics are determined by the  $(-1)^{\varepsilon_{ij}} $ signs,  their inequivalence 
can only appear in the sector of the second-excited states $v_{ij;n}^{(2)}$.\par
The $2$-particle Hamiltonian and the $2$-particle extension of the diagonal operator (\ref{diagop}) are respectively given by
\bea
H_n^{(2)} = H_n\otimes {\mathbb I}_{2^{n+1}}+ {\mathbb I}_{2^{n+1}}\otimes H_n,  &&
H_{d;n}^{(2)} = H_{d;n}\otimes {\mathbb I}_{2^{n+1}}+ {\mathbb I}_{2^{n+1}}\otimes H_{d;n}.
\eea
The $2$-particle Hilbert space $
{\cal H}_{2;n}$ is spanned by the vectors
\bea
{\cal H}_{2;n}&=& \{  |vac\rangle_n^{(2)}, v_{i;n}^{(2)}, v_{ij;n}^{(2)}\}.
\eea
\par
~\par
We now analyze the inequivalent parastatistics  for $n=2$ and $n=3$.\par
~\par
{\it The  $n=2$ cases}:\par
~\par
The total number of states of the ${\cal H}_{2;n=2}$ Hilbert spaces, depending on the graded Lie (super)algebras of
(\ref{2cases}) are
\bea
{\textrm{{\it a}) for $2_1$ and $2_3$ (para)bosonic algebras:\qquad\quad}} &~& 1+4+10 =15;\nonumber\\
{\textrm{{\it b}) for $2_2$ and $2_4$ (para)fermionic superalgebras:}} &~& 1+4+8~~ =13.
\eea
The above numbers are split into the contributions coming from the $E=0,1,2$ energy eigenstates.
The difference in the above $a$ and $b$ subcases is that, for the graded superalgebras, $v_{11;2}=v_{22;2}=0$ due to the Pauli exclusion
principle.\par
It follows that the inequivalence of, on one side, the $2_1$, $2_3$ (para)bosonic statistics versus, on the other side, the $2_2$, $2_4$ (para)fermionic statistics can be read from the degeneracy of the $2$-particle  energy  eigenstates with $E=2$.\par
Determining the further inequivalence of $2_1$ bosons versus $2_3$ parabosons and of $2_2$ fermions versus $2_4$ parafermions requires measuring another observable.\par
Due to the fact that $a_{0;2}^\dagger$ from (\ref{creation2}) is a $00$-graded operator, the further differences in signs are
encountered for the normalized vectors $w_{12}= \frac{1}{\sqrt{2}} v_{12;2}$,  $w_{13}= \frac{1}{\sqrt{2}} v_{13;2}$, $w_{23}= \frac{1}{\sqrt{2}} v_{23;2}$. It follows from (\ref{secondexcited}) and the tables (\ref{2cases}) that the $w_{12}, w_{13}, w_{23}$ states are given, for each one of the four graded (super)algebra cases, by
\bea\label{fourcases}
2_1 {\textrm{~algebra case}})\qquad ~&& w_{12}= \frac{1}{\sqrt{2}} (r_{30}+r_{44}),\quad w_{13}= \frac{1}{\sqrt{2}} (r_{32}+r_{60}),\quad w_{23}= \frac{1}{\sqrt{2}} (r_{48}+r_{62}).\nonumber\\
2_3 {\textrm{~algebra case}})\qquad ~&& w_{12}= \frac{1}{\sqrt{2}} (r_{30}-r_{44}),\quad w_{13}= \frac{1}{\sqrt{2}} (r_{32}+r_{60}),\quad w_{23}= \frac{1}{\sqrt{2}} (r_{48}-r_{62}).\nonumber\\
2_2 {\textrm{~superalgebra case}})&& w_{12}= \frac{1}{\sqrt{2}} (r_{30}-r_{44}),\quad w_{13}= \frac{1}{\sqrt{2}} (r_{32}+r_{60}),\quad w_{23}= \frac{1}{\sqrt{2}} (r_{48}-r_{62}).\nonumber\\
2_4 {\textrm{~superalgebra case}})&& w_{12}= \frac{1}{\sqrt{2}} (r_{30}+r_{44}),\quad w_{13}= \frac{1}{\sqrt{2}} (r_{32}-r_{60}),\quad w_{23}= \frac{1}{\sqrt{2}} (r_{48}-r_{62}).\nonumber\\
&&
\eea

For each one of the above cases the $w_{12}$ vector is an eigenstate
of $H_{d;2}$ with eigenvalue $x_1+x_2$:
\bea
H_{d;2}~w_{12} &=& (x_1+x_2)w_{12}.
\eea

Let us now assume that we have prepared the system under investigation in the $w_{12}$ state. In order to determine if our system is composed by ordinary particles or by paraparticles we perform the measurement of a suitable observable on $w_{12}$. A relevant observable which allows to do that is $Y_{12;2}$, which is constructed in terms of the $X_{12;2}$ exchange operator:
\bea
Y_{12;2} &=& X_{12;2}\otimes X_{12;2}.
\eea
By construction $Y_{12;2}$ is hermitian ($Y_{12;2}^\dagger= Y_{12;2}$) and $00$-graded ($[Y_{12;2}]=00$). Its admissible eigenvalues are $0, \pm 1$.  Its action on the $w_{13}$, $w_{23}$ vectors, for all  above four cases (\ref{fourcases}),
gives $Y_{12;2}~w_{13}=Y_{12;2}~w_{23}=0$. On the other hand,
\bea
{\textrm{for $2_1$:}} \qquad {w}_{12}=\frac{1}{\sqrt{2}}(r_{30}+r_{44}) &\Rightarrow& Y_{12;2}~{w}_{12}= +{w}_{12},\nonumber\\
{\textrm{for $2_3$:}} \qquad {w}_{12}=\frac{1}{\sqrt{2}}(r_{30}-r_{44}) &\Rightarrow &Y_{12;2}~{ w}_{12}= -{ w}_{12}.
\eea

Similarly, for the superalgebra cases,
\bea
{\textrm{for $2_2$:}} \qquad {w}_{12}=\frac{1}{\sqrt{2}}(r_{30}-r_{44}) &\Rightarrow& Y_{12;2}~{w}_{12}= -{w}_{12},\nonumber\\
{\textrm{for $2_4$:}} \qquad {w}_{12}=\frac{1}{\sqrt{2}}(r_{30}+r_{44}) &\Rightarrow &Y_{12;2}~{ w}_{12}= +{ w}_{12}.
\eea

Therefore, the measured $\pm 1$ eigenvalue of $Y_{12;2}$ on the $w_{12}$ state allows to discriminate the $2_1$ statistics from the $2_3$ parastatistics and the $2_2$ statistics from the $2_4$ parastatistics, completing the proof of their detectability.
\par
~\par

{\it The  $n=3$ cases}:\par
~\par
The extension of the construction to the $n=3$ cases gives the following results.
The total number of states of the ${\cal H}_{2;n=3}$ Hilbert spaces, depending on the graded Lie (super)algebras of
(\ref{inequivalent3}), are
\bea
{\textrm{{\it a}) for the $3_1$ and $3_2$ (para)bosonic algebras:\qquad\quad\quad~~}} &~& 1+8+36 =45;\nonumber\\
{\textrm{{\it b}) for the $3_3$, $3_4$ and $3_5$ (para)fermionic superalgebras:}} &~& 1+8+32 =41.
\eea
In the graded superalgebra cases four vectors are vanishing due to the Pauli exclusion principle. As before,
the inequivalence of the $3_1$, $3_2$ (para)bosonic statistics with respect to the $3_3$, $3_4$, $3_5$ (para)fermionic statistics is read from the degeneracy of the $2$-particle energy eigenstates with $E=2$.\par
The inequivalence of the parastatistics within the two subclasses of graded algebras and graded superalgebras goes as follows.
\par
Discriminating $3_1$ and $3_2$ proceeds as for the $n=2$ cases. It is sufficient to measure the eigenvalue of the
$Y_{12;3}$ observable, introduced through the position
 \bea
Y_{12;3} &=& X_{12;3}\otimes X_{12;3},
\eea
acting on the normalized state ${\overline w}_{12} =\frac{1}{\sqrt{2}}v_{12;3}$, which is an eigenstate
of $H_{d;3}$ with eigenvalue $x_1+x_2$:
\bea
H_{d;3} {\overline w}_{12} &=& (x_1+x_2){\overline w}_{12}.
\eea 
We get that
\bea
{\textrm{for $3_1$:}} \quad {\overline w}_{12}=\frac{1}{\sqrt{2}}(r_{54}+r_{84}) &\Rightarrow& Y_{12;3}~{\overline w}_{12}= +{\overline w}_{12},\nonumber\\
{\textrm{for $3_2$:}} \quad {\overline w}_{12}=\frac{1}{\sqrt{2}}(r_{54}-r_{84}) &\Rightarrow &Y_{12;3}~{\overline w}_{12}= -{\overline w}_{12}.
\eea
~\par
For the superalgebra cases we need to discriminate three different parastatistics $3_3,3_4,3_5$. In order to do that we need to perform  {\it two} separate measurements on suitably prepared states.  We can select the normalized states ${\overline w}_{24} =\frac{1}{\sqrt{2}}v_{24;3}$ and ${\overline w}_{35} =\frac{1}{\sqrt{2}}v_{35;3}$. They are eigenstates of $H_{d;3}$ with respective eigenvalues $x_2+x_4$  and $x_3+x_5$.
We measure the 
$Y_{24;3} = X_{24;3}\otimes X_{24;3}$ observable on ${\overline w}_{24}$ and the 
$Y_{35;3} = X_{35;3}\otimes X_{35;3}$ observable on ${\overline w}_{35}$. We get
\bea
{\textrm{for $3_3$:}} \quad {\overline w}_{24}=\frac{1}{\sqrt{2}}(r_{90}+r_{150}) &\Rightarrow& Y_{24;3}~{\overline w}_{24}= +{\overline w}_{24},\nonumber\\
{\textrm{for $3_4$:}} \quad {\overline w}_{24}=\frac{1}{\sqrt{2}}(r_{90}-r_{150}) &\Rightarrow &Y_{24;3}~{\overline w}_{24}= -{\overline w}_{24},\nonumber\\
{\textrm{for $3_5$:}} \quad {\overline w}_{24}=\frac{1}{\sqrt{2}}(r_{90}-r_{150}) &\Rightarrow &Y_{24;3}~{\overline w}_{24}= -{\overline w}_{24}
\eea
and
\bea
{\textrm{for $3_3$:}} \quad {\overline w}_{35}=\frac{1}{\sqrt{2}}(r_{124}+r_{184}) &\Rightarrow& Y_{35;3}~{\overline w}_{35}= +{\overline w}_{35},\nonumber\\
{\textrm{for $3_4$:}} \quad {\overline w}_{35}=\frac{1}{\sqrt{2}}(r_{124}+r_{184}) &\Rightarrow &Y_{35;3}~{\overline w}_{35}= +{\overline w}_{35},\nonumber\\
{\textrm{for $3_5$:}} \quad {\overline w}_{35}=\frac{1}{\sqrt{2}}(r_{124}-r_{184}) &\Rightarrow &Y_{35;3}~{\overline w}_{35}= -{\overline w}_{35}.
\eea
Therefore, the following ordered pairs of $Y_{24;3}, Y_{35;3}$ eigenvalues resulting from the respective measurements on
${\overline w}_{24}$, ${\overline w}_{35}$ allow to discriminate the three parastatistics induced by the graded superalgebras. We have
\bea
&\begin{array}{|c|c|c|}\hline
3_3: & 3_4: & 3_5: \\ \hline
(+1,+1)&(-1,+1)&(-1,-1)\\
\hline \end{array}&
\eea
~\\
{\it Comments}: some comments are in order. The framework to detect inequivalent parastatistics follows the approach discussed in \cite{{top1},{top2}} for ${\mathbb Z}_2^2$-graded (super)algebras. The present construction allows to extend the analysis to general ${\mathbb Z}_2^n$-graded Hamiltonians. To our knowledge we presented here the first  investigation of inequivalent parastatistics for a ${\mathbb Z}_2^3$-graded quantum Hamiltonian.\\
We should further point that this analysis is immediately applicable to quantum Hamiltonians possessing an infinite spectrum like the matrix quantum oscillator whose $H_{osc;n}$ Hamiltonian can be introduced as
\bea
H_{osc;n}&=&\frac{1}{2}(\partial_x^2+x^2)\cdot {\mathbb I}_{2^{n+1}}+ H_n.
\eea
The inequivalence of its ${\mathbb Z}_2^n$-graded parastatistics follows from the obtained results for the $H_n$ term entering the right hand side.

\section{The statistical transmutations of the ${\cal N}=1,2,4,8$-extended supersymmetric quantum mechanics}

In this Section we apply the framework of the induced graded Lie (super)algebras to the Supersymmetric Quantum Mechanics which, since its introduction in \cite{wit}  as a reformulation of the Atiyah-Singer index theorem, finds relevant applications in both physics and mathematics. To describe the different  $n$-bit parastatistics associated with the Supersymmetric Quantum Mechanics we adopt a term, {\it statistical transmutation}, which has already been employed by condensed matter physicists in a slightly different context (this topic is discussed, e.g., in \cite{mar} and references therein). In our case the notion of {\it algebraic statistical transmutation} refers to the  ${\mathbb Z}_2^n$-graded Lie (super)algebras which are induced by the supersymmetric quantum mechanics operators.\par
To further clarify this setting we mention that, when the authors of  \cite{brdu} produced an ${\cal N}=2$ supersymmetric quantum model which was invariant under a ${\mathbb Z}_2^2$-graded superalgebra, a question arised, namely which was the physical role of the ${\mathbb Z}_2^2$-graded invariant superalgebra. This question was answered in \cite{top1} by proving that, in the multiparticle sector, the ${\mathbb Z}_2^2$-graded invariance induces a detectable parafermionic statistics.  The present algebraic framework is a generalization
which is both {\it model-independent} and  applicable, for any integer $n$, to ${\mathbb Z}_2^n$ gradings. \par
~\par
For any given positive integer ${\cal N}=1,2,3,4,5,\ldots$,
the superalgebra ${\mathfrak{sqm}}_{\cal N}$ of the ${\cal N}$-extended one-dimensional supersymmetric quantum mechanics is given by the (anti)commutators
\bea\label{sqm}
&\{Q_i,Q_j\} = 2\delta_{ij} H, \qquad [H, Q_i]=0,\qquad {\textrm{ for any}}\qquad i,j=1,\ldots,{\cal N}.&
\eea
The $Q_i$ generators are the supercharges and $H$ is the Hamiltonian; they are assumed to be Hermitian, so that 
$Q_i^\dagger=Q_i$, $H^\dagger=H$. 
~\par
We denote as $U_{\cal N} := {\cal U} ({\mathfrak{sqm}}_{\cal N})$ the respective Universal Enveloping Superalgebras; they are therefore spanned, for $m=0,1,2,\ldots $, by the sets of operators
\bea\label{univenvsusy}
U_{{\cal N}=1}&=&\{ H^m, H^mQ_1\}, \nonumber\\ 
U_{{\cal N}=2}&=&\{ H^m, H^mQ_1, H^mQ_2, H^mQ_1Q_2\}, \nonumber
\\
U_{{\cal N}=3}&=&\{ H^m, H^mQ_1, H^mQ_2, H^mQ_3, H^mQ_1Q_2, H^mQ_1Q_3, H^m Q_2Q_3,H^m Q_1Q_2Q_3\},\nonumber\\
\ldots &=&\ldots .
\eea 
At any given $m$ the number of operators entering $U_{\cal N}$ is $2^{\cal N}$.\par
The further analysis is based on the matrix differential representations of the $Q_i$'s and $H$ operators entering
(\ref{sqm}). There are two types of differential representations for the ${\mathfrak{sqm}_{\cal N}}$ superalgebras:
at the classical level we have the time-dependent worldline $D$-module representations presented in \cite{{pato},{krt}} and  which are applicable to the construction of invariant, worldline sigma models; at the quantum level we have the differential representations realizing the Hamiltonian $H$ as a second-order differential operator of the space coordinates (the connection between the two types of differential representations is elucidated in \cite{cht} and, in a ${\mathbb Z}_2^2$-graded context, \cite{akt2}).
~\par
Following \cite{pato}, in  the {\it minimal}, irreducible, $D$-module representations of the ${\mathfrak{sqm}}_{\cal N}$ superalgebra, the ${\cal N}$ supercharges $Q_i$ are expressed as $d_{\cal N}\times d_{\cal N}$ matrix,
first-order differential operators in the time variable $t$, with $d_{\cal N}$ given by the following formula. \par
Let us parametrize ${\cal N}$ as
\bea
{\cal N} &=& 8k+r, \qquad {\textrm{where}}\quad k=0,1,2, \ldots \in {\mathbb N}_0\quad {\textrm{and}} \quad r=1,2,3,4,5,6,7,8,
\eea
the $d_{\cal N}$ matrix size is given by
\bea\label{susydim}
\qquad\qquad d_{\cal N} = 2^{\left(4k+z(r)+1\right)}, \qquad\qquad {\textrm{for}} \quad z(r) = \lceil\log_2 r\rceil.
\eea 
In the last formula the ceiling function appears, so that 
\bea
&z(1)=0, \qquad z(2)= 1,\qquad z(3)=z(4)= 2,\qquad z(5)=z(6)=z(7)=z(8)= 3.&
\eea
The expression $\frac{1}{2}\times d_{\cal N}$ (i.e., the dimensionality of the bosonic (fermionic) subspaces) produces the $A034583$ sequence in the OEIS (Online Encyclopedia of Integer Sequences) database at
 https://oeis.org.:
\bea
\frac{1}{2}\times d_{\cal N}& \Rightarrow & 1,2, 4,4,8,8,8,8,16,32,64,64,128,128,128,128,256,\ldots .
\eea\par
Starting from ${\cal N}\geq 4$, ${\mathfrak{sqm}}_{\cal N}$ also admits, see \cite{gkt}, reducible, but indecomposable $D$-module representations. The combinatorics of the statistical transmutations can in principle be  computed for any  $D$-module (both reducible and irreducible) representation  of the ${\cal N}$-extended ${\mathfrak{sqm}}_{\cal N}$ superalgebra.   In the irreducible cases it is convenient to introduce
\bea\label{nversusN}
n_{\cal N} &:=& 4k+z(r)+1, \qquad\qquad {\textrm{so that}}\quad d_{\cal N}\times d_{\cal N} = 2^{n_{\cal N}}\times 2^{n_{\cal N}}.
\eea
It follows that the irreducible supercharges $Q_i$ can be assigned to the nonzero graded sectors of a ${\mathbb Z}_2^{n_{\cal N}}$ grading, with the Hamiltonian $H$ assigned to the $n_{\cal N}$-bit zero vector ${\underline 0}$. \par 
From now on we focus for simplicity on the irreducible representations of the
${\cal N}=1,2,4,8$-extended one-dimensional Supersymmetric Quantum Mechanics since these values of ${\cal N}$, which are related to division algebras, are the most widely investigated in the literature.\par
~\par
Before further proceeding some comments are in order: for ${\cal N}=1,2,4,8$, the minimal $D$-module representations are respectively  given by $2\times 2$, $4\times4$, $8\times 8$, $16\times 16$ matrices both in the classical and quantum cases (examples of quantum $D$-module representations for superconformal quantum mechanics are given in Section {\bf 7}). We have,
from (\ref{nversusN}),
\bea
& n_{{\cal N}=1}=1, \quad n_{{\cal N}=2}=2,\quad n_{{\cal N}=4}=3,\quad n_{{\cal N}=8}=4.
&
\eea
It follows that, for ${\cal N}=1,2,4,8$, the inequivalent graded Lie (super)algebras induced by
different grading assignments of the supercharges $Q_i$ are respectively read from the $n=1,2,3,4$-bit tables of Appendix {\bf A}.
We are now in the position to compute the $s_{\cal N}$ inequivalent graded Lie (super)algebras induced on ${\cal U}({\mathfrak{sqm}_{\cal N}})$ by the minimal $H, Q_i$ operators satisfying (\ref{sqm}). The scheme,
with differences that will be pointed out, is the one already applied to (split-)quaternions and biquaternions in Appendices {\bf C} and {\bf E}. 

\subsection{The ${\cal N}=1$ statistical transmutations}

The Hamiltonian $H$ is assigned to the vanishing grading, so that $[H]=0$; the grading of the single supercharge $Q_1$ is therefore given by $[Q_1]=1$. The inequivalent graded Lie (super)algebras are read from the $1$-bit tables (\ref{1bitcases}). The two cases are
\bea
1_1:&\Rightarrow& {\textrm{the Lie algebra}} \qquad\qquad [H,Q_1]=0;\nonumber\\
1_2: &\Rightarrow&{\textrm{the Lie superalgebra}} \quad~~ [H,Q_1]=0,  \quad \{Q_1,Q_1\}= 2H.
\eea

The result is
\bea
s_{{\cal N}=1} &=& 2.
\eea
\subsection{The ${\cal N}=2$ statistical transmutations}

A novel feature appears for ${\cal N}=2$. The Hamiltonian $H$ is assigned to the $[H]=00$ grading. The supercharges $Q_1,Q_2$ have to be assigned to the nonvanishing gradings. We can set  $[Q_1]=\alpha$, $[Q_2]=\beta$ (with $\alpha\neq \beta$), which
take values $10,01,11$. It follows that the grading assignments of the operators entering the Universal Enveloping (Super)algebra ${\cal U}({\mathfrak{sqm}}_{{\cal N}=2})$ is 
\bea
&[H^m]=00,\quad [H^mQ_1]=\alpha, \quad [H^mQ_2]=\beta, \quad [H^mQ_1Q_2]=\alpha+\beta ~~ mod~~ 2.&
\eea
It should be noted that, for the ${\mathfrak{sqm}_{{\cal N}=2}}$ superalgebra spanned by $H, Q_1,Q_2$, an empty slot is assigned to the $\alpha+\beta$-graded sector. This empty slot, denoted as ``$\emptyset$", can be thought to be associated with a matrix with all vanishing entries. This empty slot plays the role of a marked generator. It then follows that the computation of the $s_{{\cal N}=2}$ inequivalent graded Lie (super)algebras reproduces the
computation of the split-quaternions presented in Appendix {\bf C}. Both these computations are based on the $1+2$ decompositions of the $2$-bit nonvanishing graded sectors. The results, recovered from the (\ref{2cases}) tables, are the following.
\par
~\par
$2_1 ~\Rightarrow ~ 1 ~ {\textrm{case}}$; this is an ordinary Lie algebra with infinite generators (since $[Q_1,Q_2]\neq 0$)  induced by assigning the gradings $[Q_1]=10$, $[Q_2]=01$. Since the two supercharges $Q_1, Q_2$ are now bosonic, we refer to this case as the ``$2B$" transmutation.\par
~\par
$2_2 ~\Rightarrow ~ 2 ~ {\textrm{cases}}$, which are split into $2_{2_\alpha}$ and $2_{2_\beta}$.\par
~\par
The subcase $2_{2_\alpha}$ is recovered by assigning the gradings $[Q_1]=10, ~[Q_2]= 11$.  It corresponds to an ordinary Lie superalgebra with infinite generators (since $[Q_1,Q_2]\neq 0$). Since the supercharge $Q_1$ is fermionic,
while $Q_2$ is bosonic, we refer to this subcase as the ``$1F+1B$" transmutation.\par~
\par
The subcase $2_{2_\beta}$ is recovered by assigning the gradings $[Q_1]=10, ~[Q_2]= 01$.  It corresponds to the original ${\mathfrak{sqm}}_{{\cal N}=2}$ Lie superalgebra with two fermionic supercharges  $Q_1, Q_2$. This subcase is the ``identity transmutation" which will be referred to as ``$2F$".\par
~\par
$2_3 ~\Rightarrow ~ 1 ~ {\textrm{case}}$; this is a parabosonic Lie algebra recovered from the gradings $[Q_1]=10$, $[Q_2]=01$. A finite, $3$-generator, graded-abelian Lie algebra is  obtained. It is defined by the (anti)commutators
$[H,Q_1]=[H,Q_2]=0$, $\{Q_1,Q_2\}=0$. Since the two supercharges $Q_1, Q_2$ are parabosonic, we refer to this case as the ``$2P_B$" transmutation.\par
~\par
$2_4 ~\Rightarrow ~ 2 ~ {\textrm{cases}}$, which are split into $2_{4_\alpha}$ and $2_{4_\beta}$.\par
~\par
The subcase $2_{4_\alpha}$ is a parafermionic Lie superalgebra recovered by assigning the gradings $[Q_1]=10, ~[Q_2]= 11$.  A finite, $3$-generator, graded Lie superalgebra is obtained.  It is defined by the (anti)commutators
$[H,Q_1]=[H,Q_2]=0$, $\{Q_1,Q_1\} = 2H$, $ \{Q_1,Q_2\}=0$.\\
Since the supercharge $Q_1$ is parafermionic,
while $Q_2$ is parabosonic, we refer to this subcase as the ``$1P_F+1P_B$" transmutation.\par~
\par
The subcase $2_{4_\beta}$ is recovered by assigning the gradings $[Q_1]=10, ~[Q_2]= 01$.  It produces a $4$-generator finite parafermionic Lie superalgebra (the extra generator being $Z:=Q_1Q_2$).  This graded Lie superalgebra corresponds to the ${\mathbb Z}_2^2$-graded worldline super-Poincar\'e algebra introduced in \cite{brdu}. It is defined by the (anti)commutators $[H,Q_1]=[H,Q_2]=[H,Z]=0$, $\{Q_1,Q_1\}=\{Q_2,Q_2\}= 2H$, $~ [Q_1,Q_2]= 2Z$, $~\{Q_1,Z\}=\{Q_2,Z\}=0$.\par
Since both $Q_1, Q_2$ supercharges are parafermionic, we refer to this subcase as the ``$2P_F$" transmutation.
\par
~\par
By summing the contributions from the four $2$-bit tables (\ref{2cases}) we get
\bea\label{sn2}
s_{{\cal N}=2} &=& 1+2+1+2 = 6.
\eea

\subsection{The ${\cal N}=4$ statistical transmutations}

The ${\cal N}=4$ transmutations are recovered from the $3$-bit tables  (\ref{inequivalent3}). The four supercharges
$Q_i$,  together with the three vanishing matrices representing the empty slots $\emptyset$,  have to be accommodated into the seven non-zero graded sectors. This results in two classes of equivalence for the marked generators with $4+3=7$ elements. The non-zero graded sectors are encoded, as visualzed in (\ref{fano}), in a Fano's plane. There are two important remarks.\par
~\par
{\it Remark I}: since the product of two vanishing matrices is a vanishing matrix, the three  empty slots $\emptyset$ are aligned along an edge of the Fano's plane.\par
{\it Remark II}: conversely, in the $3_1 - 3_5$ tables of (\ref{inequivalent3}), the product of two (para)bosons produces a third (para)boson which is necessarily aligned along an edge of the Fano's plane.\par ~
\par
By applying these two properties, the combinatorics which leads to the computation of $s_{{\cal N}=4}$ easily follows. We have\par
~\par

$3_1 ~\Rightarrow ~ 1 ~ {\textrm{case}}$; the seven non-zero graded sectors represent seven bosons. Any grading assignment for the four $Q_i$'s, such as $[Q_1]=001,~ [Q_2]=101,~ [Q_3] =011 , ~[Q_4] = 111$, leads to $4$ bosonic supercharges.
Therefore, we refer to this case as the ``$4B$" transmutation.
\par
~\par

$3_2 ~\Rightarrow ~ 2 ~ {\textrm{cases}}$, denoted as $3_{2_\alpha}$, $3_{2_\beta}$. They are implied by the seven non-zero graded sectors to be split into $1$ bosonic and $6$ parabosonic particles, the grading of the boson being $001$. \par
~\par
The subcase $3_{2_\alpha}$ is recovered by assigning the gradings $[Q_1]=001, ~[Q_2]= 101, ~[Q_3]=011, ~[Q_4]=111$.  The  supercharge $Q_1$ is bosonic, while the three remaining ones are parabosonic. Therefore, we
refer to this subcase as the ``$1 B+3 P_B$" transmutation.\par~
\par

The subcase $3_{2_\beta}$ is recovered by assigning the gradings $[Q_1]=110, ~[Q_2]= 010, ~[Q_3]=011, ~[Q_4]=111$.  The  four supercharges are parabosonic. Therefore, we
refer to this subcase as the ``$4 P_B$" transmutation.\par~
\par

$3_3 ~\Rightarrow ~ 2 ~ {\textrm{cases}}$, denoted as $3_{3_\alpha}$, $3_{3_\beta}$. They are implied by the seven non-zero graded sectors to be split into $3$ bosons  aligned along an edge of the Fano's plane and $4$ fermions, with the gradings of the bosons being $001, 010,011$.  From {\it{remark I}}, the four supercharges $Q_i$ need to be complementary to an edge of the Fano's plane, leaving only two inequivalent possibilities. \par
~\par
The subcase $3_{3_\alpha}$ is recovered by assigning the gradings $[Q_1]=001, ~[Q_2]= 011, ~[Q_3]=101, ~[Q_4]=111$.  The  supercharge $Q_1, Q_2$ are bosonic, while $Q_3,Q_4$ are fermionic. Therefore, we
refer to this subcase as the ``$2 B+2F$" transmutation.\par~
\par

The subcase $3_{2_\beta}$ is recovered by assigning the gradings $[Q_1]=100, ~[Q_2]= 110, ~[Q_3]=101, ~[Q_4]=111$. All  four supercharges are fermionic. This subcase reproduces the original ${\mathfrak{sqm}}_{{\cal N}=4}$ superalgebra. Therefore, we
refer to it as the ``$4 F$" identity transmutation.\par~
\par
$3_4 ~\Rightarrow ~ 3 ~ {\textrm{cases}}$, denoted as $3_{4_\alpha}$, $3_{4_\beta}$, $3_{4_\gamma}$. They are implied by the seven non-zero graded sectors to be split into $1$ boson and $2$ parabosons  aligned along an edge of the Fano's plane,  plus $4$ parafermions; the grading of the boson is  $001$, while the gradings of the parabosons are $110, 111$.  By taking into account that the four supercharges $Q_i$ are complementary to an edge of the Fano's plane, three inequivalent possibilities follow. \par
~\par

The subcase $3_{4_\alpha}$ is recovered by assigning the gradings $[Q_1]=001, ~[Q_2]= 110, ~[Q_3]=100, ~[Q_4]=011$.  The  supercharge $Q_1$ is bosonic, while $ Q_2$ is parabosonic and $Q_3,Q_4$ are parafermionic. Therefore, we
refer to this subcase as the ``$1B+ 1 P_B+2P_F$" transmutation.\par~
\par

The subcase $3_{4_\beta}$ is recovered by assigning the gradings $[Q_1]=111, ~[Q_2]= 110, ~[Q_3]=010, ~[Q_4]=011$.  The  supercharges $Q_1, Q_2$ are parabosonic, while $Q_3,Q_4$ are parafermionic. Therefore, we
refer to this subcase as the ``$2 P_B+2P_F$" transmutation.\par~
\par

The subcase $3_{4_\gamma}$ is recovered by assigning the gradings $[Q_1]=101, ~[Q_2]= 010, ~[Q_3]=100, ~[Q_4]=011$.  All four supercharges are parafermionic. Therefore, we
refer to this subcase as a ``$4P_F$" transmutation.\par~
\par
$3_5 ~\Rightarrow ~ 2 ~ {\textrm{cases}}$, denoted as $3_{5_\alpha}$, $3_{5_\beta}$. They are implied by the seven non-zero graded sectors to be split into $3$ parabosons  aligned along an edge of the Fano's plane and $4$ parafermions, with the gradings of the parabosons being $110, 101,011$.  The analysis of this case mimicks what has been done in deriving the two inequivalent cases from $3_3$. \par
~\par
The subcase $3_{5_\alpha}$ is recovered by assigning the gradings $[Q_1]=101, ~[Q_2]= 011, ~[Q_3]=010, ~[Q_4]=001$.  The  supercharge $Q_1, Q_2$ are parabosonic, while $Q_3,Q_4$ are parafermionic. Therefore, we
refer to this subcase as the ``$2 P_B+2P_F$" transmutation.\par~
\par

The subcase $3_{5_\beta}$ is recovered by assigning the gradings $[Q_1]=100, ~[Q_2]= 111, ~[Q_3]=010, ~[Q_4]=001$. All  four supercharges are parafermionic.  Therefore, we
refer to this subcase as a ``$4 P_F$" transmutation.\par~
\par
It should be pointed out that the subcases $3_{4_\gamma}$ and $3_{5_\beta}$, despite being both characterized
as  a ``$4 P_F$ transmutation", are inequivalent. Indeed, in $3_{4_\gamma}$ the descendent operators
$Q_iQ_j$ for $i<j$ produce $6$ operators which are accommodated as $2$ bosons and $4$ parabosons;  in $3_{5_\beta}$, the $6$ descendent operators are accommodated as $6$ parabosons.\par
~\par

By summing the contributions from the five $3$-bit tables (\ref{inequivalent3}) we end up with
\bea
s_{{\cal N}=4} &=& 1+2+2+3+2 = 10.
\eea

\subsection{The ${\cal N}=8$ statistical transmutations: a tale of two Fano's planes}

The ${\cal N}=8$ transmutations are recovered from the seven $4$-bit tables  (\ref{4bitcases}). The $8$ supercharges
$Q_i$,  together with the $7$ vanishing matrices representing the empty slots $\emptyset$,  have to be accommodated into the $15$ non-zero graded sectors.\par
We present at first the computations from the graded Lie algebras $4_1, ~4_2, ~4_3$ and, subsequently,
the computations from the graded Lie superalgebras $4_4,~4_5,~4_6,~4_7$.
\par
~\par
Concerning the graded Lie algebras we have\par
~\par

$4_1 ~\Rightarrow ~ 1 ~ {\textrm{case}}$; the non-zero graded sectors represent $15$ bosons. Any grading assignment for the $8$ $Q_i$'s leads to $8$ bosonic supercharges.
Therefore, we refer to this case as the ``$8B$" transmutation.\par
~\par

$4_2 ~\Rightarrow ~ 2 ~ {\textrm{cases}}$, denoted as $4_{2_\alpha}$, $4_{2_\beta}$, implied by the $15$ non-zero graded sectors being split into $3$ bosons and $12$ parabosons, with the three bosons (whose gradings are $0001,~0010,~0011$) being aligned. The derivation goes as follows. At first one notices that at least $1$ boson should be assigned to an empty slot $\emptyset$. Indeed, trying to assign all empty slots to the parabosonic sectors
leads to a contradiction. Let's say, without loss of generality, that two $\emptyset$'s are assigned to the gradings $1000$,
$1100$. The product of the vanishing matrices implies that $\emptyset$ is also assigned to the $0100$ grading. On the other hand, no other grading can be assigned to an empty slot. By assuming, e.g., that $\emptyset$ is assigned to $1001$, the product of the $1000$ and $1001$-graded vanishing matrices implies, against the assumption, that the bosonic sector $0001$ is assigned to a vanishing matrix. It follows that there is no room to accommodate all $7$ empty slots in the parabosonic sectors. As a next observation, if two bosons are assigned to an empty slot, the third boson should also be assigned to $\emptyset$  since the product of two vanishing matrices is a vanishing matrix. We end up with only two inequivalent possibilities: $1$ boson assigned to $\emptyset$ or $3$ bosons assigned to $\emptyset$. By taking the complementary viewpoint of the $Q_i$'s supercharges we get the following two subcases.\par
~\par
The subcase $4_{2_\alpha}$, recovered by taking $2$  bosonic supercharges and $6$ parabosonic supercharges. Therefore, we
refer to this subcase as the ``$2 B+6 P_B$" transmutation.\par~
\par
The subcase $4_{2_\beta}$, recovered by taking $8$ parabosonic supercharges. Therefore, we
refer to this subcase as an ``$8 P_B$" transmutation.\par~
\par

$4_3 ~\Rightarrow ~ 1 ~ {\textrm{case}}$; the non-zero graded sectors represent $15$ parabosons. Any grading assignment for the $8$ $Q_i$'s leads to $8$ parabosonic supercharges.
Therefore, we refer to this case as an ``$8 P_B$" transmutation.\par
~\par

The combinatorics of the inequivalent graded Lie superalgebras induced by the (\ref{4bitcases}) tables $4_4, 4_5, 4_6,4_7$ is obtained by intersecting two Fano's planes:\\
~\\
$~$ {\it i}) the Fano's plane of the multiplication table of the $7$ empty slots ${\emptyset}$ and\\
{\it ii}) the Fano's plane of the reduced multiplication tables of the $7$ (para)bosons entering $4_4, 4_5, 4_6,4_7$.
\par
~\par
The intersection leads to only two possibilities; either\\
~\\
$i$) ~$3$ empty slots $\emptyset$ are assigned to an edge of the (para)bosonic Fano's plane, or\\
$ii$)  all $7$ empty slots ${\emptyset}$ are assigned to the $7$ vertices of the (para)bosonic Fano's plane.\\
~\par
The proof is straightforward. The multiplication of two (para)fermions gives a (para)boson. If the two (para)fermions are assigned to the vanishing matrices of the empty slot, the (para)boson should also be assigned to the empty slot $\emptyset$. By taking two suitable pairs of (para)fermions, it is easily realized that at least two (para)bosons should be assigned to the empty slot. Next, the multiplication of the two vanishing matrices implies that the third (para)boson lying
on the edge should also be assigned to $\emptyset$; this proves that the condition leading to case $i$) is satisfied. The further assignment of
a fourth (para)boson to an empty slot implies (once more due to the multiplication of the vanishing matrices) that
all $7$ (para)bosons should be assigned to $\emptyset$, realizing the case $ii$).\par
~\par
The inequivalent contributions  from $4_4,4_5,4_6,4_7$ to the statistical transmutations are obtained by further taking into account\\
~\\
$iii$) the vertices of the $4_4$, $4_7$ (para)bosonic Fano's planes are unmarked (they are $7$ bosons for $4_4$ and $7$ parabosons for $4_7$);\\
$iv$) ~the $4_5$ (para)bosonic Fano's plane presents $3$ marked vertices corresponding to bosons lying on a edge (the $4$ remaining vertices are associated with parabosons);\\
$v$) ~~the $4_6$ (para)bosonic Fano's plane present $1$ marked vertex corresponding to the single boson, while the $6$ remaining vertices are associated with parabosons.\par
~\par
The inequivalent contributions can be visualized in terms of the following diagrams (labeled with greek letters to denote the correponding subcases) which present the respective
(para)bosonic Fano's planes  with/without marked points. The encircled vertices are the ones which are assigned to the empty slots $\emptyset$. \newpage
We have, for $4_4$ and $4_7$, the following $\alpha$ diagram on top and $\beta$ diagram on the bottom:

{\centerline{\includegraphics[width=0.4\textwidth]{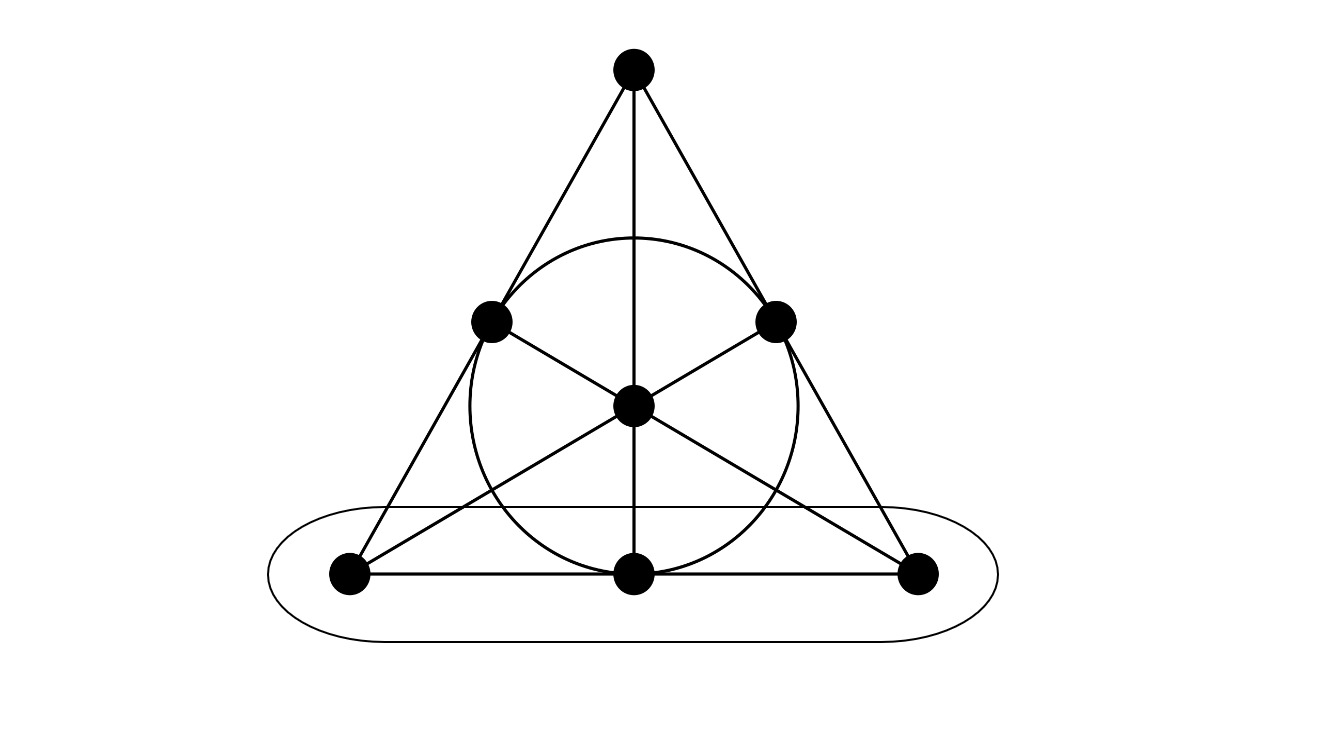}}}
{\centerline{\includegraphics[width=0.4\textwidth]{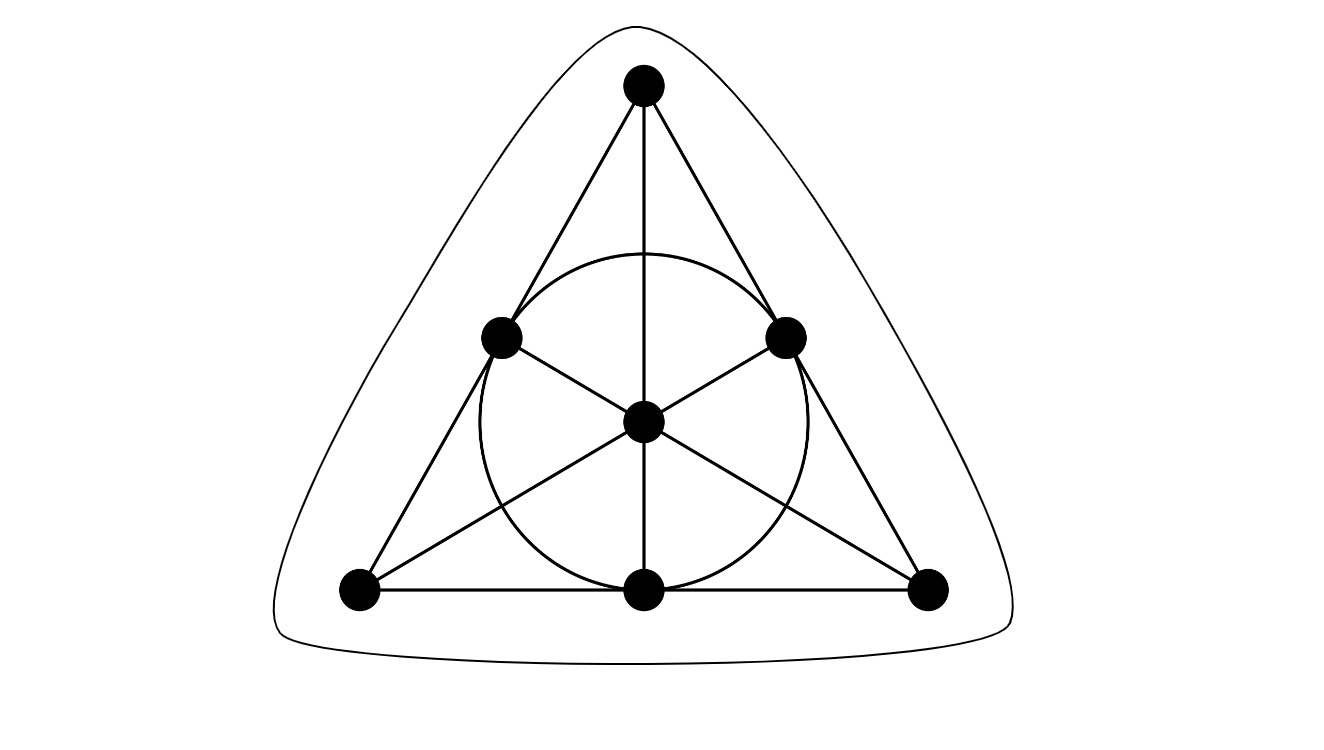}}}
\bea
&{\textrm{They give the two inequivalent contributions from unmarked vertices.}}&
\eea
\par
~\par
For $4_5$ we have, from top to bottom, the three $\alpha,\beta,\gamma$ diagrams:

{\centerline{\includegraphics[width=0.4\textwidth]{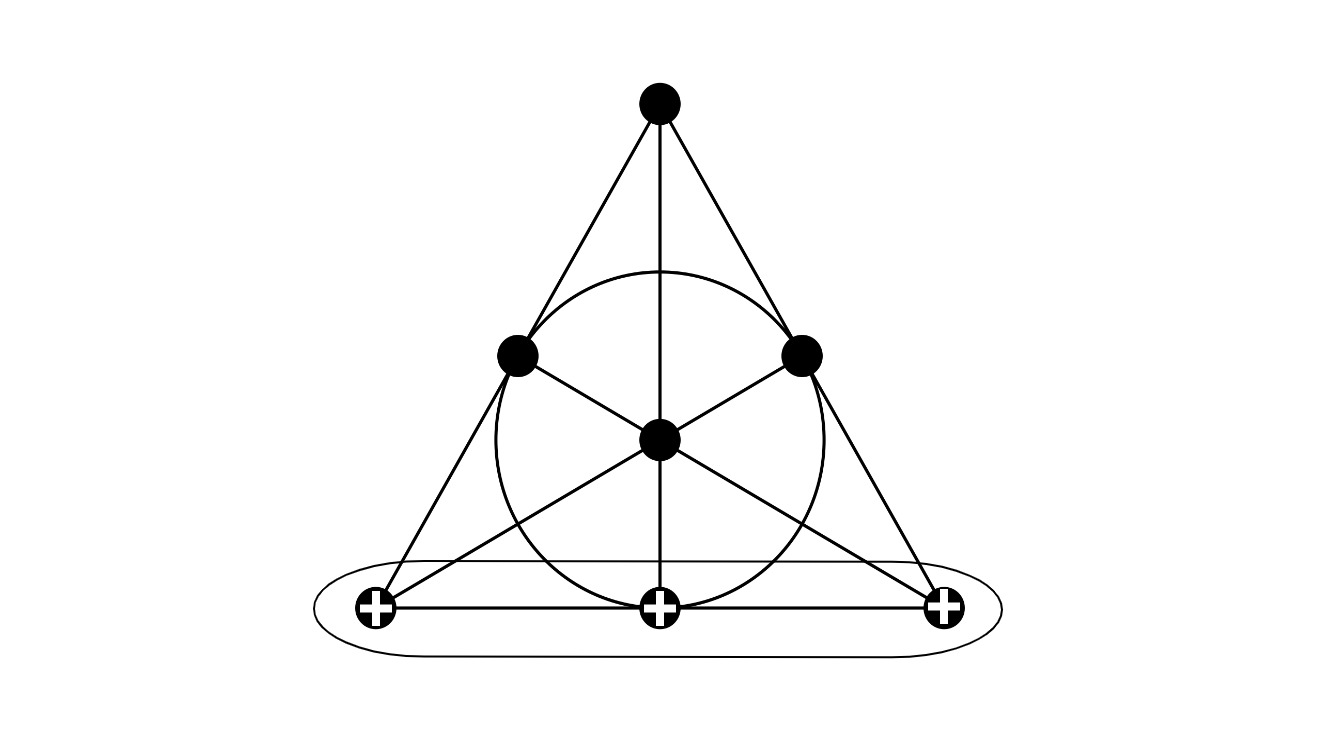}}}

{\centerline{\includegraphics[width=0.4\textwidth]{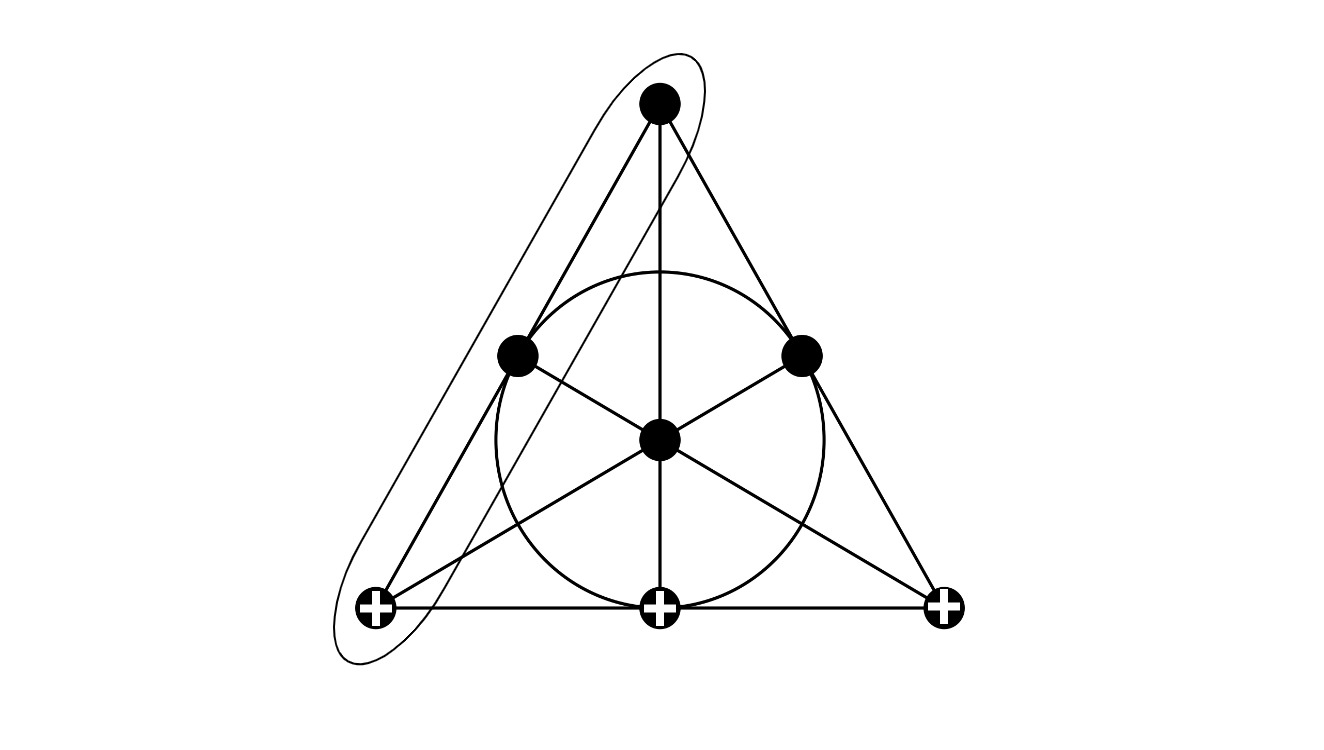}}}
{\centerline{\includegraphics[width=0.4\textwidth]{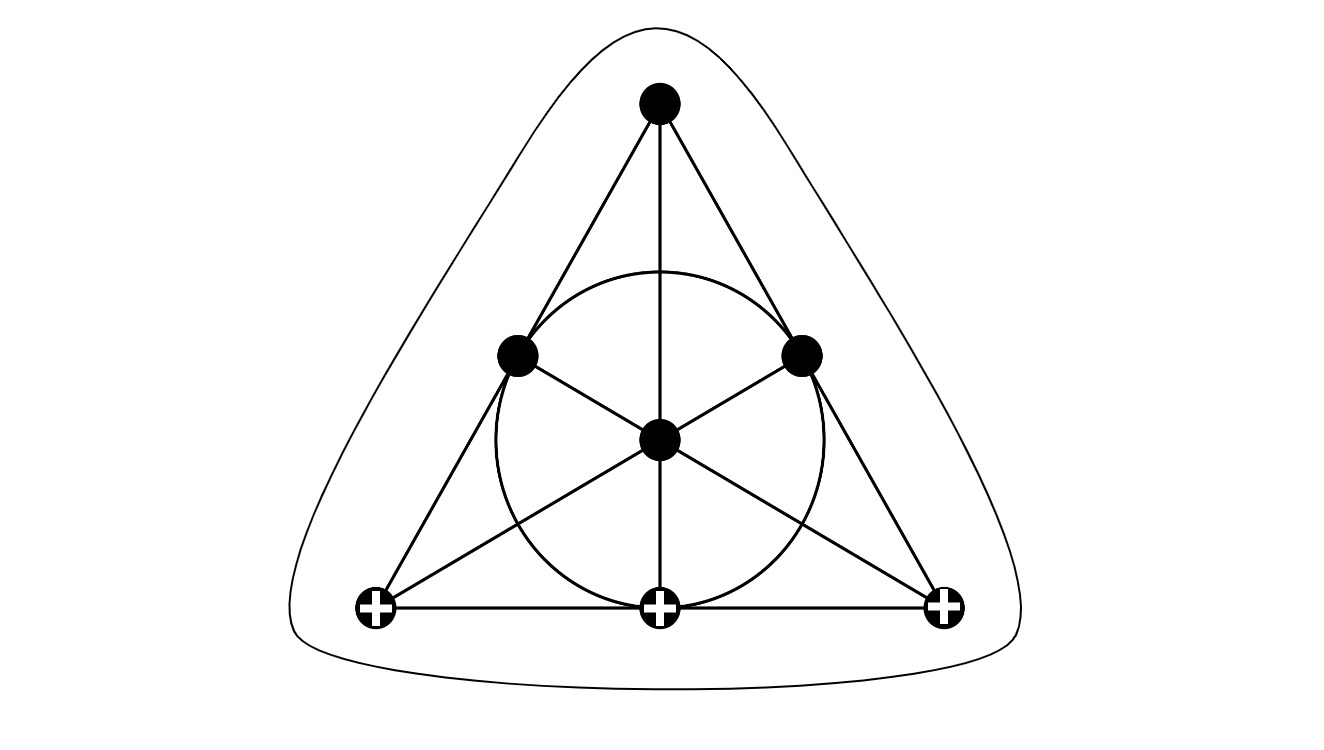}}}

\bea
&{\textrm{They give the three inequivalent contributions from $3$ marked vertices.}}&
\eea

\par
~\par
For $4_6$ we have, from top to bottom, the three $\alpha,\beta,\gamma$ diagrams:

{\centerline{\includegraphics[width=0.4\textwidth]{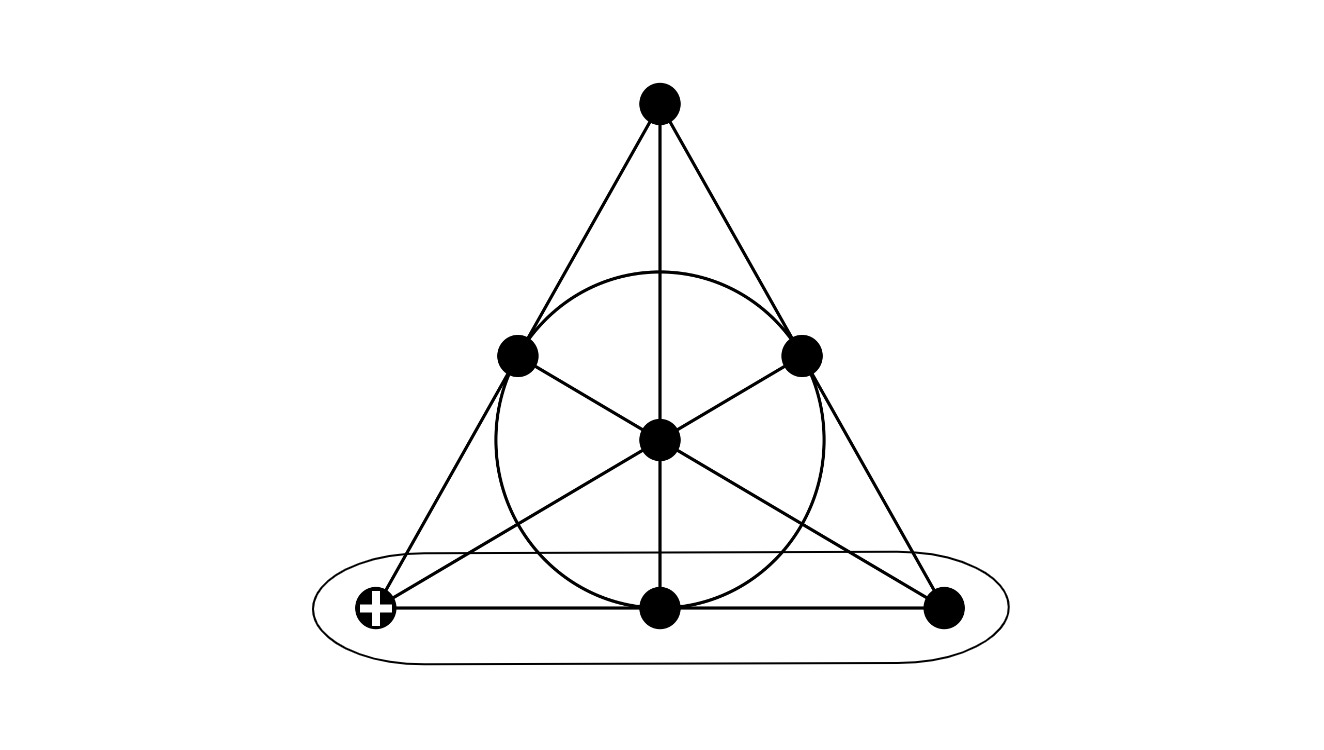}}}

{\centerline{\includegraphics[width=0.4\textwidth]{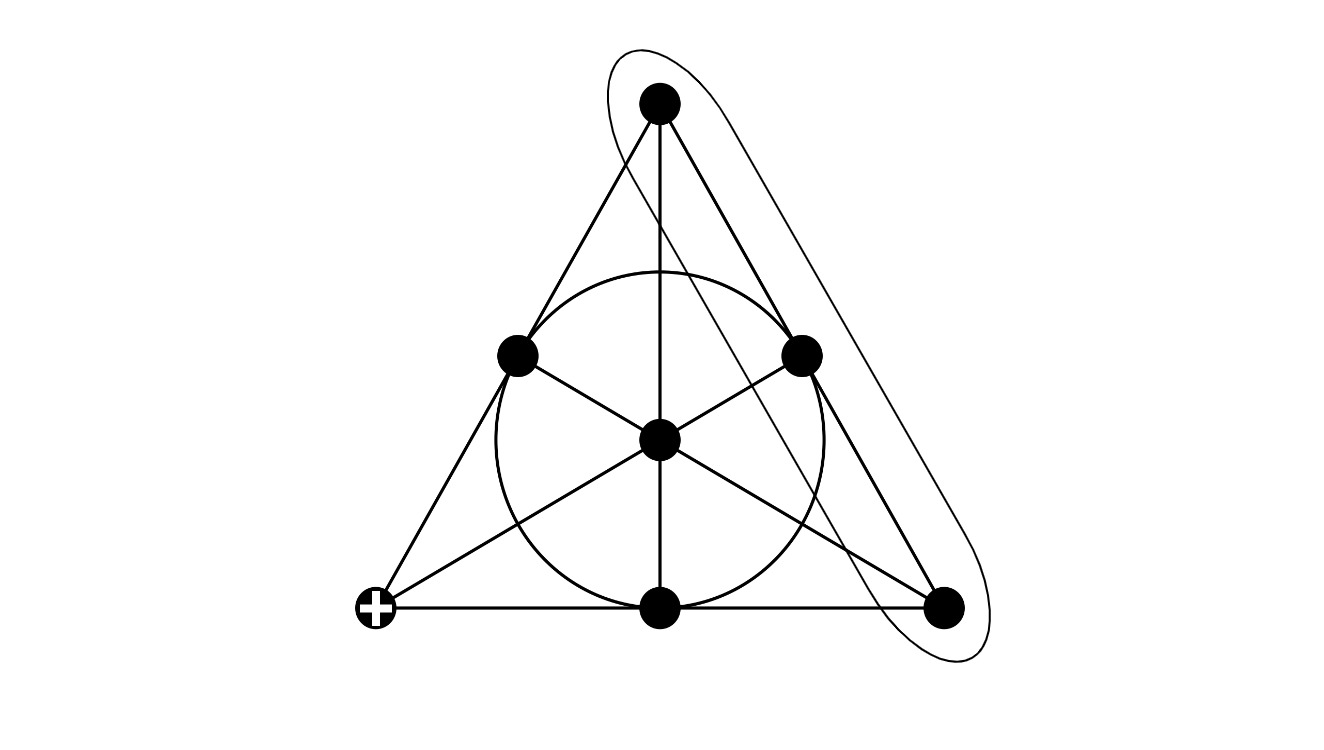}}}

{\centerline{\includegraphics[width=0.4\textwidth]{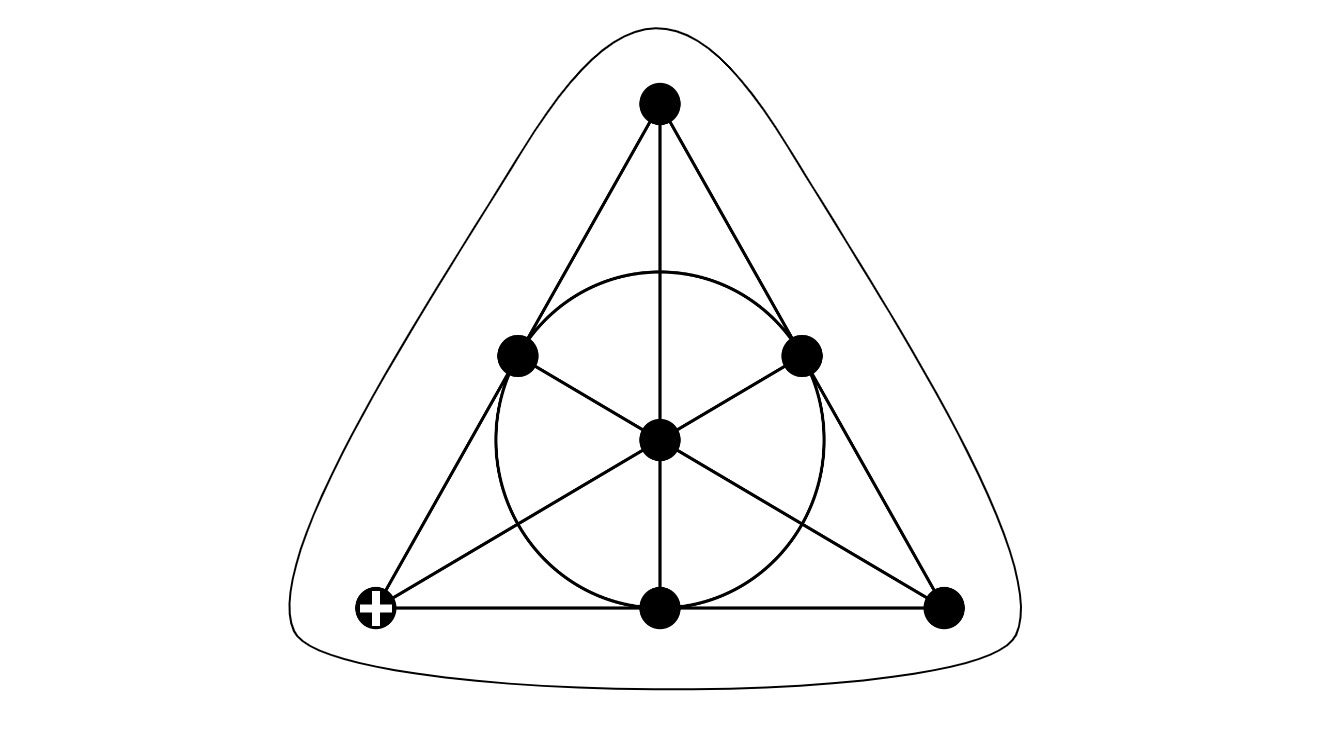}}}
\bea
&{\textrm{They give the three inequivalent contributions from $1$ marked vertex.}}&
\eea
\par
~\par
It follows that the $8$ supercharges ${Q_i}$ of the statistical transmutations are\par
~\par
$4_4 ~\Rightarrow ~ 2~ {\textrm{cases}}$, denoted as $4_{4_\alpha}$, $4_{4_\beta}$. \par
~\par
The subcase $4_{4_\alpha}$ is recovered by taking $4$  bosonic supercharges and $4$ fermionic supercharges. Therefore, we
refer to this subcase as the ``$4 B+4F$" transmutation.\par~
\par
The subcase $4_{4_\beta}$ is recovered by taking $8$ fermionic supercharges. It corresponds to the original ${\mathfrak{sqm}_{{\cal N}=8}}$ superalgebra; the ``identity transmutation" is denoted as ``$8F$".\par~
\par
$4_5 ~\Rightarrow ~ 3~ {\textrm{cases}}$, denoted as $4_{5_\alpha}$, $4_{5_\beta}$, ${4_{5_\gamma}}$. \par
~\par
The subcase $4_{5_\alpha}$ is recovered by taking $4$ parabosonic and $4$ parafermionic supercharges. Therefore, we
refer to this subcase as the ``$4P_B+4P_F$" transmutation.\par~
\par
The subcase $4_{5_\beta}$ is recovered by taking $2$ bosonic, $2$ parabosonic and $4$ parafermionic supercharges. Therefore, we
refer to this subcase as the ``$2B+2P_B+4P_F $" transmutation.\par~
\par
The subcase $4_{5_\gamma}$ is recovered by taking $8$ parafermionic supercharges. Therefore, we
refer to this subcase as a ``$8 P_F$" transmutation.\par~
\par
$4_6 ~\Rightarrow ~ 3~ {\textrm{cases}}$, denoted as $4_{6_\alpha}$, $4_{6_\beta}$, ${4_{6_\gamma}}$. \par
~\par
The subcase $4_{6_\alpha}$ is recovered by taking $4$  parabosonic and $4$ parafermionic supercharges. Therefore, we
refer to this subcase as a ``$4P_B+4 P_F$" transmutation.\par~
\par
The subcase $4_{6_\beta}$ is recovered by taking $1$ bosonic, $3$ parabosonic and $4$ fermionic supercharges. Therefore, we
refer to this subcase as the ``$1B+3P_B+4 P_F$" transmutation.\par~
\par
The subcase $4_{6_\gamma}$ is recovered by taking $8$ parafermionic supercharges. Therefore, we
refer to this subcase as an ``$8 P_B$" transmutation.\par~
\par
$4_7 ~\Rightarrow ~ 2~ {\textrm{cases}}$, denoted as $4_{7_\alpha}$, $4_{7_\beta}$. \par
~\par
The subcase $4_{7_\alpha}$ is recovered by taking $4$  parabosonic supercharges and $4$ parafermionic supercharges. Therefore, we
refer to this subcase as a ``$4 P_B+4P_F$" transmutation.\par~
\par
The subcase $4_{7_\beta}$ is recovered by taking $8$ parafermionic supercharges. Therefore, we
refer to this subcase as a ``$8P_F$" transmutation.\par~
\\
We end up with $14$ inequivalent statistical transmutations of the ${\cal N}=8$ Supersymmetric Quantum Mechanics. Their inequivalence is spotted in terms of the (para)bosonic/(para)fermionic assignments  of the $8$ supercharges $Q_i$ and the $7$ empty slots $\emptyset$; they are given by:

\bea
&\begin{array}{|c|c|c|}\hline {\textrm{Transmutations:}}&{\textrm{$8$ supercharges:}}&{\textrm{$7$ empty slots:}}\\
\hline 4_1&8B&7B\\ \hline
4_{2_\alpha}&2B+6P_B&1B+6P_B\\ \hline
4_{2_\beta}&8P_B&3B+4P_B\\ \hline
4_3&8 P_B&7P_B\\ \hline
4_{4_\alpha}&4B+4F&3B+4F\\ \hline
4_{4_\beta}&8F&7B\\ \hline
4_{5_\alpha}&4P_B+4P_F&3B+4P_F\\ \hline
4_{5_\beta}&2B+2P_B+4P_F&1B+2P_B+4P_F\\ \hline
4_{5_\gamma}&8P_F&3B+4P_B\\ \hline
4_{6_\alpha}&4P_B+4P_F&1B+2P_B+4P_F\\ \hline
4_{6_\beta}&1B+3P_B+4P_F&3P_B+4P_F\\ \hline
4_{6_\gamma}&8P_F&1B+6P_B\\ \hline
4_{7_\alpha}&4P_B+4P_F&3P_B+4P_F\\ \hline
4_{7_\beta}&8P_F&7P_B\\ \hline
\end{array}&
\eea

By summing the contributions from the seven $4$-bit tables (\ref{4bitcases}) we finally obtain
\bea
s_{{\cal N}=8} &=& 1+2+1+2+3+3+2 = 14.
\eea
\subsection{Summary}

In this Section we proved that the Universal Enveloping Superalgebras $U_{\cal N}\equiv{\cal U}({\mathfrak{sqm}_{\cal N}})$ of the one-dimensional, ${\cal N}$-extended Supersymmetric Quantum Mechanics admit alternative gradings besides the original supersymmetric grading. They define on $U_{\cal N}$  inequivalent graded Lie (super)algebras. As discussed in Section {\bf{4}}, each graded Lie (super)algebra defined on an enveloping algebra
induces its own (para)statistics. For ${\cal N}=1,2,4,8$ the respective $s_{\cal N}$ numbers of inequivalent graded Lie (super)algebras are 

\bea\label{snstattransm}
&\begin{array}{|c|c|l|} \hline {\cal N}=1:&n_1=1&s_1=1+1={\bf 2}\\ \hline {\cal N}=2:&n_2=2&s_2=1+2+1+2={\bf 6}\\ \hline
 {\cal N}=4:&n_4=3&s_4=1+2+2+3+2={\bf 10}\\ \hline {\cal N}=8:&n_8=4&s_8=1+2+1+2+3+3+2={\mathbf{14}}\\ \hline \end{array}&
\eea

~\\
In the above table the $s_{\cal N}$ numbers are partitioned into the $b_{n_{\cal N}}$ separate contributions coming from each one of the inequivalent $n_{\cal N}$-bit graded brackets presented in Appendix {\bf A}. In accordance with the discussion in Section {\bf 3}, the presence of ``marked" generators makes $s_{\cal N}\geq b_{n_{\cal N}}$.\par
The above $s_{\cal N}$ values give the total number of admissible statistical transmutations of the ${\cal N}$ supercharges $Q_i$.\par
~\par
The next relevant question is whether the $s_{\cal N}$ parastatistics imply physically observable consequences.
This question is addressed and partially answered in the next Section (the parastatistics are detected from ordinary statistics by measuring the degeneracy of the energy levels in a multiparticle sector).  For simplicity, we  discuss this issue in the
framework of the Superconformal Quantum Mechanics with the addition of the de Alfaro-Fubini-Furlan \cite{dff} oscillator term; this setting has the advantage of introducing a normalized ground state and a discrete spectrum, while preserving the spectrum-generating superconformal algebra of the original model.

\section{Detectable parastatistics in superconformal quantum mechanics with $DFF$ oscillator terms}

In this Section we address the question of the physical detectability of the statistical transmutations of the Supersymmetric Quantum Mechanics presented in Section {\bf 6}. We remind the discussion in Section {\bf 4}, namely that
each graded Lie (super)algebra defined on a Universal Enveloping Algebra, such as $U_{\cal N} := {\cal U} ({\mathfrak{sqm}}_{\cal N})$ from formula (\ref{univenvsusy}), implies its own (para)statistics in the multiparticle sector
of an associated quantum model. The inequivalent graded Lie (super)algebras applied to a single-particle sector give alternative, but {\it physically equivalent}, descriptions; the analysis of their physical equivalence/inequivalence in a multiparticle sector is model-dependent. \par
~\par
Concerning the $s_{{\cal N}}$ statistical transmutations (given in (\ref{snstattransm})) of the
${\cal N}=1,2,4,8$ Supersymmetric Quantum Mechanics (\ref{sqm}), what can
be said about their detectability?\par A complete answer to this question requires, as explained later, a deep and lengthy investigation of its own which cannot be conducted in this already long paper. Here we limit to discuss the
simplest nontrivial example (the $2$-particle sector of an ${\cal N}=2$ model). We anticipate the finding.
The $s_{{\cal N}=2}=6$ parastatistics  from (\ref{sn2}) are split into $3$ ordinary statistics (involving ordinary
bosonic/fermionic particles from the tables $2_1, 2_2$ of (\ref{2cases})) and $3$ genuine parastatistics involving paraparticles (parabosons/parafermions  from the tables $2_3, 2_4$ of (\ref{2cases})). 
The degeneracy of the $2$-particle energy levels recovered from the $3$ genuine parastatistics differs from the degeneracies produced by the $3$ ordinary statistics. Stated otherwise, in the model under consideration ordinary bosons/fermions 
{\it cannot} reproduce the degeneracy of the energy levels of the three ${\mathbb Z}_2^2$-graded parastatistics 
$2_3, 2_{4_\alpha}, 2_{4_\beta}$ introduced in subsection {\bf{6.2}}.\par
To our knowledge this is the first example where Rittenberg-Wyler's ${\mathbb Z}_2^2$-graded parabosons/\\ parafermions  are shown to affect the energy spectrum of a quantum theory. Indeed, we point out that
in the  models discussed in Section {\bf 5} and the ones in references \cite{{top1},{top2}}, a different mechanism to detect $n$-bit paraparticles from ordinary particles is at work: in those cases an observable different from the Hamiltonian has to be measured
since the same energy spectrum is produced by both ordinary particles and paraparticles.\par
~\par
The analysis of the statistical transmutations is simplified if one works in the framework of the Superconformal Quantum Mechanics, since one can use as a tool a class of simple Lie superalgebras (known as superconformal algebras) which act as spectrum-generating superalgebras. Examples of such superalgebras are $osp(1|2)$ for ${\cal N}=1$, $sl(2|1)$ for ${\cal N}=2$, the exceptional superalgebras $D(2,1;\alpha)$ for ${\cal N}=4$ (see \cite{cht}) and $F(4)$ for ${\cal N}=8$ (see \cite{quasi}).  For a review on superconformal mechanics one can consult \cite{fil} and references therein. \par The addition of the de Alfaro-Fubini-Furlan \cite{dff} oscillator term
in a Hamiltonian preserves the spectrum-generating superconformal algebra; it introduces a normalized ground state with a discrete energy spectrum. This makes particularly simple to analyze the physical consequences of the model under consideration. It is this setting that will be investigated in the following.
\subsection{On ${\cal N}=1,2,4,8$ superconformal quantum mechanics} 

A one-dimensional ${\cal N}$-extended superconformal algebra is a  simple Lie superalgebra ${\mathfrak g}$ entering the Kac's classification \cite{kac} and whose generators  satisfy the following additional properties. Besides the ${\mathbb Z}_2$-grading they possess a scaling dimension $s=-1,-\frac{1}{2},0,\frac{1}{2},1$ so that
\bea
{\mathfrak g} &=& {\mathfrak g}_{-1}\oplus {\mathfrak g}_{-\frac{1}{2}}\oplus {\mathfrak g}_0\oplus {\mathfrak g}_{\frac{1}{2}}\oplus {\mathfrak  g}_1.
\eea
The generators entering ${\mathfrak g}_{\pm \frac{1}{2}}$ are odd (fermions), while those entering ${\mathfrak g}_0, {\mathfrak g}_{\pm 1}$ are even (bosons).  The (anti)commutators are such that $[{\mathfrak g}_r, {\mathfrak g}_s\} \subset {\mathfrak g_{r+s}}$. \par
The positive subalgebra ${\mathfrak g}_{>0} = {\mathfrak g}_{\frac{1}{2}}\oplus {\mathfrak  g}_1$ is isomorphic to the ${\mathfrak {sqm}_{\cal N}}$ superalgebra (\ref{sqm}). A single generator, the Hamiltonian $H$, is accommodated in ${\mathfrak g}_1$, while the ${\cal N}$ supercharges $Q_i$ are accommodated in ${\mathfrak g}_{\frac{1}{2}}$. A single generator $K$, the conformal partner of the Hamiltonian, is accommodated in ${\mathfrak g}_{-1}$, while ${\cal N}$ generators ${\widetilde Q}_i$, the conformal partners of the supercharges $Q_i$, belong to ${\mathfrak g}_{-\frac{1}{2}}$. Finally, ${\mathfrak g}_0$ contains a dilatation operator $D$ and a subalgebra $R$,
known in the literature as the ``$R$-symmetry". The generators $H,D,K$ close an $sl(2)$ subalgebra with $D$ as the Cartan's element. 
The total number of generators of a one-dimensional ${\cal N}$-extended superconformal algebra is $d=2{\cal N}+3+r$,
where $r$ is the number of generators of the $R$-symmetry.\par
For the superconformal algebras mentioned below $osp(1|2)$ has $5$ generators (${\cal N}=1$ and $r=0$ since there is no $R$-symmetry generator), $sl(2|1)$ has $8$ generators (${\cal N}=2$ and $r=1$, the $R$-symmetry being a $u(1)$ subalgebra), $D(2,1;\alpha)$ has, for generic (see \cite{ackt}) values of $\alpha$, $17$ generators (${\cal N}=4$ and $r=6$, the $R$-symmetry being a $su(2)\oplus su(2)$ subalgebra), $F(4)$ has $40$ generators (${\cal N}=8$ and $r=21$, the $R$-symmetry being a $so(7)$ subalgebra).\par
~\par
The following differential matrix representations of superconformal algebras satisfy the irreducibility conditions,
discussed in Section {\bf 6}, for their ${\mathfrak {sqm}}_{\cal N}$ subalgebras; the presentation of these representations, in terms of the space coordinate $x$, makes use of tensor products of the $2\times 2$ matrices
{\footnotesize{$
I=\left(\begin{array}{cc}1&0\\0&1\end{array}\right),~X=\left(\begin{array}{cc}1&0\\0&-1\end{array}\right),~
Y=\left(\begin{array}{cc}0&1\\1&0\end{array}\right),~ A=\left(\begin{array}{cc}0&1\\-1&0\end{array}\right)
$
}} introduced in (\ref{letters}).\par
~
\par\par
~\par
For ${\cal N}=1$, the differential matrix representation of $osp(1|2)$ is given by
\bea
Q_1&=& \frac{1}{\sqrt{2}}\left(\partial_x\cdot A +\frac{\beta}{x}\cdot Y\right),\nonumber\\
{\widetilde Q}_1&=& \frac{i}{\sqrt{2}} x\cdot A,\nonumber\\
H&=& \frac{1}{2}\left( -\partial_x^2+\frac{\beta^2}{x^2}\right)\cdot I -\frac{\beta}{2x^2}\cdot X,\nonumber\\
D&=& -\frac{i}{2}\left( x\partial_x+\frac{1}{2}\right)\cdot I,\nonumber\\
K&=& \frac{1}{2}x^2\cdot I,
\eea
where $\beta$ is an arbitrary real parameter. The above operators are Hermitian. 
The nonvanishing $osp(1|2)$ (anti)commutators are
\bea
&\{Q_1,Q_1\} = 2H,\qquad \quad\{Q_1,{\widetilde Q}_1\}= 2 D, \qquad \quad\{{\widetilde Q}_1,{\widetilde Q}_1\} = 2K, &\nonumber\\
&[D,Q_1] ={\frac{i}{2}}Q_1,\qquad [D,{\widetilde Q}_1]= -\frac{i}{2}{\widetilde Q}_1, \qquad 
[K, Q_1]= i {\widetilde Q}_1,\qquad [K,{\widetilde Q}_1\} = -iQ_1, &\nonumber\\
&[D,H] = iH,~~~ \qquad  [D,K] = - iK, \qquad ~~~\quad [H,K] = -2i D.&
\eea
\par
~\par
For ${\cal N}=2$, the differential matrix representation of $sl(2|1)$ is given by
\bea
Q_1&=& \frac{1}{\sqrt{2}}\left(\partial_x\cdot A\otimes I +\frac{\beta}{x}\cdot Y\otimes I\right),\nonumber\\
Q_2&=& \frac{1}{\sqrt{2}}\left(\partial_x\cdot Y\otimes A +\frac{\beta}{x}\cdot A\otimes A\right),\nonumber\\
{\widetilde Q}_1&=& \frac{i}{\sqrt{2}} x\cdot A\otimes I,\nonumber\\
{\widetilde Q}_2&=& \frac{i}{\sqrt{2}} x\cdot Y\otimes  A,\nonumber\\
H&=& \frac{1}{2}\left( -\partial_x^2+\frac{\beta^2}{x^2}\right)\cdot I\otimes I -\frac{\beta}{2x^2}\cdot X\otimes I,\nonumber\\
D&=& -\frac{i}{2}\left( x\partial_x+\frac{1}{2}\right)\cdot I\otimes I,\nonumber\\
K&=& \frac{1}{2}x^2\cdot I\otimes I,\nonumber\\
W&=&\frac{i}{4}\left(X\otimes A +2\beta\cdot I\otimes A\right),
\eea
where $W$ is the $R$-symmetry generator. As before, the operators are Hermitian and $\beta $ is an arbitrary real parameter. 
The nonvanishing $sl(2|1)$ (anti)commutators are, for $j=1,2$,
\bea\label{sl21algebra}
&\{Q_1,Q_1\} = \{Q_2,Q_2\}= 2H,\qquad\{{\widetilde Q}_1,{\widetilde Q}_1\}= \{{\widetilde Q}_2, {\widetilde Q}_2\} = 2K, &\nonumber\\
&~\{Q_1,{\widetilde Q}_1\}= \{Q_2,{\widetilde Q}_2\}= 2D, \qquad \{Q_1,{\widetilde Q}_2\}=-\{Q_2,{\widetilde Q}_1\}= 2W,\nonumber\\
&[D,Q_j] ={\frac{i}{2}}Q_j,\quad [D,{\widetilde Q}_j]= -\frac{i}{2}{\widetilde Q}_j, \quad 
[K, Q_j]= i {\widetilde Q}_j,\quad [K,{\widetilde Q}_j\} = -iQ_j, &\nonumber\\
&[W,Q_1] = \frac{i}{2}Q_2, \quad [W,Q_2]= -\frac{i}{2}Q_1,\quad [W, {\widetilde Q}_1] = \frac{i}{2}{\widetilde Q}_2, \quad [W, {\widetilde Q}_2] = -\frac{i}{2}{\widetilde Q}_1,&\nonumber\\
&[D,H] = iH,~~~ \qquad  [D,K] = - iK, \qquad ~~~\quad [H,K] = -2i D.&
\eea

For ${\cal N}=4$, the differential matrix representation of $D(2,1;\alpha)$  is recovered by the repeated (anti)commutators of the $4 ~Q_i$ supercharges and the $K$ generator, given by
\bea
Q_1&=& \frac{1}{\sqrt{2}}\left(\partial_x\cdot A\otimes I\otimes I +\frac{\beta}{x}\cdot Y\otimes I\otimes I\right),\nonumber\\
Q_2&=& \frac{1}{\sqrt{2}}\left(\partial_x\cdot Y\otimes A\otimes X +\frac{\beta}{x}\cdot A\otimes A\otimes X\right),\nonumber\\
Q_3&=& \frac{1}{\sqrt{2}}\left(\partial_x\cdot Y\otimes A\otimes Y+\frac{\beta}{x}\cdot A\otimes A\otimes Y\right),\nonumber\\
Q_4&=& \frac{1}{\sqrt{2}}\left(\partial_x\cdot Y\otimes I\otimes A +\frac{\beta}{x}\cdot A\otimes  I\otimes A\right),\nonumber\\
K&=& \frac{1}{2}x^2\cdot I\otimes I\otimes I.
\eea

The above operators realize, see \cite{ackt}, the  $D(2,1;\alpha)$ superalgebra with  the identification $\alpha= \beta-\frac{1}{2}$.\par
~\par

For ${\cal N}=8$, the differential matrix representation of $F(4)$ is presented in \cite{quasi}.\par
~\par
When applying the statistical transmutations of the supersymmetry, the $D,H,K$ operators belong to the vanishing graded sector ${\underline 0}$.  For a given $i$, the supercharge $Q_i$ and its conformal superpartner ${\widetilde Q}_i$ belong to the same nonvanishing graded sector. In particular, for ${\cal N}=2$ we have 
\bea
&[H]=[K]=[D]= 00, \quad [Q_1]=[{\widetilde Q}_1]=\mu,\quad  [Q_2]=[{\widetilde Q}_2]=\nu, \quad [W]=\mu+\nu~~mod~2,&\nonumber\\
&&
\eea
for $\mu\neq\nu$ taking values $10,01,11$.\par
The de Alfaro-Fubini-Furlan \cite{dff} Hamiltonian $H_{DFF}$ is introduced through the position
\bea
H_{DFF} &=& H+K.
\eea
It corresponds to a $\beta$-deformation of a matrix quantum oscillator.\par
From now on we specialize to the ${\cal N}=2$ case, so that
\bea
H_{DFF} &=&  \frac{1}{2}\left( -\partial_x^2+\frac{\beta^2}{x^2} +x^2\right)\cdot I \otimes I-\frac{\beta}{2x^2}\cdot X
\otimes I, \qquad {\textrm{with}}\quad H_{DFF}^\dagger = H_{DFF}. \eea
We can introduce the $j=1,2$ pairs of creation/annihilation operators $a_j^\dagger, a_j$, defined through
\bea
a_j := Q_j-i {\widetilde Q}_j,  &\qquad &  a_j^\dagger := Q_j+i{\widetilde Q}_j.
\eea
 They satisfy
\bea
[H_{DFF}, a_j] = -a_j, &\quad & [H_{DFF},a_j^\dagger] = a_j^\dagger.
\eea 
Furthermore, we have
\bea
&\{a_1,a_1^\dagger\} =\{a_2,a_2^\dagger\} = 2 H_{DFF}.&
\eea
Each creation/annihilation pair satisfies a $\beta$-deformed Heisenberg algebra
\bea\label{klein}
[a_1,a_1^\dagger]=[a_2,a_2^\dagger] &=& {\mathbb I}_4 - 2\beta {\overline K}\qquad {\textrm{for ~~$ {\overline K} = X\otimes I$,}}
\eea
where ${\overline K}$ is a Klein operator satisfying
\bea
&{\overline K}^2={\mathbb I}_4, \qquad \{a_j,{\overline K}\} = \{a_j^\dagger,{\overline K}\}=0, \qquad {\textrm{with ~~ $j=1,2$.}}&
\eea
The creation operators $a_j^\dagger$ satisfy
\bea\label{n2creation}
&\{a_i^\dagger,a_j^\dagger\} = 2\delta_{ij} Z, \qquad [Z, a_j^\dagger ] = 0,\qquad {\textrm{for $i,j=1,2$,}}&
\eea
with 
\bea
Z&=& H+K+2iD.
\eea
The superalgebra (\ref{n2creation}) reproduces the ${\mathfrak{sqm}}_{{\cal N}=2}$ (anti)commutators; the difference is that the operators $a_j^\dagger, Z$ are not Hermitian.\par
~\par
An exhaustive analysis of the admissible Hilbert spaces, depending on the range of $\beta$, of the $H_{DFF}$ Hamiltonian is presented in \cite{ackt}.
For $\beta>-\frac{1}{2}$ we can introduce a single-particle Hilbert space ${\cal H}_{\beta}^{(1)}$ through the bosonic Fock vacuum  $\Psi_{\beta}$:
\bea
\Psi_{\beta}(x) &=& \frac{1}{\sqrt{\Gamma(\beta+\frac{1}{2})}}x^\beta e^{-\frac{1}{2}x^2}\left(\begin{array}{c}1\\0\\0\\0\end{array}\right), \qquad {\textrm{with}}\quad a_1\Psi_\beta(x)=a_2\Psi_\beta(x)=0.
\eea
The Gamma function ensures, see \cite{ackt}, that $\Psi_\beta(x)$ is normalized to satisfy $\int_{-\infty}^\infty dx Tr(\Psi_\beta^\dagger\Psi_\beta) = 1$.\par
~\par
The Hilbert space ${\cal H}_{\beta}^{(1)}$ is spanned by the vectors $|m;r,s\rangle$:
\bea
{\textrm{Given}}\qquad |m;r,s\rangle &:=& Z^m (a_1^\dagger)^r(a_2^\dagger)^s\Psi_\beta(x), \quad {\textrm{for $~r,s=0,1~$ and $~m=0,1,2,\ldots$,}}
\nonumber\\
{\textrm{~ we have}}\qquad {\cal H}_{\beta}^{(1)} &=& \{| m; r,s\rangle \}.
\eea

The vectors $|m;r,s\rangle$ are energy eigenstates with energy eigenvalues
\bea
H_{DFF} |m;r,s\rangle &=& E_{m;r,s}|m;r,s\rangle, \qquad E_{m;r,s}= \frac{1}{2}+\beta+ 2m+r+s.
\eea
The energy spectrum of the model is therefore given by $\frac{1}{2}+\beta + n$, with $n=0,1,2,\ldots$.\\
Apart from the vacuum energy $E_{vac}:=E_{0;0,0}=\frac{1}{2}+\beta$, all other energy levels are doubly degenerate, producing the $(1,2,2,2,\ldots )$ tower.

\subsection{Statistical transmutations of the ${\cal N}=2$ $DFF$ deformed oscillator} 

We now apply the framework discussed in Section {\bf 4} to the construction of the $2$-particle Hilbert spaces
for each one of the six ${\cal N}=2$ parastatistics introduced in subsection {\bf 6.2}.\par
For a given $\beta>-\frac{1}{2}$, the $2$-particle Hilbert spaces ${\cal H}_{\beta,\ast}^{(2)}$ are Fock spaces (the asterisk denotes the corresponding parastatististics), such that
\bea
{\cal H}_{\beta,\ast}^{(2)}&\subset &{\cal H}_{\beta}^{(1)}\otimes {\cal H}_{\beta}^{(1)}.
\eea
The $2$-particle creation/annihilation operators and Hamiltonian are respectively given by
\bea
\Delta(a_j^\dagger) &=& a_{j}^\dagger\otimes_{br}{\mathbb I}_4+{\mathbb I}_4\otimes_{br} a_{j}^\dagger ,
\qquad \qquad {\textrm{for $j=1,2$.}}\nonumber\\
\Delta(a_j) &=&a_{j}\otimes_{br}{\mathbb I}_4+{\mathbb I}_4\otimes_{br} a_{j}, 
\qquad \qquad {\textrm{for $j=1,2$.}}\nonumber\\
\Delta(H_{DFF}) &=& H_{DFF}\otimes_{br}{\mathbb I}_4+{\mathbb I}_4\otimes_{br} H_{DFF}.
\eea
The $2$-particle Fock vacuum $\Psi_{\beta;0}(x,y)$ is the $16$-component vector
\bea
\Psi_{\beta;0}(x,y)&=&\frac{1}{\Gamma(\beta+\frac{1}{2})}(xy)^\beta e^{-\frac{1}{2}(x^2+y^2)}\rho_1,
\eea
where $\rho_1$ is the $16$-component vector with entry $1$ in the first position and $0$ otherwise. \\
The normalization is chosen so that $\int\int_{-\infty}^\infty dxdy Tr (\Psi_{\beta;0}^\dagger \Psi_{\beta;0}) =1$.\par
The vacuum satisfies the conditions
\bea
&\Delta(a_1)\Psi_{\beta;0}=
\Delta(a_2)\Psi_{\beta;0}=0.&
\eea
The Hilbert spaces 
${\cal H}_{\beta,\ast}^{(2)}$ are spanned by the vectors obtained by repeatedly applying the creation operators 
$\Delta(a_1^\dagger), ~
\Delta(a_2^\dagger)$ on $\Psi_{\beta;0}(x,y)$.\par
The $2$-particle energy spectrum is given by the discrete $E_{\beta;n}^{(2)} $ energy eigenvalues
\bea
E_{\beta;n}^{(2)}&=& 1+2\beta +n, \qquad {\textrm{with $n=0,1,2,\ldots$,}}
\eea
where $E_{\beta;0}^{(2)} = 1+2\beta$ is the vacuum energy.\par
~\par
Let us introduce the symbols
\bea
&\Delta_1=\Delta(a_1^\dagger), \qquad \Delta_2=\Delta(a_2^\dagger)\qquad {\textrm{and} }\quad&\nonumber\\
&\Delta_{11}= \Delta_1\cdot \Delta_1, ~~\Delta_{22}= \Delta_2\cdot \Delta_2, ~~\Delta_{12}= \Delta_1\cdot \Delta_2, ~~\Delta_{21}= \Delta_2\cdot \Delta_1.&
\eea
Up to the second excited states, the $2$-particle Hilbert space ${\cal H}_\beta^{(2)}$ is spanned by the energy eigenvectors
\bea
E_{\beta;0}^{(2)}= 1+2\beta:&& \Psi_{\beta;0}; \nonumber\\
E_{\beta;1}^{(2)}= 2+2\beta:&& \Psi_{\beta;1}= \Delta_1\Psi_{\beta;0}, ~~\Psi_{\beta;2}= \Delta_2\Psi_{\beta;0};\nonumber\\
E_{\beta;2}^{(2)}= 3+2\beta:&& \Psi_{\beta;11}= \Delta_{11}\Psi_{\beta;0}, ~\Psi_{\beta;{22}}= \Delta_{22}\Psi_{\beta;0},
\Psi_{\beta;12}= \Delta_{12}\Psi_{\beta;0}, ~\Psi_{\beta;{21}}= \Delta_{21}\Psi_{\beta;0}.\nonumber\\&&
\eea
It follows that the vacuum state is not degenerate, the degeneracy of the first excited energy level $E_{\beta;1}^{(2)}$ is $2$, while the degeneracy of the second excited energy level $E_{\beta;2}^{(2)}$ depends on the parastatistics. \par
Indeed we get,
in terms of the $\pm 1$ signs $\delta_{11}$, $\delta_{22}$ and $\delta_{12}=\delta_{21}$:
\bea\label{deltas}
\Delta_{11} &=& Z\otimes_{br}{\mathbb I}_4+ {\mathbb I}_4\otimes_{br} Z+ (1+\delta_{11}) a_1^\dagger\otimes_{br} a_1^\dagger,\nonumber\\
\Delta_{22} &=& Z\otimes_{br}{\mathbb I}_4+ {\mathbb I}_4\otimes_{br} Z+ (1+\delta_{22}) a_2^\dagger\otimes_{br} a_2^\dagger,\nonumber\\
\Delta_{12} &=& V\otimes_{br}{\mathbb I}_4+ {\mathbb I}_4\otimes_{br} V+ a_1^\dagger\otimes_{br} a_2^\dagger+
\delta_{12} \cdot a_2^\dagger\otimes_{br} a_1^\dagger,\nonumber\\
\Delta_{21} &=& -V\otimes_{br}{\mathbb I}_4- {\mathbb I}_4\otimes_{br} V+ a_2^\dagger\otimes_{br} a_1^\dagger+
\delta_{21} \cdot a_1^\dagger\otimes_{br} a_2^\dagger.
\eea
In the above formulas we set $Z=a_1^\dagger a_1^\dagger= a_2^\dagger a_2^\dagger$ and $V= a_1^\dagger a_2^\dagger$.\par
~\par
The three independent $\pm 1$ signs $\delta_{11}, \delta_{22}, \delta_{12}$ are recovered from the  $\varepsilon_{ij}= 0,1$ entries (associated with the given graded operators) presented in the  (\ref{2cases}) tables, through the position
\bea
\delta_{ij} &=& (-1)^{\varepsilon_{ij}}.
\eea
The connection of the three signs with the six ${\cal N}=2$ parastatistics introduced in subsection {\bf 6.2} is
given by
\bea
&\begin{array}{|c|c|c|c|} \hline $parastatistics$&\delta_{11}&\delta_{22}& \delta_{12} \\ \hline
~2_1:&+1&+1&+1\\ \hline 
 ~2_3:&+1&+1&-1\\ \hline
 2_{2_\alpha}:&+1&-1&+1\\ \hline
 2_{2_\alpha}:&-1&+1&+1\\ \hline
 2_{4_\alpha}:&+1&-1&-1\\ \hline
 2_{4_\alpha}:&-1&+1&-1\\ \hline
 2_{4_\beta}:&-1&-1&+1\\ \hline
 2_{2_\beta}:&-1&-1&-1\\ \hline
\end{array}&
\eea

It follows from (\ref{deltas}) that 
\bea
\Delta_{11}=\Delta_{22}~~  &{\textrm{for}}& \delta_{11}=\delta_{22}=-1 \qquad {\textrm{($\Delta_{11}$ is not proportional to $\Delta_{22}$ otherwise),}}\nonumber\\
\Delta_{12}=-\Delta_{21}  &{\textrm{for}}& \delta_{12}=-1 \qquad \qquad ~~ {\textrm{($\Delta_{12}$ is not proportional to $\Delta_{21}$ otherwise).}}\nonumber\\
&&
\eea 
For $ \delta_{11}=\delta_{22}=-1$, the same ray vector is individuated by  $ \Psi_{\beta;11},  \Psi_{\beta;22} $; 
for $\delta_{12}=-1$ the same ray vector is individuated by $ \Psi_{\beta;12},  \Psi_{\beta;21} $.\par
~\par
The degeneracy $d_g$ of the energy levels, depending on the parastatistics, is  reported in the table below
\bea\label{n2energylevels}
&\begin{array}{|c|c|c|c|c|} \hline $parastatistics$&E=1+2\beta&E=2+2\beta&E=3+2\beta& $excitations$\\ \hline\hline
~ 2_1:&1&2&4& 2B \\ \hline
2_{2_\alpha}:&1&2&4& 1F+1B \\ \hline\hline
2_{2\beta}:&1&2&2& 2F\\ \hline\hline
~ 2_3:&1&2&3&2 P_B \\ \hline 
2_{4_\alpha}:&1&2&3&1P_F+1P_B \\ \hline
2_{4_\beta}:&1&2&3&2P_F \\ \hline
\end{array}&
\eea
The last column indicates the type of particles, (para)bosons/(para)fermions, entering the model.\par
Measuring the energy degeneracy of the second-excited level ($E=3+2\beta$) does not allow to discriminate
between the $2_1$ and $2_{2_\alpha}$ statistics ($d_g=2$) and does not allow to discriminate the three parastatistics $2_3, 2_{4_\alpha}, 2_{4_\beta}$, with $d_g=3$,  among themselves. It is nevertheless sufficient to prove, for the model under consideration, that the ${\mathbb Z}_2^2$-graded parastatistics imply a different type
of physics than the ordinary bosons/fermions statistics.

\subsection{Summary and comments}

In this Section we presented the preliminary investigation of the consequences of the algebraic statistical transmutations of supersymmetry (applied to Superconformal Quantum Mechanics with the addition 
of the de Alfaro-Fubini-Furlan oscillator term). We produced the first evidence that the ${\mathbb Z}_2^2$-graded parastatistics directly affect the energy spectrum of a given quantum model. This is a new mechanism to detect parastatistics than the one at work for the quantum models of Section {\bf 5} and the quantum models discussed in \cite{{top1},{top2}}; in those cases new observables, constructed in terms of exchange operators, have to be measured.\par
What makes the difference is the fact that the creation operators of the quantum models presented in Section {\bf 5} and in references \cite{{top1},{top2}} are nilpotent, while the creation operators of the $DFF$ deformed oscillator, as shown in formula (\ref{n2creation}), are not nilpotent.\par
In this Section we laid the ground for a systematic investigation, which will be presented in a forthcoming paper, of the detectability of the inequivalent parastatistics induced by the algebraic statistical transmutations of supersymmetry. In the Conclusions we will give more comments about ongoing and planned future investigations.
\par
~\par

\section{Conclusions}

In this paper we studied the consequences of assuming a ${\mathbb Z}_2^n$ grading for an associative ring of operators. We pointed out that the $c_n$ number of inequivalent induced compatible graded Lie (super)algebras and $n$-bit parastatistics satisfies $c_n\geq b_n$,  where the lower bound $b_n$ is given in formula (\ref{introfloor}). The equality is satisfied 
if the original set of operators, belonging to a single class of equivalence,  can be interchanged without affecting their algebraic properties. In the converse case,  when the original set of operators is split into two or more classes of equivalence, the strict inequality for $c_n$ holds. \par
We computed the $c_n$ numbers in some illustrative cases: table (\ref{introquat}) presents the results for the $2$-bit quaternions, split-quaternions  and $3$-bit biquaternions;  the results obtained for the  one-dimensional ${\cal N}=1,2,4,8$-extended Supersymmetric Quantum Mechanics (with respective values of $n$ being $n=1,2,3,4$) are reported in table (\ref{introsusy}).\par
This analysis has been the starting point to proceed to physical applications. In Section {\bf 5} we produced a class of ${\mathbb Z}_2^n$-graded quantum Hamiltonians which satisfy the $b_n$ lower bound. By using the \cite{{top1},{top2}} techniques based on graded Hopf algebras endowed with braided tensor products, we proved  
that all $b_n$ induced parastatistics are {\it physically detectable} in the multi-particle sector of these quantum models; this is so due to the fact that the presence of paraparticles can be deduced by measuring the eigenvalues of certain observables. We point out that, to our knowledge, this is the first example of $b_3=5$ inequivalent ${\mathbb Z}_2^3$-graded quantum Hamiltonians being presented.\par
As a major physical application we started investigating the {\it algebraic statistical transmutations} of the Supersymmetric and Superconformal Quantum Mechanics. The framework of the Superconformal Quantum Mechanics with the addition of the de Alfaro-Fubini-Furlan \cite{dff} oscillator term is a favourite playground to present such analysis due to the neatness of the results that it produces (uniqueness of the vacuum, discreteness of the energy spectrum, etc.).
\par
Already the simplest setting, namely the ${\cal N}=2$ model with $sl(2|1)$ spectrum-generating superalgebra, produces a new mechanism to detect ${\mathbb  Z}_2^2$-graded paraparticles that, to our knowledge, has not been discussed in the literature: in the $2$-particle sector the degeneracy of the energy eigenstates which is recovered from ${\mathbb Z}_2^2$-graded (para)bosons/(para)fermions cannot be reproduced, see table (\ref{n2energylevels}), by assigning ordinary bosonic/fermionic statistics to the particle creation operators.\par
The presented results prove that, in well-defined theoretical settings, the $n$-bit parastatistics of ${\mathbb Z}_2^n$-graded quantum Hamiltonians produce inequivocal, physically measurable, consequences. This outcome provides further motivation for the recent boost of activity, pursued by different groups and recalled in the Introduction, on physical and mathematical aspects of
${\mathbb Z}_2^n$-graded color Lie algebras and superalgebras.\par
It is in the light of these considerations that we
introduced, in Appendix {\bf B}, a Boolean logic gates presentation of the inequivalent ${\mathbb Z}_2^n$-graded brackets
of Lie type. This presentation can be regarded as a possible blueprint to help experimentalists to either simulate or engineer in
the laboratory the $n$-bit paraparticles (in the Introduction we already recalled, for a different type of parastatistics, the simulated/engineered paraparticles dicussed in \cite{{parasim},{paraexp}}).~\par
Possibly, the most promising applications would come from condensed matter; it was pointed out in \cite {z2z2sdiv} that  ${\mathbb Z}_2^2$-graded superdivision algebras  allow to go beyond the celebrated \cite{{kit},{rsfl}} $10$-fold way of the periodic table of topological insulators and superconductors (in \cite {z2z2sdiv} the connection of certain parafermionic Hamiltonians with ${\mathbb Z}_2^2$-graded superdivision algebras has been illustrated).
~\par
Section {\bf 7} lays the ground for a systematic analysis, which will be presented in a forthcoming paper, of the statistical transmutations of the one-dimensional ${\cal N}$-extended Superconformal Quantum Mechanics. We  sketch the main issues that will be postponed to this future work; they include the derivation of the statistical transmutations of
the models with spectrum-generating superalgebras  $D(2,1;\alpha)$ (for ${\cal N}=4$) and $F(4)$ (for ${\cal N}=8$), the degeneracies of the multiparticle energy spectra and the analysis of the possible role of extra observables related with the $R$-symmetry generators. A fascinating topics concerns the {\it statistically transmuted}  spectrum-generating graded (super)algebras.  Each parastatistics defines its own spectrum-generating (super)algebra which controls the energy degeneracy of the multiparticle sector. The {\it transmuted} spectrum-generated (super)algebras are finite, but not necessarily linear (super)algebras (alternatively, in the unfolded formalism, they are presented as  infinitely generated, linear, graded (super)algebras). In this dynamical setting the role of the {\it simple} ${\mathbb Z}_2\times{\mathbb Z}_2$-graded Lie superalgebras, as the ones introduced in \cite{{tol1},{stvj1},{stvj2}}, should be properly understood. Their connection with parastatistics was motivated by the \cite{{gapa},{pal}} interpretation that the Green's trilinear relations \cite{gre} are implemented as graded Jacobi identities of superalgebras. On the other hand,
the famous Wigner's quantization of the harmonic oscillator presented in \cite{wig} can be rephrased, with a modern terminology \cite{top0}, as a case of statistical transmutation. On the bosonic side we have the non-semisimple Schr\"odinger algebra as dynamical symmetry of the Schr\"odinger equation \cite{{nie},{nie2}} and which acts as spectrum-generating algebra
of the harmonic oscillator; on the fermionic side of the picture the  spectrum of the harmonic oscillator is recovered from an irreducible representation of the $osp(1|2)$ superalgebra. The ${\mathbb Z}_2^n$-graded extensions of this
construction to $n$-bit parastatistics require a careful investigation.\par
We finally mention that in this paper Klein operators, such as the one entering formula (\ref{klein}), have been derived but not directly used in the construction of parastatistics. Klein operators were employed in \cite{ara} to switch statistics. The \cite{que} construction, which makes use of them to describe a ${\mathbb Z}_2^2$-graded Lie superalgebra, can be extended to any ${\mathbb Z}_2^n$-graded compatible Lie (super)algebra inducing an $n$-bit parastatistics.

\appendix 

\titleformat{name=\section}[display] 
 {\normalfont\Large\bfseries}{}{0pt}{}

\renewcommand{\theequation}{A.\arabic{equation}} 
 
\section{Appendix A: tables (up to $n\leq 4$) of the inequivalent graded brackets compatible with $n$-bit particles' assignments}
	
In this Appendix we present the tables (up to $n\leq 4$) of the inequivalent graded Lie brackets which are compatible with the $n$-bit assignments of particles (para)statistics. They are given by ${\mathbb Z}_2^n\times{\mathbb Z}_2^n\rightarrow {\mathbb Z}_2$ mappings. Following our conventions the $0$ entries correspond to commutators, while the $1$ entries correspond to anticommutators. The tables are symmetric in the exchange of rows and columns. 
The tables with {\it all} vanishing elements in the diagonal define the (para)bosonic graded algebras; in \cite{{riwy1},{riwy2},{sch}},
by taking into account the $mod~2$ property, they were expressed as antisymmetric matrices (one can just flip the signs of the lower triangular entries, so that $+1\mapsto -1$); to make contact with the Boolean logic we prefer here to work with non-negative entries only. A $0$ diagonal entry implies that the corresponding particle is a (para)boson, while a $1$ diagonal entry implies that the corresponding particle is a (para)fermion satisfying the Pauli's exclusion principle (see \cite{top2} for a discussion).\par
The following tables are labeled as ``$n_k$", denoting the $k$-th graded bracket of the $n$-bit case; $k$ is restricted to the values $k=1,\ldots, b_n$, where the maximal value  $b_n$ is given in (\ref{floor}). The representatives of the inequivalent brackets are:\par
~
\\
- {\it The $b_1=2$ inequivalent brackets of $n=1$}:

\bea \label{1bitcases}
{\textrm{$1_1$ case:}}\qquad &&\begin{array}{c|cc}&0&1\\ \hline 0&0&0\\1&0&0
\end{array} \qquad {\textrm{(comment: an ordinary Lie algebra),}}
\nonumber\\
&&\nonumber\\
{\textrm{$1_2$ case:}}\qquad&& \begin{array}{c|cc}&0&1\\ \hline 0&0&0\\1&0&1
\end{array} \qquad {\textrm{(comment: an ordinary Lie superalgebra);}}
\nonumber\\
\eea
~\\
- {\it The $b_2=4$ inequivalent brackets of $n=2$}:

\bea \label{2cases}
{\textrm{$2_1$ case:}}\qquad &&\begin{array}{c|cccc}&00&10&01&11\\ \hline 00&0&0&0&0\\10&0&0&0&0\\01&0&0&0&0\\11&0&0&0&0
\end{array} \qquad {\textrm{(comment:  an ordinary Lie algebra),}}
\nonumber\\
&&\nonumber\\
{\textrm{$2_2$ case:}}\qquad&& 
\begin{array}{c|cccc}&00&10&01&11\\ \hline 00&0&0&0&0\\10&0&1&1&0\\01&0&1&1&0\\11&0&0&0&0
\end{array}\qquad
 {\textrm{(comment: an ordinary ${\mathbb Z}_2$-graded Lie superalgebra),}}
\nonumber\\&&\nonumber\\
{\textrm{$2_3$ case:}}\qquad&& 
\begin{array}{c|cccc}&00&10&01&11\\ \hline 00&0&0&0&0\\10&0&0&1&1\\01&0&1&0&1\\11&0&1&1&0
\end{array}\qquad
 {\textrm{(comment: the ${\mathbb Z}_2^2$ color Lie algebra),}}\nonumber\\
&&\nonumber\\
{\textrm{$2_4$ case:}}\qquad&& 
\begin{array}{c|cccc}&00&10&01&11\\ \hline 00&0&0&0&0\\10&0&1&0&1\\01&0&0&1&1\\11&0&1&1&0
\end{array}\qquad
 {\textrm{(comment: the ${\mathbb Z}_2^2$ color Lie superalgebra);}}\nonumber\\&&
\eea

{\it Comment}: for each one of the above four cases equivalent presentations are obtained by rearranging (permuting) rows and columns.\\
~\\
- {\it The $b_3=5$ inequivalent brackets of $n=3$}:\\
~\\
the rows (columns) are labeled by $3$-bit, $\alpha_1,\alpha_2,\alpha_3$ (and, respectively, $\beta_1,\beta_2,\beta_3$). The $1$-bit entries are expressed as $mod~ 2$ formulas.\\
~\\
{\footnotesize \bea \label{inequivalent3}
{\textrm{{\normalsize{$3_1$ case:}}}}\qquad &&\begin{array}{c|cccccccc}&000&100&010&001&110&101&011&111\\ \hline 000&0&0&0&0&0&0&0&0\\
100&0&0&0&0&0&0&0&0\\
010&0&0&0&0&0&0&0&0\\
001&0&0&0&0&0&0&0&0\\
110&0&0&0&0&0&0&0&0\\
101&0&0&0&0&0&0&0&0\\
011&0&0&0&0&0&0&0&0\\
111&0&0&0&0&0&0&0&0
\end{array} \qquad {\textrm{{\normalsize{(the Lie algebra with all $0$ entries),}}}}\nonumber
\\&&\nonumber\\
{\textrm{{\normalsize{$3_2$ case:}}}}\qquad &&\begin{array}{c|cccccccc}&000&100&010&001&110&101&011&111\\ \hline 000&0&0&0&0&0&0&0&0\\
100&0&0&1&0&1&0&1&1\\
010&0&1&0&0&1&1&0&1\\
001&0&0&0&0&0&0&0&0\\
110&0&1&1&0&0&1&1&0\\
101&0&0&1&0&1&0&1&1\\
011&0&1&0&0&1&1&0&1\\
111&0&1&1&0&0&1&1&0
\end{array} \qquad {\textrm{{\normalsize{(from $\alpha_1\beta_2+\alpha_2\beta_1$ mod $2$),}}}}\nonumber
\\&&\nonumber\\
{\textrm{{\normalsize{$3_3$ case:}}}}\qquad &&\begin{array}{c|cccccccc}&000&100&010&001&110&101&011&111\\ \hline 000&0&0&0&0&0&0&0&0\\
100&0&1&0&0&1&1&0&1\\
010&0&0&0&0&0&0&0&0\\
001&0&0&0&0&0&0&0&0\\
110&0&1&0&0&1&1&0&1\\
101&0&1&0&0&1&1&0&1\\
011&0&0&0&0&0&0&0&0\\
111&0&1&0&0&1&1&0&1
\end{array} \qquad {\textrm{{\normalsize{(from $\alpha_1\beta_1$ mod $2$),}}}}\nonumber
\\&&\nonumber\\
{\textrm{{\normalsize{$3_4$ case:}}}}\qquad &&\begin{array}{c|cccccccc}&000&100&010&001&110&101&011&111\\ \hline 000&0&0&0&0&0&0&0&0\\
100&0&1&0&0&1&1&0&1\\
010&0&0&1&0&1&0&1&1\\
001&0&0&0&0&0&0&0&0\\
110&0&1&1&0&0&1&1&0\\
101&0&1&0&0&1&1&0&1\\
011&0&0&1&0&1&0&1&1\\
111&0&1&1&0&0&1&1&0
\end{array} \qquad {\textrm{{\normalsize{(from $\alpha_1\beta_1+\alpha_2\beta_2$ mod $2$),}}}}\nonumber
\\&&\nonumber\\
{\textrm{{\normalsize{$3_5$ case:}}}}\qquad &&\begin{array}{c|cccccccc}&000&100&010&001&110&101&011&111\\ \hline 
000&0&0&0&0&0&0&0&0\\
100&0&1&0&0&1&1&0&1\\
010&0&0&1&0&1&0&1&1\\
001&0&0&0&1&0&1&1&1\\
110&0&1&1&0&0&1&1&0\\
101&0&1&0&1&1&0&1&0\\
011&0&0&1&1&1&1&0&0\\
111&0&1&1&1&0&0&0&1
\end{array} \qquad {\textrm{{\normalsize{(from $\alpha_1\beta_1+\alpha_2\beta_2+\alpha_3\beta_3$ mod $2$).}}}}
\nonumber\\
 &&
\eea}}
The inequivalence of the $5$ graded brackets is spotted in terms of:\par
~{\it i}) the number $R(n_k)$ of nonvanishing rows and \par
{\it ii}) the trace  $Tr(n_k)$ of the above matrices.\par
We have
\bea
Tr(3_1)=Tr(3_2)=0, && Tr(3_3)=Tr(3_4)=Tr(3_5) =4,
\eea
which implies that the cases $3_1$ and $3_2$ correspond to (para)bosonic Lie algebras.\par
The numbers of nonvanishing rows are given by
\bea
&R(3_1)=0, \quad R(3_2)= 6, \quad R(3_3)=4, \quad R(3_4)=6, \quad R(3_5) =7.&
\eea
~\\
-
{\it The $b_4=7$ inequivalent brackets of $n=4$}:
\\~\\
the rows (columns) are labeled by $4$-bit, $\alpha_1,\alpha_2,\alpha_3,\alpha_4$ (and, respectively, $\beta_1,\beta_2,\beta_3,\beta_4$). The $1$-bit entries are expressed as $mod~ 2$ formulas.
{\footnotesize \bea 
&{\textrm{{\normalsize{$4_1$ case:}}}}&\nonumber\\&\begin{array}{c|cccccccccccccccc}&0000&1000&0100&0010&0001&1100&1010&1001&0110&0101&0011&1110&1101&1011&0111&1111\\ \hline 
0000&0&0&0&0&0&0&0&0&0&0&0&0&0&0&0&0\\ 
1000&0&0&0&0&0&0&0&0&0&0&0&0&0&0&0&0\\ 
0100&0&0&0&0&0&0&0&0&0&0&0&0&0&0&0&0\\ 
0010&0&0&0&0&0&0&0&0&0&0&0&0&0&0&0&0\\ 
0001&0&0&0&0&0&0&0&0&0&0&0&0&0&0&0&0\\ 
1100&0&0&0&0&0&0&0&0&0&0&0&0&0&0&0&0\\ 
1010&0&0&0&0&0&0&0&0&0&0&0&0&0&0&0&0\\ 
1001&0&0&0&0&0&0&0&0&0&0&0&0&0&0&0&0\\ 
0110&0&0&0&0&0&0&0&0&0&0&0&0&0&0&0&0\\ 
0101&0&0&0&0&0&0&0&0&0&0&0&0&0&0&0&0\\ 
0011&0&0&0&0&0&0&0&0&0&0&0&0&0&0&0&0\\ 
1110&0&0&0&0&0&0&0&0&0&0&0&0&0&0&0&0\\ 
1101&0&0&0&0&0&0&0&0&0&0&0&0&0&0&0&0\\ 
1011&0&0&0&0&0&0&0&0&0&0&0&0&0&0&0&0\\ 
0111&0&0&0&0&0&0&0&0&0&0&0&0&0&0&0&0\\ 
1111&0&0&0&0&0&0&0&0&0&0&0&0&0&0&0&0\\ 
\end{array} &\nonumber
\eea
~
\bea
&{\textrm{{\normalsize{$4_2$ case:}}}}&\nonumber\\&\begin{array}{c|cccccccccccccccc}&0000&1000&0100&0010&0001&1100&1010&1001&0110&0101&0011&1110&1101&1011&0111&1111\\ \hline 
0000&0&0&0&0&0&0&0&0&0&0&0&0&0&0&0&0\\ 
1000&0&0&1&0&0&1&0&0&1&1&0&1&1&0&1&1\\ 
0100&0&1&0&0&0&1&1&1&0&0&0&1&1&1&0&1\\ 
0010&0&0&0&0&0&0&0&0&0&0&0&0&0&0&0&0\\ 
0001&0&0&0&0&0&0&0&0&0&0&0&0&0&0&0&0\\ 
1100&0&1&1&0&0&0&1&1&1&1&0&0&0&1&1&0\\ 
1010&0&0&1&0&0&1&0&0&1&1&0&1&1&0&1&1\\ 
1001&0&0&1&0&0&1&0&0&1&1&0&1&1&0&1&1\\ 
0110&0&1&0&0&0&1&1&1&0&0&0&1&1&1&0&1\\ 
0101&0&1&0&0&0&1&1&1&0&0&0&1&1&1&0&1\\ 
0011&0&0&0&0&0&0&0&0&0&0&0&0&0&0&0&0\\ 
1110&0&1&1&0&0&0&1&1&1&1&0&0&0&1&1&0\\ 
1101&0&1&1&0&0&0&1&1&1&1&0&0&0&1&1&0\\ 
1011&0&0&1&0&0&1&0&0&1&1&0&1&1&0&1&1\\ 
0111&0&1&0&0&0&1&1&1&0&0&0&1&1&1&0&1\\ 
1111&0&1&1&0&0&0&1&1&1&1&0&0&0&1&1&0\\ 
\end{array} &\nonumber
\eea
~
\bea
&{\textrm{{\normalsize{$4_3$ case:}}}}&\nonumber\\&\begin{array}{c|cccccccccccccccc}&0000&1000&0100&0010&0001&1100&1010&1001&0110&0101&0011&1110&1101&1011&0111&1111\\ \hline 
0000&0&0&0&0&0&0&0&0&0&0&0&0&0&0&0&0\\ 
1000&0&0&1&0&0&1&0&0&1&1&0&1&1&0&1&1\\ 
0100&0&1&0&0&0&1&1&1&0&0&0&1&1&1&0&1\\ 
0010&0&0&0&0&1&0&0&1&0&1&1&0&1&1&1&1\\ 
0001&0&0&0&1&0&0&1&0&1&0&1&1&0&1&1&1\\ 
1100&0&1&1&0&0&0&1&1&1&1&0&0&0&1&1&0\\ 
1010&0&0&1&0&1&1&0&1&1&0&1&1&0&1&0&0\\ 
1001&0&0&1&1&0&1&1&0&0&1&1&0&1&1&0&0\\ 
0110&0&1&0&0&1&1&1&0&0&1&1&1&0&0&1&0\\ 
0101&0&1&0&1&0&1&0&1&1&0&1&0&1&0&1&0\\ 
0011&0&0&0&1&1&0&1&1&1&1&0&1&1&0&0&0\\ 
1110&0&1&1&0&1&0&1&0&1&0&1&0&1&0&0&1\\ 
1101&0&1&1&1&0&0&0&1&0&1&1&1&0&0&0&1\\ 
1011&0&0&1&1&1&1&1&1&0&0&0&0&0&0&1&1\\ 
0111&0&1&0&1&1&1&0&0&1&1&0&0&0&1&0&1\\ 
1111&0&1&1&1&1&0&0&0&0&0&0&1&1&1&1&0\\ 
\end{array} &\nonumber
\eea
~
\bea
&{\textrm{{\normalsize{$4_4$ case:}}}}&\nonumber\\&\begin{array}{c|cccccccccccccccc}&0000&1000&0100&0010&0001&1100&1010&1001&0110&0101&0011&1110&1101&1011&0111&1111\\ \hline 
0000&0&0&0&0&0&0&0&0&0&0&0&0&0&0&0&0\\ 
1000&0&1&0&0&0&1&1&1&0&0&0&1&1&1&0&1\\ 
0100&0&0&0&0&0&0&0&0&0&0&0&0&0&0&0&0\\ 
0010&0&0&0&0&0&0&0&0&0&0&0&0&0&0&0&0\\ 
0001&0&0&0&0&0&0&0&0&0&0&0&0&0&0&0&0\\ 
1100&0&1&0&0&0&1&1&1&0&0&0&1&1&1&0&1\\ 
1010&0&1&0&0&0&1&1&1&0&0&0&1&1&1&0&1\\ 
1001&0&1&0&0&0&1&1&1&0&0&0&1&1&1&0&1\\ 
0110&0&0&0&0&0&0&0&0&0&0&0&0&0&0&0&0\\ 
0101&0&0&0&0&0&0&0&0&0&0&0&0&0&0&0&0\\ 
0011&0&0&0&0&0&0&0&0&0&0&0&0&0&0&0&0\\ 
1110&0&1&0&0&0&1&1&1&0&0&0&1&1&1&0&1\\ 
1101&0&1&0&0&0&1&1&1&0&0&0&1&1&1&0&1\\ 
1011&0&1&0&0&0&1&1&1&0&0&0&1&1&1&0&1\\ 
0111&0&0&0&0&0&0&0&0&0&0&0&0&0&0&0&0\\ 
1111&0&1&0&0&0&1&1&1&0&0&0&1&1&1&0&1\\ 
\end{array} &\nonumber
\eea
~
\bea
&{\textrm{{\normalsize{$4_5$ case:}}}}&\nonumber\\&\begin{array}{c|cccccccccccccccc}&0000&1000&0100&0010&0001&1100&1010&1001&0110&0101&0011&1110&1101&1011&0111&1111\\ \hline 
0000&0&0&0&0&0&0&0&0&0&0&0&0&0&0&0&0\\ 
1000&0&1&0&0&0&1&1&1&0&0&0&1&1&1&0&1\\ 
0100&0&0&1&0&0&1&0&0&1&0&0&1&1&0&1&1\\ 
0010&0&0&0&0&0&0&0&0&0&0&0&0&0&0&0&0\\ 
0001&0&0&0&0&0&0&0&0&0&0&0&0&0&0&0&0\\ 
1100&0&1&1&0&0&0&1&1&1&1&0&0&0&1&1&0\\ 
1010&0&1&0&0&0&1&1&1&0&0&0&1&1&1&0&1\\ 
1001&0&1&0&0&0&1&1&1&0&0&0&1&1&1&0&1\\ 
0110&0&0&1&0&0&1&0&0&1&1&0&1&1&0&1&1\\ 
0101&0&0&1&0&0&1&0&0&1&1&0&1&1&0&1&1\\ 
0011&0&0&0&0&0&0&0&0&0&0&0&0&0&0&0&0\\ 
1110&0&1&1&0&0&0&1&1&1&1&0&0&0&1&1&0\\ 
1101&0&1&1&0&0&0&1&1&1&1&0&0&0&1&1&0\\ 
1011&0&1&0&0&0&1&1&1&0&0&0&1&1&1&0&1\\ 
0111&0&0&1&0&0&1&0&0&1&1&0&1&1&0&1&1\\ 
1111&0&1&1&0&0&0&1&1&1&1&0&0&0&1&1&0\\ 
\end{array} &\nonumber
\eea
~
\bea
&{\textrm{{\normalsize{$4_6$ case:}}}}&\nonumber\\&\begin{array}{c|cccccccccccccccc}&0000&1000&0100&0010&0001&1100&1010&1001&0110&0101&0011&1110&1101&1011&0111&1111\\ \hline 
0000&0&0&0&0&0&0&0&0&0&0&0&0&0&0&0&0\\ 
1000&0&1&0&0&0&1&1&1&0&0&0&1&1&1&0&1\\ 
0100&0&0&1&0&0&1&0&0&1&1&0&1&1&0&1&1\\ 
0010&0&0&0&1&0&0&1&0&1&0&1&1&0&1&1&1\\ 
0001&0&0&0&0&0&0&0&0&0&0&0&0&0&0&0&0\\ 
1100&0&1&1&0&0&0&1&1&1&1&0&0&0&1&1&0\\ 
1010&0&1&0&1&0&1&0&1&1&0&1&0&1&0&1&0\\ 
1001&0&1&0&0&0&1&1&1&0&0&0&1&1&1&0&1\\ 
0110&0&0&1&1&0&1&1&0&0&1&1&0&1&1&0&0\\ 
0101&0&0&1&0&0&1&0&0&1&1&0&1&1&0&1&1\\ 
0011&0&0&0&1&0&0&1&0&1&0&1&1&0&1&1&1\\ 
1110&0&1&1&1&0&0&0&1&0&1&1&1&0&0&0&1\\ 
1101&0&1&1&0&0&0&1&1&1&1&0&0&0&1&1&0\\ 
1011&0&1&0&1&0&1&0&1&1&0&1&0&1&0&1&0\\ 
0111&0&0&1&1&0&1&1&0&0&1&1&0&1&1&0&0\\ 
1111&0&1&1&1&0&0&0&1&0&1&1&1&0&0&0&1\\ 
\end{array} &\nonumber
\eea
~
\bea
&{\textrm{{\normalsize{$4_7$ case:}}}}&\nonumber\\&\begin{array}{c|cccccccccccccccc}&0000&1000&0100&0010&0001&1100&1010&1001&0110&0101&0011&1110&1101&1011&0111&1111\\ \hline 
0000&0&0&0&0&0&0&0&0&0&0&0&0&0&0&0&0\\ 
1000&0&1&0&0&0&1&1&1&0&0&0&1&1&1&0&1\\ 
0100&0&0&1&0&0&1&0&0&1&1&0&1&1&0&1&1\\ 
0010&0&0&0&1&0&0&1&0&1&0&1&1&0&1&1&1\\ 
0001&0&0&0&0&1&0&0&1&0&1&1&0&1&1&1&1\\ 
1100&0&1&1&0&0&0&1&1&1&1&0&0&0&1&1&0\\ 
1010&0&1&0&1&0&1&0&1&1&0&1&0&1&0&1&0\\ 
1001&0&1&0&0&1&1&1&0&0&1&1&1&0&0&1&0\\ 
0110&0&0&1&1&0&1&1&0&0&1&1&0&1&1&0&0\\ 
0101&0&0&1&0&1&1&0&1&1&0&1&1&0&1&0&0\\ 
0011&0&0&0&1&1&0&1&1&1&1&0&1&1&0&0&0\\ 
1110&0&1&1&1&0&0&0&1&0&1&1&1&0&0&0&1\\ 
1101&0&1&1&0&1&0&1&0&1&0&1&0&1&0&0&1\\ 
1011&0&1&0&1&1&1&0&0&1&1&0&0&0&1&0&1\\ 
0111&0&0&1&1&1&1&1&1&0&0&0&0&0&0&1&1\\ 
1111&0&1&1&1&1&0&0&0&0&0&0&1&1&1&1&0\\ 
\end{array} &\nonumber
\eea
}

\bea\label{4bitcases}
&&\eea
The entries of the corresponding cases are given by\par
$4_1$: $0$,\par
$4_2$: $\alpha_1\beta_2+\alpha_2\beta_1$ mod $2$ ,\par
$4_3$: $\alpha_1\beta_2+\alpha_2\beta_1+\alpha_3\beta_4+\alpha_4\beta_3$ mod $2$,\par 
$4_4$: $\alpha_1\beta_1$ mod $2$,\par
$4_5$: $\alpha_1\beta_1+\alpha_2\beta_2$ mod $2$,\par $4_6$: $\alpha_1\beta_1+\alpha_2\beta_2+\alpha_3\beta_3$ mod $2$,\par $4_7$: $\alpha_1\beta_1+\alpha_2\beta_2+\alpha_3\beta_3+\alpha_4\beta_4$ mod $2$.

\bea&&\eea
The inequivalence of the $7$ graded brackets is spotted in terms of:\par
~{\it i}) the number $R(n_k)$ of nonvanishing rows and \par
{\it ii}) the trace  $Tr(n_k)$ of the above matrices.\par
We have
\bea
Tr(4_1)=Tr(4_2)=Tr(4_3)=0, && Tr(4_4)=Tr(4_5)=Tr(4_6) =Tr(4_7)=8,
\eea
which implies that the cases $4_1$, $4_2$ and $4_3$ correspond to (para)bosonic Lie algebras.\par
The numbers of nonvanishing rows are given by
\bea
&R(4_1)=0, ~~ R(4_2)= 12, ~~ R(4_3)=15, ~~ R(4_4)=8, ~~ R(4_5) =12,~~ R(4_6)=14,~~ R(4_7)=15.&\nonumber\\
&&
\eea

\titleformat{name=\section}[display] 
{\normalfont\Large\bfseries}{}{0pt}{}
\renewcommand{\theequation}{B.\arabic{equation}}
\section{Appendix B: The logic gates presentation of graded Lie brackets}

In this Appendix we illustrate how the tables of the inequivalent graded Lie brackets are reformulated in terms of Boolean logic gates (the motivations to introduce this reformulation have been discussed in Subsection {\bf 2.2}). The Boolean logic presentation requires a few steps. At first the graded sectors entering  the tables given in Appendix {\bf A} are rearranged in a Gray code presentation (only one bit  changes from one graded sector to the next one). This rearrangement offers the possibility to use Karnaugh maps which, further simplified, allow to express the graded-bracket tables in terms of the logical gates ``AND", ``OR",``XOR" and of the ``NOT" operation.  In their tables of truth, the outputs of these operations acting on  $a,b= 0,1$ are
{\footnotesize{
\bea
&{\textrm{NOT:}}~
 \begin{array}{|ccc|}  
\hline
0&\Rightarrow & 1\\
1& \Rightarrow &0 \\
\hline
\end{array}~, \qquad
{\textrm{AND:}}~
 \begin{array}{|ccc|}  
\hline
0,0 &\Rightarrow & 0\\
0,1& \Rightarrow &0\\1,0&\Rightarrow &0\\1,1&\Rightarrow&1 \\
\hline
\end{array} ~,
\qquad {\textrm{OR:}}~
 \begin{array}{|ccc|}  
\hline
0,0 &\Rightarrow & 0\\
0,1& \Rightarrow &1\\1,0&\Rightarrow &1\\1,1&\Rightarrow&1\\
\hline
\end{array}~ ,
\qquad
{\textrm{XOR:}}~
 \begin{array}{|ccc|}  
\hline
0,0 &\Rightarrow & 0\\
0,1& \Rightarrow &1\\1,0&\Rightarrow &1\\1,1&\Rightarrow&0 \\
\hline
\end{array}~ .
 &\nonumber\\
&&
\eea
}}
The symbols used to denote the above operations are
\bea
&{\textrm{NOT:}} ~ a\mapsto {\overline{a}},\quad
{\textrm{AND:}} ~ a,b\mapsto {{a\cdot b}},\quad 
{\textrm{OR:}} ~ a,b\mapsto {{a+b}},\quad 
{\textrm{XOR:}} ~ a,b \mapsto {{a\oplus b}}. &
\eea

For $n=2$, Gray code presentations of the $2_2$, $2_3$, $2_4$ cases of Appendix {\bf A} are given by \par
~\par
- $2_2$ case:
\bea &&
{\textrm{
{\centerline{\includegraphics[width=0.6\textwidth]{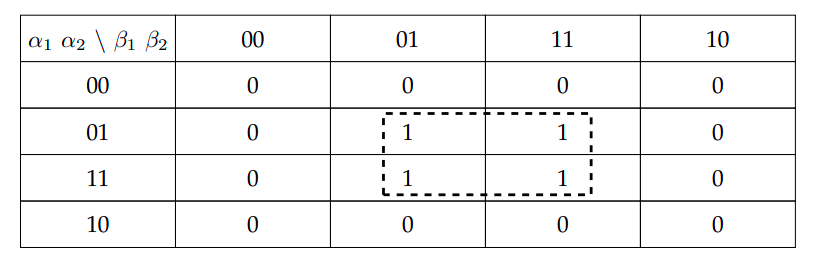}}}
}}
\eea

- $2_3$ case:
\bea &&
{\textrm{
{\centerline{\includegraphics[width=0.6\textwidth]{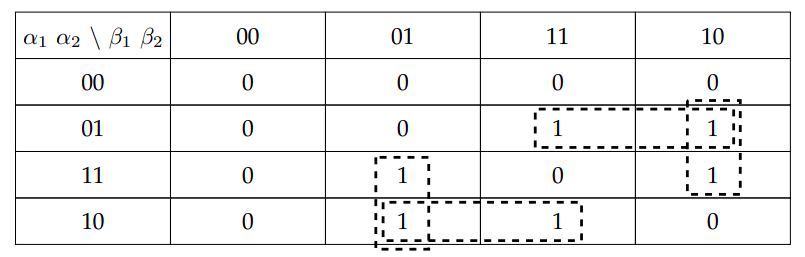}}}
}}
\eea

- $2_4$ case:
\bea \label{kar2s4}&&
{\textrm{
{\centerline{\includegraphics[width=0.6\textwidth]{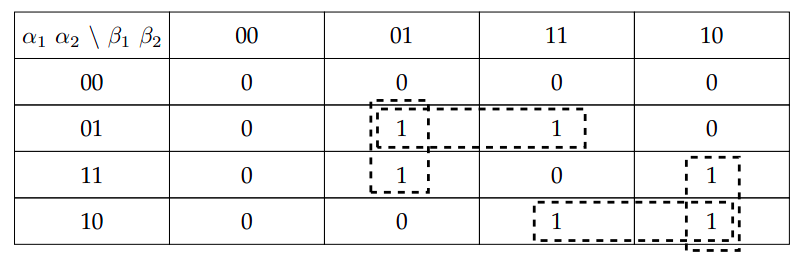}}}
}}
\eea

The $1$ entries in the above tables are encircled to indicate how they are grouped together in a Karnaugh map. 
The horizontal/vertical neighbouring $1$ entries are grouped, when possible, in even numbers (the possible exception is an isolated ``$1$" in a table). After grouping the entries in pairs, one checks which bit (they are $\alpha_1,\alpha_2,\beta_1,\beta_2$ in the above tables) varies in passing from one encircled row (column) to another row (column). Let us consider, as an example, in table
(\ref{kar2s4}) the encircled row from $\alpha_1=0, ~\alpha_2=1$. The $\beta_2$ bit remains constant ($\beta_2=1$),
while $\beta_1$ changes from $0$ to $1$. The non-varying bits in this encircled row are $\alpha_1,\alpha_2,\beta_2$.
The Karnaugh map associates with this row the following $mod ~2$ value obtained from the non-varying bits:
\bea
&{\overline\alpha}_1\cdot\alpha_2\cdot  \beta_2.&
\eea
The bar over $\alpha_1$ denotes the ${\textrm{NOT}}$ operation; it is due to the fact that, in the encircled row,
$\alpha_1=0$ (while $\alpha_2=\beta_2=1$). The different encircled pairs are added with the ${\textrm{OR}}$ operation. As a result, the (\ref{kar2s4}) table is encoded in the $mod~2$ equality
\bea
\langle\alpha,\beta\rangle &=& {\overline \alpha}_1\cdot\alpha_2\cdot\beta_2+\alpha_1\cdot {\overline\alpha}_2\cdot \beta_1+\alpha_2\cdot{\overline\beta}_1\cdot\beta_2+\alpha_1\cdot \beta_1\cdot{\overline\beta}_2.
\eea
Further simplifications are allowed by grouping $\langle\alpha,\beta\rangle$ as
\bea
\langle\alpha,\beta\rangle &=& \alpha_2\cdot\beta_2\cdot({\overline\alpha}_1+{\overline\beta}_1)+ \alpha_1\cdot\beta_1\cdot ({\overline\alpha}_2+{\overline\beta}_2).
\eea
The next step makes use of the De Morgan's theorem (i.e., ${\overline a}+{\overline b} = {\overline{a\cdot b}}$), so that we can write
\bea
\langle\alpha,\beta\rangle &=& \alpha_2\cdot\beta_2\cdot({\overline{\alpha_1\cdot\beta_1}}) +\alpha_1\cdot\beta_1\cdot({\overline{\alpha_2\cdot\beta_2}}).
\eea
The final simplification makes use of the ${\textrm{XOR}}$ operation which satisfies, $mod ~2$,
\bea
a\oplus b &=& a\cdot{\overline b}+ {\overline a}\cdot b.
\eea
The (\ref{kar2s4}) table is then rendered as
\bea\label{simple}
\langle\alpha,\beta\rangle &=& (\alpha_1\cdot\beta_1)\oplus (\alpha_2\cdot\beta_2).
\eea
By using the standard graphic presentation of the logic gates,
the simplification leading to (\ref{simple}) can be expressed as
\bea 
&
{{\includegraphics[width=0.8\textwidth]{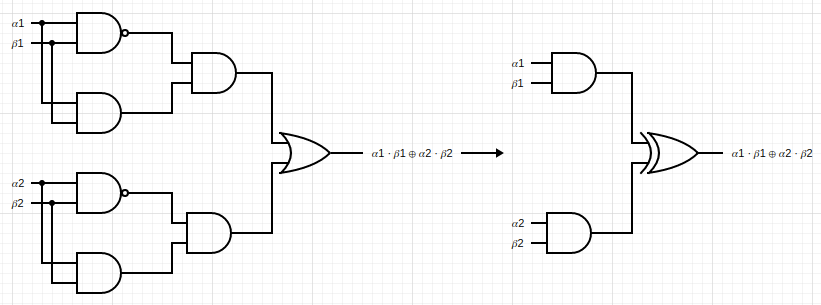}}}&
\eea
One should note that the symbol of the ${\textrm{NAND}}$ operation (the combination of ${\textrm{NOT}}$ with ${\textrm{AND}}$ leading to $a,b\mapsto {\overline{a\cdot b}}$) enters the graph on the left.\par
The above construction can be repeated for the other $n=2$ (and also  $n>2$) tables.
In the Boolean graphical presentation, the $\langle\alpha,\beta\rangle$ scalar products of the $2_2,~2_3,~2_4$ cases
are given by
\bea
\qquad 2_2: \quad \langle\alpha,\beta\rangle =\alpha_2\cdot\beta_2\quad  &\Rightarrow&
\begin{array}{c}\\
{{\includegraphics[width=0.35\textwidth]{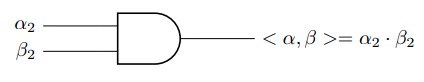}}}\\
\end{array}\qquad\quad\qquad
\eea
\bea
2_3: \quad \langle\alpha,\beta\rangle =(\alpha_1\cdot\beta_2)\oplus (\alpha_2\cdot\beta_1) &\Rightarrow&
\begin{array}{c}\\
{{\includegraphics[width=0.55\textwidth]{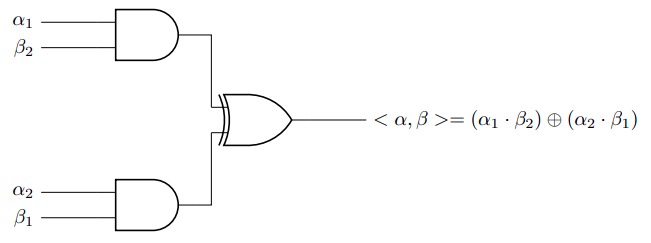}}}\\
\end{array}\qquad\quad
\eea
\bea
2_4: \quad \langle\alpha,\beta\rangle =(\alpha_1\cdot\beta_1)\oplus (\alpha_2\cdot\beta_2) &\Rightarrow&
\begin{array}{c}\\
{{\includegraphics[width=0.55\textwidth]{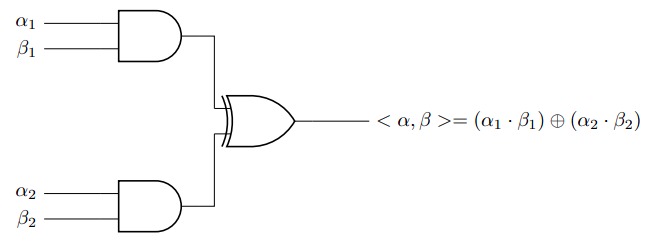}}}\\
\end{array}\qquad\quad
\eea
In terms of the Boolean logic operators, the $\langle\alpha,\beta\rangle$ scalar products of the five $3$-bit cases
(\ref{inequivalent3}) are expressed as
\bea
3_1: && 
\langle\alpha,\beta\rangle=0,\nonumber\\
3_2: && 
\langle\alpha,\beta\rangle=(\alpha_1\cdot\beta_2)\oplus(\alpha_2\cdot\beta_1),\nonumber\\
3_3: && 
\langle\alpha,\beta\rangle=\alpha_1\cdot\beta_1,\nonumber\\
3_4: && 
\langle\alpha,\beta\rangle=(\alpha_1\cdot\beta_1)\oplus(\alpha_2\cdot\beta_2),\nonumber\\
3_5: && 
\langle\alpha,\beta\rangle=(\alpha_1\cdot\beta_1)\oplus(\alpha_2\cdot\beta_2)\oplus(\alpha_3\cdot\beta_3).
\eea
The extension of this presentation to the $n$-bit scalar products with $n>3$ is straightforward.

\renewcommand{\theequation}{C.\arabic{equation}}
  
\section{Appendix C: the graded Lie (super)algebras of (split-)quaternions}

As a propaedeutic to the construction and classification in Appendix {\bf E} of the $3$-bit compatible ${\mathbb Z}_2^3$-graded Lie (super)algebras of biquaternions, we discuss in detail the recovering of the $2$-bit, ${\mathbb Z}_2^2$-graded compatible Lie (super)algebras over the reals induced by quaternions and split-quaternions.\par
Quaternions and split-quaternions are obtained from an $\varepsilon$-dependent Cayley-Dickson's doubling of the complex numbers (for $\varepsilon =\pm 1$); the sign assignment $\varepsilon = -1$ produces the division algebra of the quaternions, while the $\varepsilon =+1$ value produces its split version (see \cite{{split,ktsplit}} for details of the construction). \par
The four generators of the quaternions will be denoted as $e_0, e_i$, while the four generators of the
split-quaternions will be denoted as ${\widetilde e}_0, {\widetilde e}_i$. In both cases $i=1,2,3$.
The generators $e_0, {\widetilde e}_0$ are the respective identity elements. The ${\mathbb Z}_2^2$ multiplicative grading of quaternions and split-quaternions is preserved by faithful $4\times 4$ real matrix representations.  \par
Without loss of generality, in terms of the $2\times 2$ real matrices denoted as
{\small{\bea\label{letters}
&I=\left(\begin{array}{cc}1&0\\0&1\end{array}\right),\qquad X=\left(\begin{array}{cc}1&0\\0&-1\end{array}\right),\qquad
Y=\left(\begin{array}{cc}0&1\\1&0\end{array}\right),\qquad A=\left(\begin{array}{cc}0&1\\-1&0\end{array}\right),\qquad&
\eea
}}
 we can set for the quaternions:
\bea
&e_0 = I\otimes I, \quad e_1= A\otimes I,\quad e_2 = Y\otimes A, \quad e_3 = X\otimes A, &
\eea
so that
{\footnotesize{
\bea
&e_0=\left(\begin{array}{cccc}1&0&0&0\\0&1&0&0\\0&0&1&0\\0&0&0&1\end{array}\right),~~
e_1=\left(\begin{array}{cccc}0&0&1&0\\0&0&0&1\\-1&0&0&0\\0&-1&0&0\end{array}\right),~~
e_2=\left(\begin{array}{cccc}0&0&0&1\\0&0&-1&0\\0&1&0&0\\-1&0&0&0\end{array}\right),~~
e_3=\left(\begin{array}{cccc}0&1&0&0\\-1&0&0&0\\0&0&0&-1\\0&0&1&0\end{array}\right)
&\nonumber
\eea}}
\bea
&&
\eea
and, for the split-quaternions:
\bea
&{\widetilde e}_0 = I\otimes I, \quad {\widetilde e}_1= A\otimes I,\quad {\widetilde e}_2 = Y\otimes Y, \quad {\widetilde e}_3 = X\otimes Y, &
\eea
so that
{\footnotesize{
\bea
&{\widetilde e}_0=\left(\begin{array}{cccc}1&0&0&0\\0&1&0&0\\0&0&1&0\\0&0&0&1\end{array}\right),\quad
{\widetilde e}_1=\left(\begin{array}{cccc}0&0&1&0\\0&0&0&1\\-1&0&0&0\\0&-1&0&0\end{array}\right),\quad
{\widetilde e}_2=\left(\begin{array}{cccc}0&0&0&1\\0&0&1&0\\0&1&0&0\\1&0&0&0\end{array}\right),\quad
{\widetilde e}_3=\left(\begin{array}{cccc}0&1&0&0\\1&0&0&0\\0&0&0&-1\\0&0&-1&0\end{array}\right).
&\nonumber
\eea}}
\bea\label{split}
&&
\eea
The identity operators $e_0, {\widetilde e}_0$ are always associated with the $2$-bit $00$-graded sector,
while the remaining operators are accommodated into the $10$, $01$ and $11$-graded sectors. The $4$ compatible
graded Lie (super)algebras (denoted as $2_1, 2_2, 2_3, 2_4$) are given in Appendix ${\bf A}$, formula (\ref{2cases}). \par
~\\
{\it Remark}:
we point out that in cases $2_1$ and $2_3$ the graded sectors $10,01,11$ are all on equal footing. On the other hand, in cases
$2_2$ and $2_4$, the $11$-graded sector is singled out with respect to the $10$ and $01$ graded sectors. Indeed,  the $11$ sector of the $2_2$ case corresponds to a boson, while  $10$ and $01$ correspond to fermions; in the $2_4$ case $11$ corresponds to an ``exotic boson" (see the discussion in \cite{kuto}), while $10$ and $01$ correspond to parafermions.\par
~\par
The above remark is irrelevant for the graded (super)algebras induced by the quaternions. This is so because the three imaginary quaternions $e_1,e_2,e_3$ are on equal footing and can be interchanged. \par
On the other hand the remark becomes relevant in the construction of the graded (super)algebras induced by the split-quaternions; in this case one of the generators (given  in formula (\ref{split}) by ${\widetilde e}_1$) is singled out with respect to ${\widetilde e}_2, {\widetilde e}_3$:  by noticing that 
${\widetilde e}_1^2=-{\widetilde e}_0$, while ${\widetilde e}_2^2={\widetilde e}_3^2={\widetilde e}_0$,
we can state that ${\widetilde e}_1$ is a ``marked" generator. \par
The above observation implies that we recover $4$ inequivalent graded Lie (super)algebras from quaternions and $6$ inequivalent graded Lie (super)algebras from split-quaternions. These graded Lie (super)algebras are defined by the respective sets of (anti)commutators.
\par
~\par
The $4$ inequivalent quaternionic graded Lie (super)algebras will be denoted as ${\mathfrak q}_1$, ${\mathfrak q}_2$, ${\mathfrak q}_3$, ${\mathfrak q}_4$. Their presentation is as follows:
\par
~\par
${\mathfrak q}_1$ from $2_1$ with sectors assignments $e_0\in [00],~ e_{1}\in [10], ~e_2\in [01], ~e_3\in [11]$, so that
\bea
&[e_0,e_1]=[e_0,e_2]=[e_0,e_3]=0, \quad [e_1,e_2]=2e_3,\quad [e_2,e_3]=2e_1,\quad [e_3,e_1]=2e_2;&
\eea

${\mathfrak q}_2$ from $2_2$ with sectors assignments $e_0\in [00],~ e_{1}\in [10], ~e_2\in [01], ~e_3\in [11]$, so that
\bea
&[e_0,e_1]=[e_0,e_2]=[e_0,e_3]=0, \quad [e_1,e_3]=-2e_2,\quad [e_2,e_3]=2e_1,&\nonumber\\
&\{e_1,e_1\}=\{e_2,e_2\}=-2e_0, \quad \{e_1,e_2\}=0;&
\eea

${\mathfrak q}_3$ from $2_3$ with sectors assignments $e_0\in [00],~ e_{1}\in [10], ~e_2\in [01], ~e_3\in [11]$, so that
\bea
&[e_0,e_1]=[e_0,e_2]=[e_0,e_3]=0, \quad \{e_1,e_2\}=\{e_2,e_3\}=\{e_3,e_1\}=0;&
\eea

${\mathfrak q}_4$ from $2_4$ with sectors assignments $e_0\in [00],~ e_{1}\in [10], ~e_2\in [01], ~e_3\in [11]$, so that
\bea
&[e_0,e_1]=[e_0,e_2]=[e_0,e_3]=0, \quad [e_1,e_2]=2e_3,&\nonumber\\
&\{e_1,e_1\}=\{e_2,e_2\}=-2e_0, \quad \{e_1,e_3\}=\{e_2,e_3\}=0.&
\eea
~\\
{\it Comment}:  ${\mathfrak q}_3$  and  ${\mathfrak q}_4$ enter the \cite{kuto} classification of minimal ${\mathbb Z}_2^2$-graded Lie (super)algebras \\(${\mathfrak q}_3$ corresponds to the algebra $A7$ and
${\mathfrak q}_4$ to the superalgebra $S10_{\varepsilon=+1}$).
~\par
~\par
The $6$ inequivalent split-quaternionic graded Lie (super)algebras will be denoted as ${\widetilde {\mathfrak q}}_1$, ${\widetilde {\mathfrak q}}_{2\alpha}$, ${\widetilde {\mathfrak q}}_{2\beta}$, ${\widetilde {\mathfrak q}}_{3}$, ${\widetilde {\mathfrak q}}_{4\alpha}$, ${\widetilde {\mathfrak q}}_{4\beta}$. Their presentation is as follows:
\par
~\par
${\widetilde {\mathfrak q}}_{1}$ from $2_1$ with sectors assignments ${\widetilde e}_0\in [00],~{\widetilde  e}_{1}\in [10], ~{\widetilde e}_2\in [01], ~{\widetilde e}_3\in [11]$, so that
\bea
&[{\widetilde e}_0,{\widetilde e}_1]=[{\widetilde e}_0,{\widetilde e}_2]=[{\widetilde e}_0,{\widetilde e}_3]=0, \quad [{\widetilde e}_1,{\widetilde e}_2]=2{\widetilde e}_3,\quad [{\widetilde e}_2,{\widetilde e}_3]=-2{\widetilde e}_1,\quad [{\widetilde e}_3,{\widetilde e}_1]=2{\widetilde e}_2;&
\eea

${\widetilde {\mathfrak q}}_{2\alpha}$ from $2_2$ with sectors assignments ${\widetilde e}_0\in [00],~ {\widetilde e}_{1}\in [10], ~{\widetilde e}_2\in [01], ~{\widetilde e}_3\in [11]$, so that
\bea
&[{\widetilde e}_0,{\widetilde e}_1]=[{\widetilde e}_0,{\widetilde e}_2]=[{\widetilde e}_0,{\widetilde e}_3]=0, \quad [{\widetilde e}_1,{\widetilde e}_3]=-2{\widetilde e}_2,\quad [{\widetilde e}_2,{\widetilde e}_3]=-2{\widetilde e}_1,&\nonumber\\
&\{{\widetilde e}_1,{\widetilde e}_1\}=-2{\widetilde e}_0,\quad \{{\widetilde e}_2,{\widetilde e}_2\}=2{\widetilde e}_0, \quad \{{\widetilde e}_1,{\widetilde e}_2\}=0;&
\eea

${\widetilde {\mathfrak q}}_{2\beta}$ from $2_2$ with sectors assignments ${\widetilde e}_0\in [00],~ {\widetilde e}_{1}\in [11], ~{\widetilde e}_2\in [10], ~{\widetilde e}_3\in [01]$, so that
\bea
&[{\widetilde e}_0,{\widetilde e}_1]=[{\widetilde e}_0,{\widetilde e}_2]=[{\widetilde e}_0,{\widetilde e}_3]=0, \quad [{\widetilde e}_1,{\widetilde e}_2]=2{\widetilde e}_3,\quad [{\widetilde e}_1,{\widetilde e}_3]=-2{\widetilde e}_2,&\nonumber\\
&\{{\widetilde e}_2,{\widetilde e}_2\}=\{{\widetilde e}_3,{\widetilde e}_3\}=2{\widetilde e}_0, \quad \{{\widetilde e}_2,{\widetilde e}_3\}=0;&
\eea

${\widetilde {\mathfrak q}}_3$ from $2_3$ with sectors assignments ${\widetilde e}_0\in [00],~{\widetilde e}_{1}\in [10], ~{\widetilde e}_2\in [01], ~{\widetilde e}_3\in [11]$, so that
\bea
&[{\widetilde e}_0,{\widetilde e}_1]=[{\widetilde e}_0,{\widetilde e}_2]=[{\widetilde e}_0,{\widetilde e}_3]=0, \quad 
\{{\widetilde e}_1,{\widetilde e}_2\}=\{{\widetilde e}_2,{\widetilde e}_3\}=\{{\widetilde e}_3,{\widetilde e}_1\}=0;&
\eea

${\widetilde{\mathfrak q}}_{4\alpha}$ from $2_4$ with sectors assignments ${\widetilde e}_0\in [00],~ {\widetilde e}_{1}\in [10], ~{\widetilde e}_2\in [01], ~{\widetilde e}_3\in [11]$, so that
\bea
&[{\widetilde e}_0,{\widetilde e}_1]=[{\widetilde e}_0,{\widetilde e}_2]=[{\widetilde e}_0,{\widetilde e}_3]=0, \quad [{\widetilde e}_1,{\widetilde e}_2]=2{\widetilde e}_3,&\nonumber\\
&\{{\widetilde e}_1,{\widetilde e}_1\}=-2{\widetilde e}_0,\quad \{{\widetilde e}_2,{\widetilde e}_2\}=2{\widetilde e}_0, \quad \{{\widetilde e}_1,{\widetilde e}_3\}=\{{\widetilde e}_2,{\widetilde e}_3\}=0;&
\eea

${\widetilde{\mathfrak q}}_{4\beta}$ from $2_4$ with sectors assignments ${\widetilde e}_0\in [00],~ {\widetilde e}_{1}\in [11], ~{\widetilde e}_2\in [10], ~{\widetilde e}_3\in [01]$, so that
\bea
&[{\widetilde e}_0,{\widetilde e}_1]=[{\widetilde e}_0,{\widetilde e}_2]=[{\widetilde e}_0,{\widetilde e}_3]=0, \quad [{\widetilde e}_2,{\widetilde e}_3]=-2{\widetilde e}_1,&\nonumber\\
&\{{\widetilde e}_2,{\widetilde e}_2\}= \{{\widetilde e}_3,{\widetilde e}_3\}=2{\widetilde e}_0, \quad \{{\widetilde e}_1,{\widetilde e}_2\}=\{{\widetilde e}_1,{\widetilde e}_3\}=0.&
\eea

~\\
{\it Comment}:  ${\widetilde {\mathfrak q}}_3,~ {\widetilde{\mathfrak q}}_{4\alpha},~{\widetilde {\mathfrak q}}_{4\beta}$   enter the \cite{kuto} classification of minimal ${\mathbb Z}_2^2$-graded Lie (super)algebras \\(${\widetilde {\mathfrak q}}_3$ corresponds to the algebra $A7$, 
${\widetilde {\mathfrak q}}_{4\alpha}$ to the superalgebra $S10_{\varepsilon=-1}$
and
${\widetilde {\mathfrak q}}_{4\beta}$ to the superalgebra $S10_{\varepsilon=+1}$).

\renewcommand{\theequation}{D.\arabic{equation}}
 
\section{Appendix D: on the assignments of ${\mathbb Z}_2^3$ graded sectors}

The combinatorics which are used to derive the $16$ inequivalent graded Lie (super)algebras of the biquaternions and the $10$ statistical transmutations of the ${\cal N}=4$ supersymmetric quantum mechanics are based on the admissible $3$-bit assignments of the ${\mathbb Z}_2^3$ graded sectors. Here we present two tables which clarify this feature. \par
The ${\mathbb Z}_2^3$ grading of $8\times 8 $ matrices implies that the ${\underline 0}=000$ (i.e., zero-vector) graded elements belong to the diagonal, while the $7$ remaining sectors are expressed in terms of any choice of three fundamental gradings $\alpha,\beta,\gamma$ according to the  following $mod ~2$ relations.  The $\ast$ symbol denotes, for each graded sector, which entries of the $8\times 8$ matrices can be nonvanishing:
{\footnotesize{\bea\label{3bit8x8}
{\underline{0}}\equiv 000: \left(\begin{array}{cccccccc}\ast&&&&&&&\\
&\ast&&&&&&\\
&&\ast&&&&&\\
&&&\ast&&&&\\
&&&&\ast&&&\\
&&&&&\ast&&\\
&&&&&&\ast&\\
&&&&&&&\ast\\
\end{array}\right),&&~~~~~~~~~{{\alpha}}: \left(\begin{array}{cccccccc}&&\ast&&&&&\\
&&&\ast&&&&\\
\ast&&&&&&&\\
&\ast&&&&&&\\
&&&&&&\ast&\\
&&&&&&&\ast\\
&&&&\ast&&&\\
&&&&&\ast&&\\
\end{array}\right),\nonumber\\
{{\beta}}: \left(\begin{array}{cccccccc}
&&&\ast&&&&\\
&&\ast&&&&&\\
&\ast&&&&&&\\
\ast&&&&&&&\\
&&&&&&&\ast\\
&&&&&&\ast&\\
&&&&&\ast&&\\
&&&&\ast&&&\\
\end{array}\right),&&~~~~~~~~~{{\gamma}}: \left(\begin{array}{cccccccc}&&&&\ast&&&\\
&&&&&\ast&&\\
&&&&&&\ast&\\
&&&&&&&\ast\\
\ast&&&&&&&\\
&\ast&&&&&&\\
&&\ast&&&&&\\
&&&\ast&&&&\\
\end{array}\right),\nonumber\\
{{\alpha}+{\beta}}: \left(\begin{array}{cccccccc}
&\ast&&&&&&\\
\ast&&&&&&&\\
&&&\ast&&&&\\
&&\ast&&&&&\\
&&&&&\ast&&\\
&&&&\ast&&&\\
&&&&&&&\ast\\
&&&&&&\ast&\\
\end{array}\right),&&~~~~~{{\alpha}+{\gamma}}: \left(\begin{array}{cccccccc}
&&&&&&\ast&\\
&&&&&&&\ast\\
&&&&\ast&&&\\
&&&&&\ast&&\\
&&\ast&&&&&\\
&&&\ast&&&&\\
\ast&&&&&&&\\
&\ast&&&&&&\\
\end{array}\right),\nonumber\\
{{\beta}+{\gamma}}: \left(\begin{array}{cccccccc}
&&&&&&&\ast\\
&&&&&&\ast&\\
&&&&&\ast&&\\
&&&&\ast&&&\\
&&&\ast&&&&\\
&&\ast&&&&&\\
&\ast&&&&&&\\
\ast&&&&&&&\\
\end{array}\right),&&{{\alpha}+{\beta}+{\gamma}}: \left(\begin{array}{cccccccc}&&&&&\ast&&\\
&&&&\ast&&&\\
&&&&&&&\ast\\
&&&&&&\ast&\\
&\ast&&&&&&\\
\ast&&&&&&&\\
&&&\ast&&&&\\
&&\ast&&&&&\\
\end{array}\right).
\eea}}
In another schematical presentation, the $3$-bit nonzero vectors can be assigned to the $7$ vertices of the Fano's plane. For each one of the $7$ edges, the sum $mod ~2$ of the vectors of any two vertices gives the $3$-bit vector of the third vertex lying on the edge:

{\centerline{\includegraphics[width=0.8\textwidth]{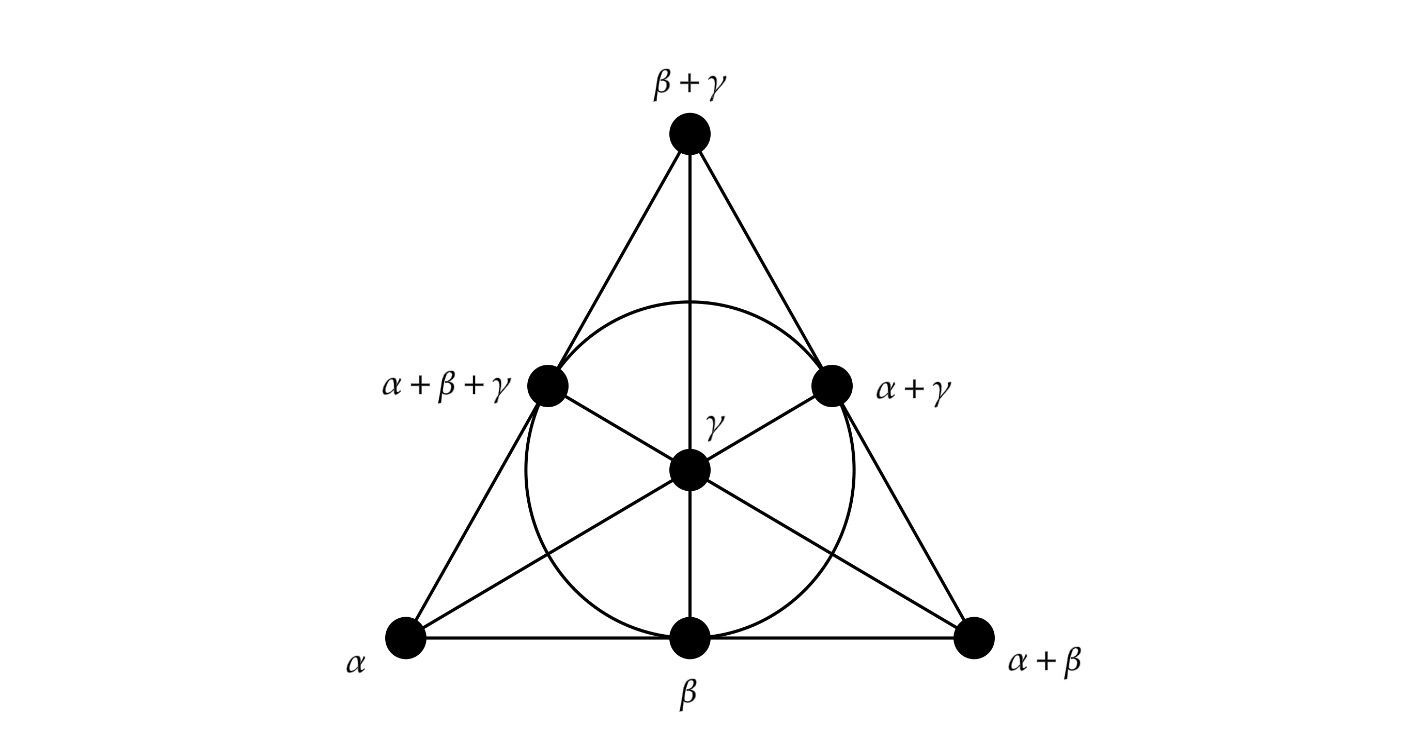}}}
\bea\label{fano}&&\eea

\renewcommand{\theequation}{E.\arabic{equation}}
\section{Appendix E: the graded Lie (super)algebras of biquaternions}

The algebra ${\mathbb{ H}}_B$ of the biquaternions can be regarded as the tensor product ${\mathbb C}\times {\mathbb H}$ of the complex numbers with the quaternions (over the reals). It admits $8$ generators which can be accommodated into a
${\mathbb Z}_2^3$ multiplicative grading (therefore, a $3$-bit assignment). A faithful $8\times 8$ matrix representation which respects the grading can be expressed by tensoring the $2\times 2$ matrices $I,X,Y,A$ introduced in (\ref{letters}). We can set
\bea
&f_0=I\otimes I\otimes I,~\quad f_1=I\otimes A\otimes I,~\quad f_2=I\otimes Y \otimes A,~ \quad f_3=I\otimes X\otimes A,&\nonumber\\
&g_0=A\otimes I\otimes I,\quad g_1=A\otimes A\otimes I,\quad g_2=A\otimes Y \otimes A, \quad g_3=A\otimes X\otimes A.&
\eea
By taking into account the definitions of the $I,X,Y,A$ matrices, the following multiplication table can be derived:
\bea\label{biquatmult}
&&
\begin{array}{c|ccccccccc}&f_0&f_1&f_2&f_3&g_0&g_1&g_2&g_3\\ \hline 
f_0&f_0&f_1&f_2&f_3&g_0&g_1&g_2&g_3\\
f_1&f_1&-f_0&f_3&-f_2&g_1&-g_0&g_3&-g_2\\
f_2&f_2&-f_3&-f_0&f_1&g_2&-g_3&-g_0&g_1\\
f_3&f_3&f_2&-f_1&-f_0&g_3&g_2&-g_1&-g_0\\
g_0&g_0&g_1&g_2&g_3&-f_0&-f_1&-f_2&-f_3\\
g_1&g_1&-g_0&g_3&-g_2&-f_1&f_0&-f_3&f_2\\
g_2&g_2&-g_3&-g_0&g_1&-f_2&f_3&f_0&-f_1\\
g_3&g_3&g_2&-g_1&-g_0&-f_3&-f_2&f_1&f_0
\end{array}
\eea
The entries in the above table give the left action of a row generator on a column generator.\par
The generator $f_0$ corresponds to the identity operator, while $g_0$ corresponds to the imaginary unit
($g_0$ commutes with all other generators). The quaternionic subalgebra is spanned by $f_0,f_1,f_2,f_3$.\par
The $7$ extra generators, besides the identity, fall into $3$ different  classes of equivalence ($A,B,C$) which, extending the analysis of Appendix {\bf C}, ``mark" them.  The classes are determined by the operations which leave invariant the  multiplication table (\ref{biquatmult}); they are the permutations of the generators in a given class
possibly  coupled with  a $\pm 1$ sign normalization of the generators. We have
\bea\label{markedbiquat}
&f_1,f_2,f_3\in A, \qquad g_0\in B,\qquad g_1,g_2,g_3\in C.&
\eea
As usual, the identity operator $f_0$ is assigned to the $000$-grading, so that
\bea
[f_0]&= &000.
\eea
The $3$-bit grading assignments of the remaining generators are recovered from the grading assignments of $f_1,f_2$ and $g_0$; let's express $[f_1]=\alpha, ~[f_2]=\beta$, $[g_0]=\gamma$. The ${\mathbb Z}_2^3$-grading consistency requires
\bea
&\begin{array}{llll} [f_0] = 000,\quad&[f_1]= \alpha,\quad & [f_2]= \beta,\quad & [f_3] = \alpha+\beta,\\
\relax [g_0]= \gamma,\quad&[g_1] = \alpha+\gamma,\quad & [g_2]=\beta+\gamma, \quad &[g_3]=\alpha+\beta+\gamma,
\end{array}&
\eea
where, in the above formulas, the sums are $mod$ $2$.\par
The inequivalent graded Lie (super)algebras which are compatible with the multiplicative ${\mathbb Z}_2^3$ grading assignments are the $5$ ones listed in (\ref{inequivalent3}) and named as $3_1, 3_2, 3_3, 3_4, 3_5$.
\par
We can extend the analysis done in Appendix {\bf C} for the $2$-bit assignments and present which graded sectors (besides $000$) are singled out in each one of the $5$ tables given in (\ref{inequivalent3}). The results are the following:\par
~\par
{\it $3_1$ case} - the seven sectors $100,010,001,110,101,011,111$ are all on equal footing and correspond to bosonic particles;\par
{\it $3_2$ case} - the sector $001$ is singled out because it corresponds to a bosonic particle, while the six remaining sectors $100,010,110,101,011,111$ correspond to parabosons and are on equal footing;\par
{\it $3_3$ case} - the three sectors $010,001,011$ correspond to bosonic particles, while the remaining four sectors $100,110,101,111$ correspond to fermions;\par
{\it $3_4$ case} - the $7$ graded sectors are divided into $3$ types: $001$ corresponds to a bosonic particle, $110$ and $111$ to parabosons, $100, 010,101,011$ to parafermions;\par
{\it $3_5$ case} - the three sectors $110,101,011$ correspond to parabosons, while the remaining four sectors $100,010,001,111$ correspond to parafermions.
\bea&&
\eea\par
~\par
Therefore, the $7$ extra graded sectors (besides $000$) are split into the following classes:
\bea
&3_1: ~7,\qquad 3_2: ~1+6,\qquad 3_3: ~3+4,\qquad 3_4: ~1+2+4,\qquad 3_5:~ 3+4.&
\eea
Inequivalent graded Lie (super)algebras are obtained by assigning the $7$ marked generators (\ref{markedbiquat})
belonging to $A,B,C$ to the above different classes of graded sectors. The computation gives, for each one of the $5$ cases, the following subcases:\\
~\par
{\it $3_1$ case} - $1$ graded Lie algebra which can be presented as 
\bea&{\textrm{$\qquad\qquad 3_{1,i}$:~~~ $\alpha =100,~\beta=010,~\gamma=001$}}&
\eea
 (all other presentations are equivalent);
\par
~
\par
{\it $3_2$ case} - $3$ inequivalent graded Lie algebras which can be presented as
\bea
&{\textrm{$\qquad\qquad 3_{2,i}$:~~~ $\alpha = 100,~\beta=010,~\gamma={\underline {001}}$,}}&\nonumber\\
&{\textrm{$\qquad\qquad 3_{2,ii}$:~~ $\alpha = {\underline{001}},~\beta=010,~\gamma={{100}}$,}}&\nonumber\\
&{\textrm{$\qquad\qquad  3_{2,iii}$:~ $\alpha = 111,~\beta=010,~\gamma={{100}}$}}&
\eea
(the singled-out sector $001$ has been underlined; in the $3_{2,iii}$ case it is assigned to the $C$ generator
 $g_3$ whose grading is $\alpha+\beta+\gamma$);\par
~\par
{\it $3_3$ case} - $3$ inequivalent graded Lie superalgebras which, taking into account the
$3+4$ ($010,001,011/100,110,101,111$) split, can be presented as
\bea&{\textrm{
$\qquad\qquad 3_{3,i}$:~~~ $\alpha = 010,~\beta=001,~\gamma={{111}}$,}}&\nonumber\\
&{\textrm{$\qquad\qquad 3_{3,ii}$:~~ $\alpha = {{010}},~\beta=100,~\gamma={{111}}$,}}&\nonumber\\
&{\textrm{$\qquad\qquad  3_{3,iii}$:~ $\alpha = 010,~\beta=100,~\gamma={{011}}$}}&
\eea
(either the three $A$ generators $f_1,f_2,f_3$ can all be accommodated in the $010,001,011$ sectors, or just one of them; the last scenario  leaves two  inequivalent possibilities for the $\gamma$ grading of $g_0$: either it is accommodated into the $010,001,011$ sectors or not);\par
~\par
{\it $3_4$ case} - $6$ inequivalent graded Lie superalgebras which, taking into account the
$1+2+4$ ($001/110,111/100,010,101,011$) split, can be presented as
\bea
&{\textrm{$\qquad\qquad 3_{4,i}$:~~~ $\alpha = 001,~\beta=110,~\gamma={{100}}$,}}&\nonumber\\
&{\textrm{$\qquad\qquad 3_{4,ii}$:~~ $\alpha = {{001}},~\beta=100,~\gamma={{110}}$,}}&\nonumber\\
&{\textrm{$\qquad\qquad  3_{4,iii}$:~ $\alpha = 001,~\beta=100,~\gamma={{011}}$,}}&\nonumber\\
&{\textrm{$\qquad\qquad 3_{4,iv}$:~~~$\alpha = 110,~\beta=100,~\gamma={{001}}$,}}&\nonumber\\
&{\textrm{$\qquad\qquad 3_{4,v}$:~~ $\alpha = {{110}},~\beta=100,~\gamma={{111}}$,}}&\nonumber\\
&{\textrm{$\qquad\qquad  3_{4,vi}$:~ $\alpha = 110,~\beta=100,~\gamma={{011}}$}}&
\eea
(let us denote as $a,b,c$ the respective graded sectors of the $1+2+4$ decomposition; either the three $A$ generators $f_1,f_2,f_3$ are all accommodated in the $a,b$ sectors, implying that $\gamma$ must be accommodated into $c$ or, alternatively, only one of the three $A$ generators can be accommodated in $a$ or $b$. 
If this generator is accommodated in $a$, it leaves two possibilities for $\gamma$ entering $b$ or $c$. If this generator is accommodated in $c$, $\gamma$  has three  inequivalent possibilities: it can enter $a$, $b$ or $c$).\par
~\par
{\it $3_5$ case} - $3$ inequivalent graded Lie superalgebras which, taking into account the
$3+4$ ($110,101,011/100,010,001,111$) split, can be presented as
\bea
&{\textrm{$\qquad\qquad 3_{5,i}$:~~~ $\alpha = 110,~\beta=101,~\gamma={{100}}$,}}&\nonumber\\
&{\textrm{$\qquad\qquad 3_{5,ii}$:~~ $\alpha = {{110}},~\beta=100,~\gamma={{111}}$,}}&\nonumber\\
&{\textrm{$\qquad\qquad  3_{5,iii}$:~ $\alpha = 110,~\beta=100,~\gamma={{011}}$}}&
\eea
(one repeats the analysis of the $3_3$ case which is also based on a $3+4$ split of the graded sectors).\par
~\par
Therefore, the total number $n_B$ of inequivalent graded Lie (super)algebras which are compatible with the 
${\mathbb Z}_2^3$ multiplicative grading of the biquaternions is
\bea
&n_B= 1+3+3+6+3 = 16.&
\eea
To save space, we limit here to present the sets of defining (anti)commutators for the three inequivalent superalgebras recovered from the $3_5$ case.  All these three subcases are defined in terms of $32$
(anti)commutators. The $7$ common ones are
\bea
&[f_0, z] =0\quad {\textrm{for any $z\in {\mathbb H}_B$}}.&
\eea
The remaining (anti)commutators are the following.\par
~\\
{\it{For $3_{5,i}$}}, $9$ of the remaining defining brackets are vanishing,
\bea
&[g_0,g_1]=[g_0,g_2]=[g_0,f_3]=\{g_1,f_3\}=\{g_2,f_3\}=\{f_1,f_2\}=\{f_1,f_3\}=\{f_2,f_3\}=[f_3,g_3]=0&\nonumber
\eea
and $16$ are nonvanishing:
\bea
&\{g_0,g_0\}=-2f_0,~ \{g_0,f_1\}=2g_1,~\{g_0,f_2\}=2g_2,~\{g_0,g_3\}= - 2f_3,~ \{g_1,g_1\}= 2f_0,&\nonumber\\
& ~[g_1,g_2]= -2f_3,~\{g_1,f_1\}= -2g_0,~[g_1,f_2]=2g_3,~\{g_1,g_3\}=-2f_2,~\{g_2,g_2\}=2f_0,&\nonumber\\
&[g_2,f_1]=-2g_3,~\{g_2,f_2\}=-2g_0,~\{g_2,g_3\}=-2f_1,~ [f_1,g_3]=-2g_2,~[f_2,g_3]=2g_1,&\nonumber\\&~\{g_3,g_3\}=2f_0.&
\eea
~\\
{For $3_{5,ii}$}, $15$ of the remaining defining brackets are vanishing,
\bea
&\{f_2,f_1\}=\{f_2,g_3\}=[f_2,g_2]=\{f_3,f_1\}=[f_3,g_3]=\{f_3,g_2\}=[g_1,f_1]=\{g_1,g_3\}=\nonumber\\
&=\{g_1,g_2\}=\{f_1,g_3\}=\{f_1,g_2\}=[f_1,g_0]=\{g_3,g_2\}=[g_3,g_0]=[g_2,g_0]=0&\nonumber
\eea
and $10$ are nonvanishing:
\bea
&\{f_2,f_2\}=-2f_0,~[f_2,f_3]=2f_1,~[f_2,g_1]=-2g_3,~\{f_2,g_0\}=2g_2,~\{f_3,f_3\}=-2f_0,&\nonumber\\
&[f_3,g_1]=2g_2,~\{f_3,g_0\}=2g_3,~\{g_1,g_1\}=2f_0,~\{g_1,g_0\}=-2f_1,~\{g_0,g_0\}=-2f_0.&\eea
{\it {For $3_{5,iii}$}}, $9$ of the remaining defining brackets are vanishing,
\bea
&\{f_2,f_1\}=\{f_2,g_1\}=[f_2,g_0]=[f_3,g_3]=\{f_3,f_1\}=[f_3,g_2]=\{g_3,g_1\}=\{g_3,g_2\}=[g_0,g_2]=0&\nonumber
\eea
and $16$ are nonvanishing:
\bea
&\{f_2,f_2\}=-2f_0,~[f_2,f_3]=2f_1,~[f_2,g_3]=2g_1,~\{f_2,g_2\}=-2g_0, \{f_3,f_3\}=-2f_0,&\nonumber\\
&[f_3,g_1]=2g_2,~\{f_3,g_0\}=2g_3,~\{g_3,g_3\}=2f_0,~[g_3,f_1]=2g_2,~\{g_3,g_0\}=-2f_3,&\nonumber\\
&\{f_1,g_1\}=-2g_0,~\{f_1,g_0\}=2g_1,~[f_1,g_2]=2g_3,~\{g_1,g_0\}=-2f_1,~[g_1,g_2]=-2f_3,&\nonumber\\
&\{g_2,g_2\}=2f_0.&
\eea

{\it Comment}: the difference between the ${3_{5,i}}$ and the $3_{5,iii}$ subcases is related to the diagonal signature of the parafermionic generators. Since $f_0$ is the identity operator, in the $3_{5,i}$ graded superalgebra case the squares of $g_0, g_1, g_2,g_3$ produce the $(-1,+1,+1,+1)$ signature, while in the $3_{5,iii}$ graded superalgebra case the squares of $f_2,f_3,g_3,g_2$ produce the $(-1,-1,+1,+1)$ signature. On the basis of this analysis $3_{5,i}$ and $3_{5,iii}$ are different real forms of a graded superalgebra.

~\par
~
\\ {\Large{\bf Acknowledgments}}
{}~\par{}~\\
We are grateful to Zhanna Kuznetsova for discussions and suggestions.\\
This work was supported by CNPq (PQ grant 308846/2021-4).


\begin{thebibliography}{99}
\bibitem{riwy1} V. Rittenberg and D. Wyler, 
{\it Generalized Superalgebras}, 
{Nucl. Phys.} {\bf B 139}, 189 (1978).
\bibitem{riwy2} V. Rittenberg and D. Wyler, 
{\it Sequences of $Z_2\otimes Z_2$ graded Lie algebras and superalgebras}, 
{J. Math. Phys.} {\bf 19}, 2193 (1978).
\bibitem{sch} M. Scheunert, 
{\it Generalized Lie algebras}, 
{J. Math. Phys.} {\bf 20}, 712 (1979).
\bibitem{kac}  V. G. Kac, {\it Lie Superalgebras}, Adv. in Math. {\bf 26}, 8 (1977).
\bibitem{brdu} A. J. Bruce and S. Duplij, {\it Double-graded supersymmetric quantum mechanics}, {J. Math. Phys.} {\bf 61}, 063503 (2020); arXiv:1904.06975 [math-ph].
\bibitem{top1} F. Toppan, {\it \zzg parastatistics in multiparticle quantum Hamiltonians}, J. Phys. A: Math. Theor. {\bf 54}, 115203 (2021); arXiv:2008.11554[hep-th].
\bibitem{maj}
S. Majid, {\it Foundations of Quantum Group Theory}, Cambridge University Press, Cambridge (1995).
\bibitem{top2} F. Toppan, {\it Inequivalent quantizations from gradings and \zzg parabosons}, J. Phys. A: Math. Theor.  {\bf 54}, 355202 (2021); arXiv:2104.09692[hep-th].
\bibitem{parasim} C. Huerta Alderete and B. M. Rodr\'\i guez-Lara, {\it Quantum simulation of driven para-Bose oscillators}, Phys. Rev. {\bf A 95}, 013820 (2017); arXiv:1609.09166[quant-ph].
\bibitem{paraexp} C. Huerta Alderete, A. M. Greene, N. H. Nguyen, Y. Zhu, B. M. Rodr\'\i guez-Lara and N. M. Linke,
{\it Experimental realization of para-particle oscillators}, arXiv:2108.05471[quant-ph].
\bibitem{luri} J. Lukierski and V. Rittenberg, {\it Color-De Sitter and Color-Conformal Superalgebras}, {{Phys. Rev.}} {\bf D 18},  385 (1978).
\bibitem{vas} M. A. Vasiliev, {\it de Sitter supergravity with positive cosmological constant and generalized Lie superalgebras}, {{Class. Quantum Grav.}} {\bf 2},  645 (1985).
\bibitem{jyw} P. D. Jarvis, M. Yang and B. G. Wybourne, {\it Generalized quasispin for supergroups}, {{J. Math. Phys.}} {\bf 28},  1192 (1987).
\bibitem{aktt1} N. Aizawa, Z. Kuznetsova, H. Tanaka and F. Toppan, {\it $ \mathbb{Z}_2 \times \mathbb{Z}_2$-graded Lie symmetries of the L\'evy-Leblond equations}, {Prog. Theor. Exp. Phys.} {\bf{2016}}, 123A01 (2016); arXiv:1609.08224[math-ph].
\bibitem{aktt2} N. Aizawa, Z. Kuznetsova, H. Tanaka and F. Toppan, {\it Generalized supersymmetry and L\'evy-Leblond equation}, in S. Duarte {et al} (eds), {Physical and Mathematical Aspects of Symmetries}, Springer, Cham, p. 79 (2017); arXiv:1609.08760[math-ph].
\bibitem{akt1} N. Aizawa, Z. Kuznetsova and F. Toppan, {\it ${\mathbb Z}_2\times{\mathbb Z}_2$-graded mechanics: the classical theory}, Eur. J. Phys. {\bf C 80}, 668 (2020); arXiv:2003.06470[hep-th].
\bibitem{brusigma} A. J. Bruce, {\it ${\mathbb Z}_2\times{\mathbb Z}_2$-graded supersymmetry: 2-d sigma models}, {{ J. Phys. A: Math. Theor. {\bf 53}, 455201 (2020)}}; arXiv:2006.08169[math-ph].
\bibitem{akt2} N. Aizawa, Z. Kuznetsova and F. Toppan, {\it ${\mathbb Z}_2\times {\mathbb Z}_2$-graded mechanics: the quantization}, Nucl. Phys. {\bf B 967}, 115426 (2021); arXiv:2005.10759[hep-th].
\bibitem{aad} N. Aizawa, K. Amakawa, S. Doi, {\it $\mathcal{N}$-Extension of double-graded supersymmetric and superconformal quantum mechanics}, J. Phys. A: Math. Theor. {\bf 53}, 065205 (2020); arXiv:1905.06548 [math-ph].
\bibitem{kuto} Z. Kuznetsova and F. Toppan, {\it 
Classification of minimal \zzg Lie (super)algebras and some applications},
J. Math. Phys. {\bf 62}, 063512 (2021); arXiv:2103.04385[math-ph].
\bibitem{que} C. Quesne, {\it Minimal bosonization of double-graded supersymmetric quantum mechanics}, Mod. Phys. Lett. {\bf A 36}, 2150238 (2021); arXiv:2108.06243 [math-ph]. 
\bibitem{pon} N. Poncin, {\it Towards integration on colored supermanifolds}, Banach Center Publication {\bf 110}, 201 (2016).
\bibitem{brusg} A. J. Bruce, {\it Is the $\mathbb{Z}_2 \times \mathbb{Z}_2$-graded sine-Gordon equation integrable?}, Nucl. Phys. {\bf B 971}, 115514 (2021); arXiv:2106.06372 [math-ph].
\bibitem{aido1} S. Doi and N. Aizawa, {\it Comments on ${\mathbb Z}_2^2$-graded supersymmetry in superfield formalism}, Nucl. Phys. {\bf B 974}, 115641 (2022); arXiv:2109.14227[math-ph].
\bibitem{aido2} N. Aizawa and S. Doi, {\it Irreducible representations of ${\mathbb Z}_2^2$-graded ${\cal N}=2$ supersymmetry algebra and ${\mathbb Z}_2^2$-graded supermechanics},  J. Math. Phys. {\bf 63}, 091704 (2022);  arXiv:2205.07263[math-ph].
\bibitem{aikt} N. Aizawa, R. Ito, Z. Kuznetsova and F. Toppan, {\it New aspects of the ${\mathbb Z}_2\times{\mathbb Z}_2$-graded $1D$ superspace: closed strings and $2D$ relativistic models}, Nucl. Phys. {\bf B 991}, 116202 (2023); arXiv:2301.06089[hep-th].
\bibitem{yaji} W. M. Yang and S. C. Jing, {\it A new kind of graded Lie algebra and parastatistical supersymmetry}, Sci. in China (Series A) {\bf 44}, 9 (2001); arXiv:math-ph/0212004. 
\bibitem{tol1} V. N. Tolstoy, {\it Once more on parastatistics}, {Phys. Part. Nucl. Lett.} {\bf{11}},  933 (2014); arXiv:1610.01628[math-ph].
\bibitem{stvj1} N. I. Stoilova and J. Van der Jeugt, {\it The ${\mathbb Z}_2\times{\mathbb Z}_2$-graded Lie superalgebra $pso(2m+1|2n)$ and new parastatistics representations},  J. Phys. {A}: Math. Theor. {\bf 51}, 135201 (2018); arXiv:1711.02136[math-ph].
\bibitem{stvj2} N. I. Stoilova and J. Van der Jeugt, \textit{The ${\mathbb{Z}}_{2}\times {\mathbb{Z}}_{2}$-graded Lie superalgebras $\mathfrak{p}\mathfrak{s}\mathfrak{o}(2n+1\vert 2n)$ and $\mathfrak{p}\mathfrak{s}\mathfrak{o}(\infty \vert \infty )$, and parastatistics Fock spaces,}  J. Phys. A: Math. Theor. \textbf{55}, 045201 (2022); arXiv:2112.12811 [math-ph]. 
\bibitem{z2z2sdiv} Z. Kuznetsova and F. Toppan, {\it Beyond the 10-fold way: 13 associative ${\mathbb Z}_2\times{\mathbb Z}_2$-graded superdivision algebras}, Adv. in Appl. Cliff. Alg. {\bf 33}, art. 24 (2023); arXiv:2112.00840[math-ph].
\bibitem{isv} P. S. Isaac, N. I. Stoilova and J. Van der Jeugt,{\it The $Z_2\times Z_2$-graded general linear Lie superalgebra}, J. Math. Phys. {\bf 61}, 011702 (2020); 
arXiv:1912.08636[math-ph].
\bibitem{stvdjclass} N. I. Stoilova and J. Van der Jeugt, {\it 
On classical $Z_2\times Z_2$-graded Lie algebras}, arXiv:2305.18604[math-ph].
\bibitem{ckp} T. Covolo, S. Kwok and N. Poncin, {\it{Differential calculus on $\mathbb{Z}_2^n$-supermanifolds}}, arXiv:1608.00949 [math.DG]. 
\bibitem{brgr1} A. J. Bruce and J. Grabowski, {\it Riemannian structures on ${\mathbb Z}_2^n$-manifolds}, 
Mathematics {\bf 8}, 1469 (2020); arXiv:2007.07666.
\bibitem{brgr2} A. J. Bruce and J. Grabowski, {\it Symplectic ${\mathbb Z}_2^n$-manifolds}, J. Geom. Mech. {\bf  13}, 285 (2021); arXiv:2103.00249[math-ph].
\bibitem{luta} R. Lu and Y. J. Tan, {\it Construction of color Lie algebras from homomorphisms of modules of Lie algebras}, J. Algebra {\bf 620}, 1 (2023). 
\bibitem{doai} S. Doi and N. Aizawa, {\it $\mathbb{Z}_2^3$-Graded extensions of Lie superalgebras and superconformal quantum mechanics}, SIGMA {\bf 17}, 071 (2021); arXiv:2103.10638 [math-ph]. 
\bibitem{aad2} N. Aizawa, K. Amakawa and S. Doi, {\it ${\mathbb Z}_2^n$-graded extensions of supersymmetric quantum mechanics}, J. Math. Phys. {\bf 61}, 052105 (2020); arXiv:1912.11195[math-ph]. 
\bibitem{kar} M. Karnaugh, {\it The Map Method for Synthesis of Combinational Logic Circuits}, Trans. of the Amer. 
Inst. of Elec. Engin., Part I: Commun. and Electr. {\bf 72}, paper 53-217, 593 (1953).
\bibitem{gre} H. S. Green, {\it A Generalized Method of Field Quantization}, Phys. Rev. {\bf 90}, 270 (1953).
\bibitem{anpo} B. Aneva and T. Popov, {\it Hopf Structure and Green Ansatz of Deformed Parastatistics Algebras},
J. Phys {\bf A}: Math. Gen. {\bf 38}, 6473 (2005); arXiv:math-ph/0412016.
\bibitem{kada} K. Kanakoglou and C. Daskaloyannis, {\it Parabosons quotients. A braided look at Green's ansatz and a generalization}, J. Math. Phys. {\bf  48}, 113516 (2007); arXiv:0901.04320[math-ph].
\bibitem{dff} V. de Alfaro, S. Fubini and G. Furlan, {\it Conformal invariance in quantum mechanics}, Nuovo Cim. {\bf A 34}, 569 (1976).
\bibitem{wit} E. Witten, {\it{Constraints on Supersymmetry Breaking}}, Nucl. Phys. {\bf B 202}, 253 (1982).
\bibitem{mar} E. Marino, {\it Quantum Field Theory Approach to Condensed Matter Physics}, Cambridge Univ. Press, Cambridge (2017).
\bibitem{pato} A. Pashnev and F. Toppan, {\it On the classification of N-extended supersymmetric quantum mechanical systems}, {J. Math. Phys.} {\bf 42}, 5257 (2001); arXiv:hep-th/0010135.
\bibitem{krt} Z. Kuznetsova, M. Rojas and F. Toppan, {\it Classification of irreps and invariants of the $N$-extended supersymmetric quantum mechanics}, JHEP  {\bf{0603}}, 098 (2006); arXiv:hep-th/0511274.
\bibitem{cht} I. E. Cunha, N. L. Holanda and F. Toppan, {\it From worldline to quantum superconformal mechanics with and without oscillatorial terms: $D(2,1;\alpha)$ and $sl(2|1)$ models}, {Phys. Rev.} {\bf{D 96}}, 065014 (2017); arXiv:1610.07205[hep-th].
\bibitem{gkt} M. Gonzales, S. Khodaee and F. Toppan, {\it On nonminimal ${\cal N}=4$ supermultiplets in $1D$ and their associated $\sigma$-models}, J. Math. Phys. {\bf 52}, 013514 (2011); arXiv:1006.4678[hep-th].
\bibitem{quasi} N. Aizawa, Z. Kuznetsova, and F. Toppan, {\it The quasi-nonassociative exceptional $F(4)$ deformed quantum oscillator}, J. Math. Phys. {\bf 59}, 022101 (2018); arXiv:1711.02923[math-ph].
\bibitem{fil} S. Fedoruk, E. Ivanov and O. Lechtenfeld, {\it Superconformal mechanics}, J. Phys. A: Math. Theor. {\bf 45}, 173001 (2012); arXiv:1112.1947[hep-th]. 
\bibitem{ackt} N. Aizawa, I. E. Cunha, Z. Kuznetsova and F. Toppan, {\it On the spectrum-generating superalgebras of the deformed one-dimensional quantum oscillators}, J. Math. Phys. {\bf 60}, 042102 (2019); arXiv:1812.00873[math-ph].
\bibitem{kit} A. Kitaev, {\it Periodic table for topological insulators and superconductors}, AIP Conf. Proceed.  {\bf 1134}, 22 (2009); arXiv:0901.2686[cond-mat.mes-hall].
\bibitem{rsfl} S. Ryu, A. P. Schnyder, A. Furusaki and A. W. W. Ludwig, {\it Topological insulators and superconductors:
tenfold way and dimensional hierarchy}, New J. Phys. {\bf 12}, 065010 (2010); arXiv:0912.2157[cond-mat.mes-hall].
\bibitem{gapa} A. Ch. Ganchev and T. D. Palev, {\it A Lie superalgebraic interpretation of the para-Bose statistics},
J. Math. Phys. {\bf 21}, 797 (1980).
\bibitem{pal} T. D. Palev, {\it Para-Bose and Para-Fermi operators as generators of orthosymplectic Lie superalgebras}, J. Math. Phys. {\bf 23}, 1100 (1982).
\bibitem{wig} E. P. Wigner, {\it Do the Equations of Motion Determine the Quantum Mechanical Equations of Motion?}, Phys. Rev. {\bf 77}, 711 (1950).
\bibitem{top0} F. Toppan, {\it Symmetries of the Schr\"{o}dinger Equation and Algebra/Superalgebra Duality}, 
J. of Phys.: Conf. Ser. {\bf 597}, 012071 (2015); arXiv:1411.7867[math-ph].
\bibitem{nie} U. Niederer, {\it The maximal kinematical invariance group of the free Schr\"odinger equation}, Helv. Phys. Acta {\bf 45}, 802 (1972).
\bibitem{nie2} U. Niederer, {\it The maximal kinematical invariance group of the harmonic oscillator}, Helv. Phys. Acta
{\bf 46}, 191 (1973).
\bibitem{ara} H. Araki, {\it On the Connection of Spin and Commutation Relations between Different Fields},
J. Math. Phys. {\bf 2}, 267 (1961).
\bibitem{split} K. McCrimmon, {\it A Taste of Jordan Algebras},  Universitext Springer, New York (2004).
\bibitem{ktsplit} Z. Kuznetsova and F. Toppan, {\it Superalgebras of (split-)division algebras and the split octonionic M-theory in $(6,5)$-signature}, arXiv:hep-th/0610122.
\end{thebibliography}
\end{document}